\RequirePackage{etex}
\documentclass{amsart}
\usepackage{amsmath, amsthm, amssymb, mathrsfs, epsfig}
\usepackage{dcpic, pictex, pinlabel, setspace, rotating}
\usepackage{array, mathdots, bbm, faktor}

\newtheorem{theorem}{Theorem}[section]

\newtheorem{conjecture}[theorem]{Conjecture}

\theoremstyle{definition}

\newtheorem*{warning}{Warning}
\newtheorem*{discussion}{Discussion}

\theoremstyle{remark}
\newtheorem{remark}[theorem]{Remark}
\newtheorem{example}[theorem]{Example}
\newtheorem{note}[theorem]{Note}

\numberwithin{equation}{section}
\numberwithin{figure}{section}
\numberwithin{table}{section}

\newcommand{\innerprod}[2]{\langle {#1} , {#2} \rangle}
\newcommand{\bra}[1]{\langle {#1} |}
\newcommand{\ket}[1]{| {#1} \rangle}

\newcommand{\R}{{\mathbb R}}

\newcommand{\mcH}{{\mathcal H}}

\newcommand{\M}{{\mathcal M}}

\newcommand{\Owe}{{\mathcal O}} 

\newcommand{\C}{{\mathbb C}}
\newcommand{\bbar}{\overline}
\newcommand{\hhat}{\widehat}

\newcommand{\bbk}{\mathbbm{k}}

\newcommand{\psitwo}{\raisebox{-0.2em}{\includegraphics[height=0.9em]{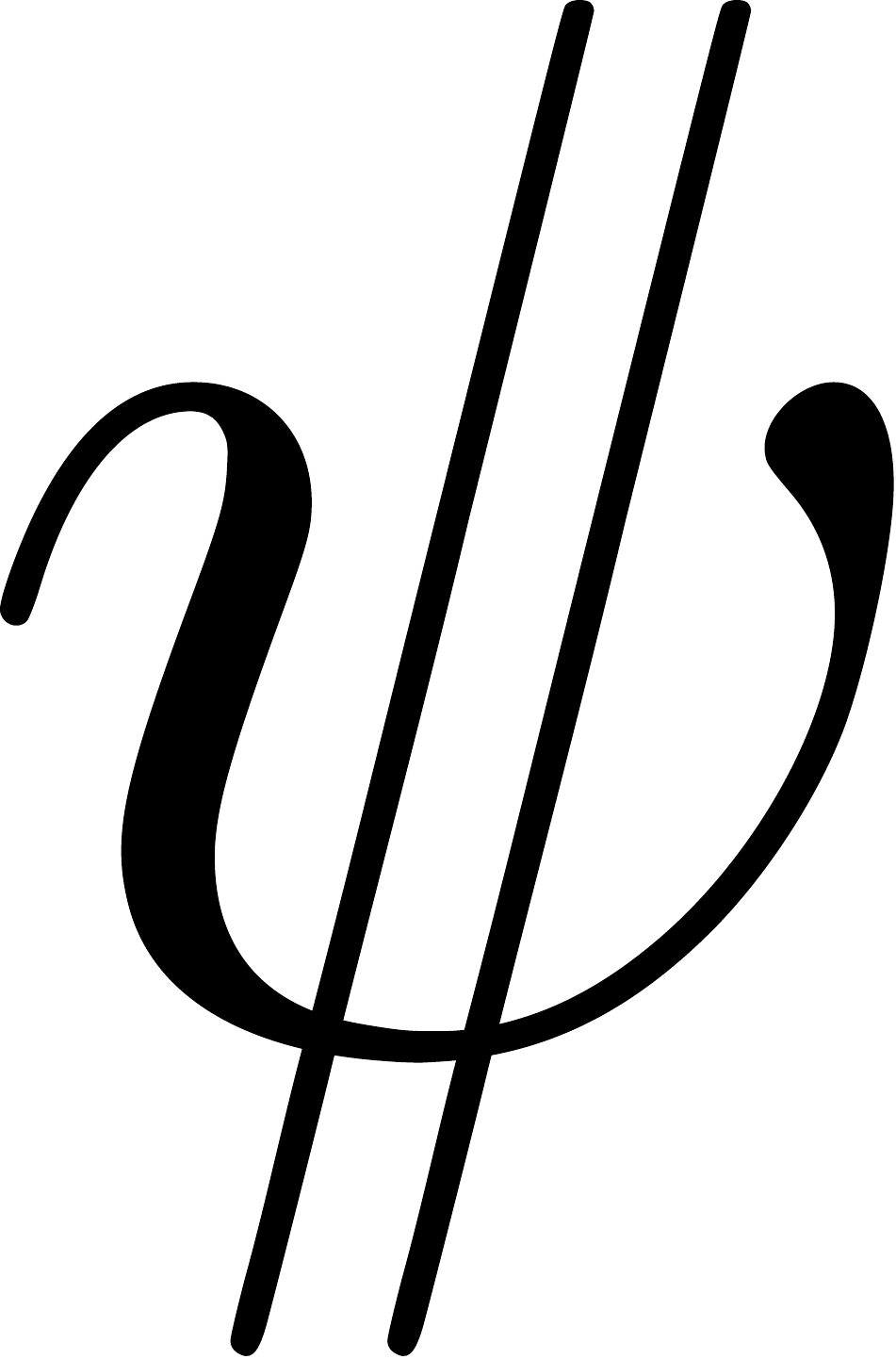}}}
\newcommand{\psithree}{\raisebox{-0.2em}{\includegraphics[height=0.9em]{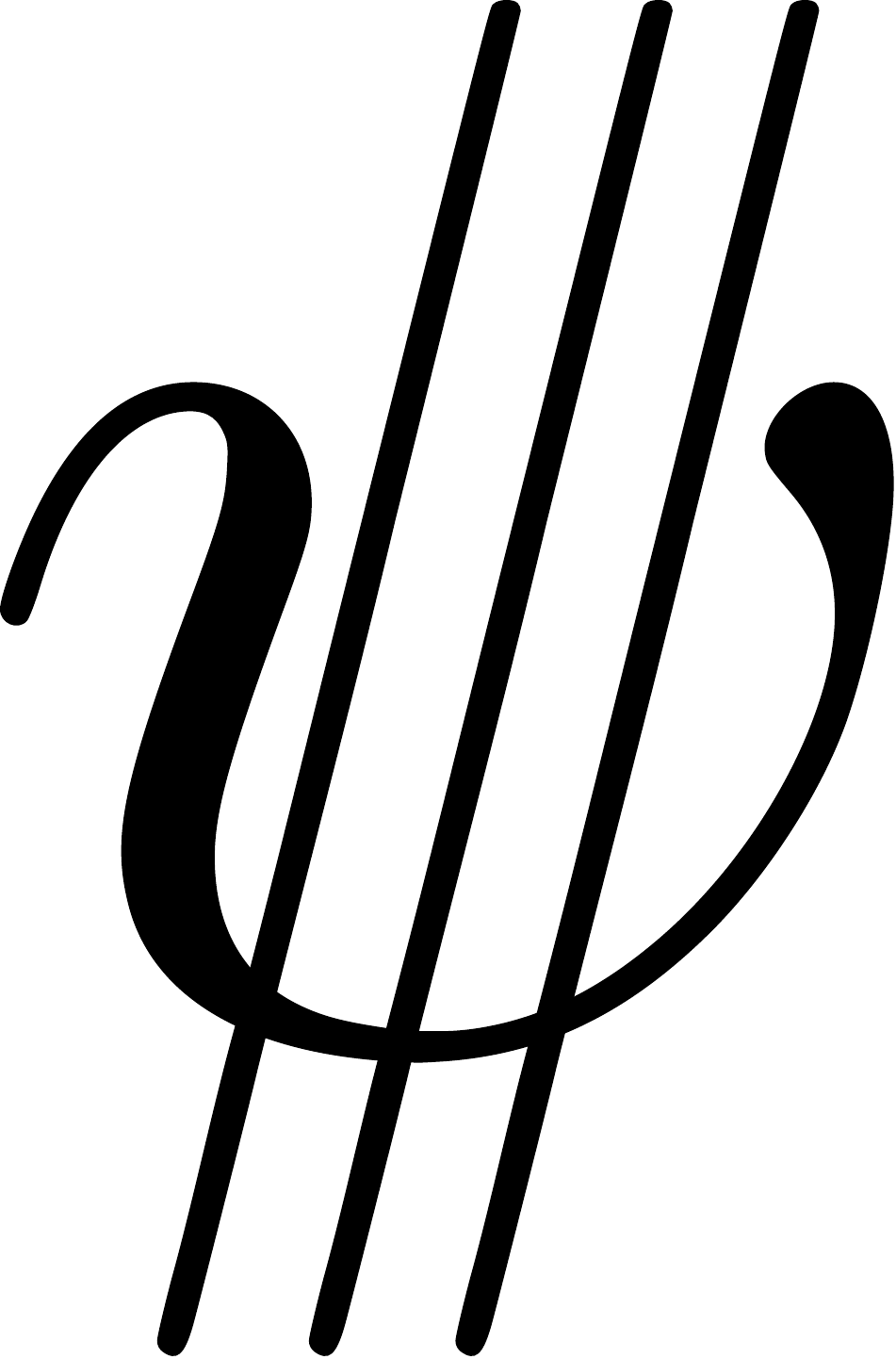}}}

\newcommand{\phitwo}{\raisebox{-0.2em}{\includegraphics[height=0.9em]{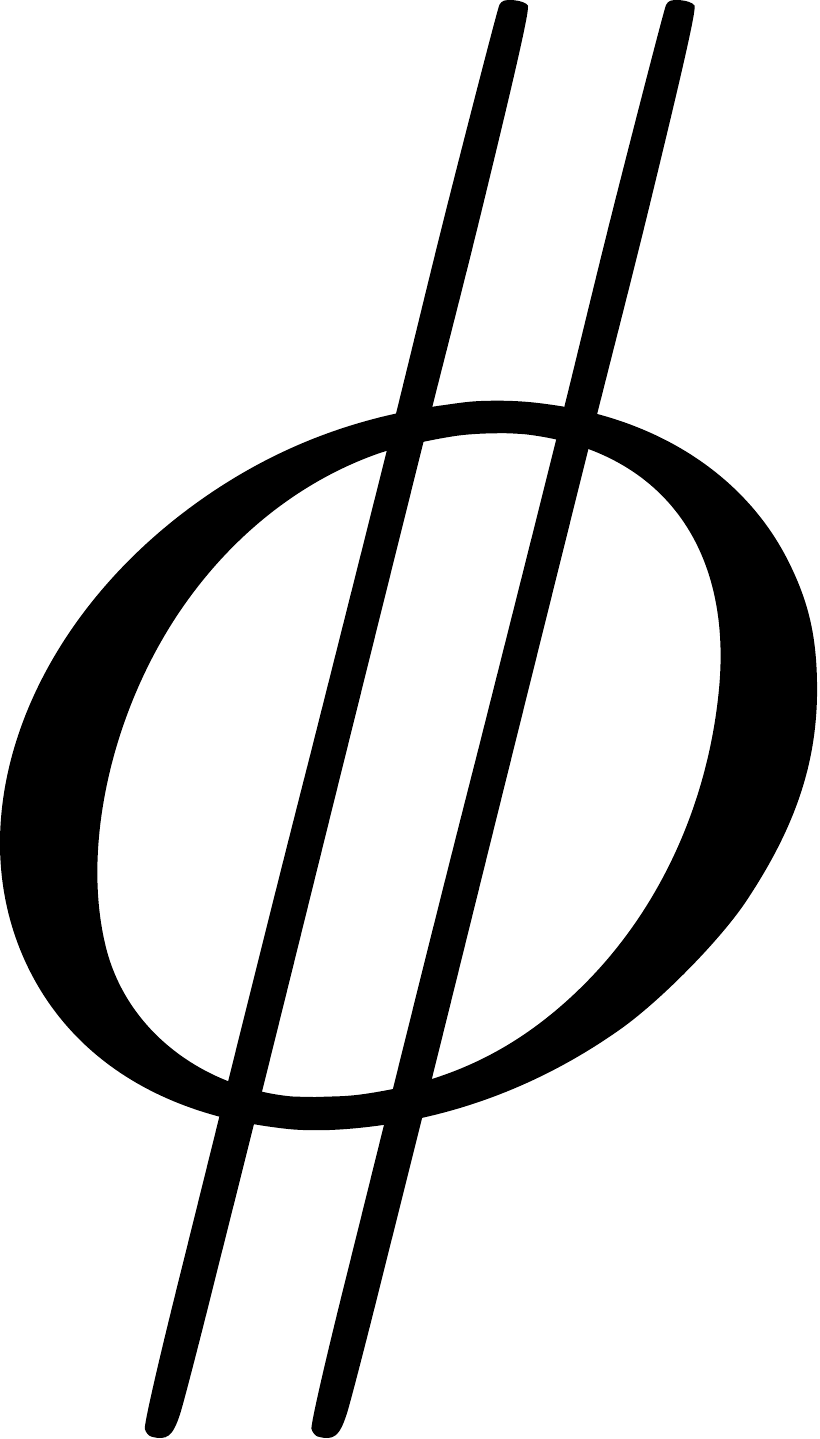}}}
\newcommand{\phithree}{\raisebox{-0.2em}{\includegraphics[height=0.9em]{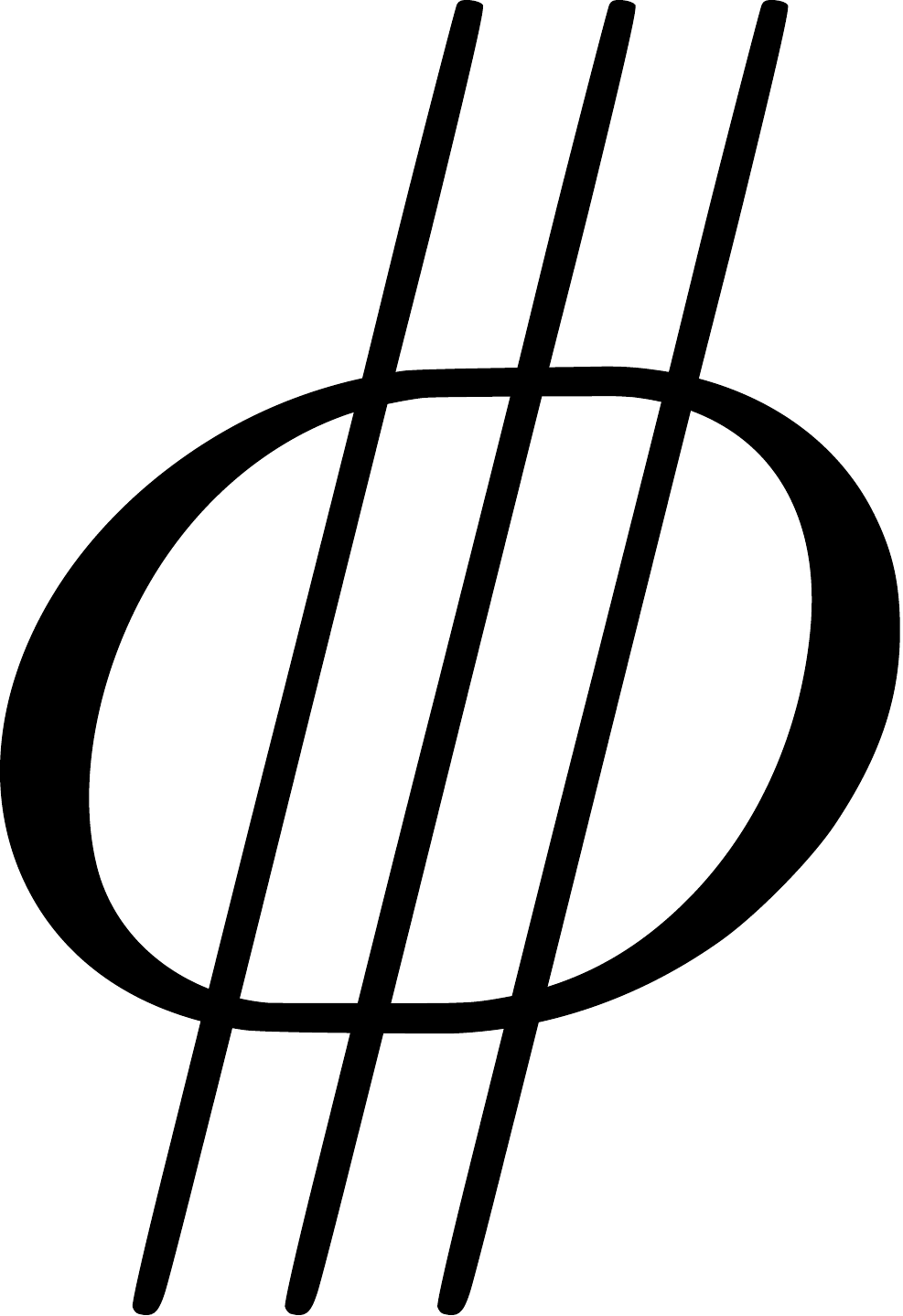}}}

\newcommand{\nablabar}{\includegraphics[height=0.8em]{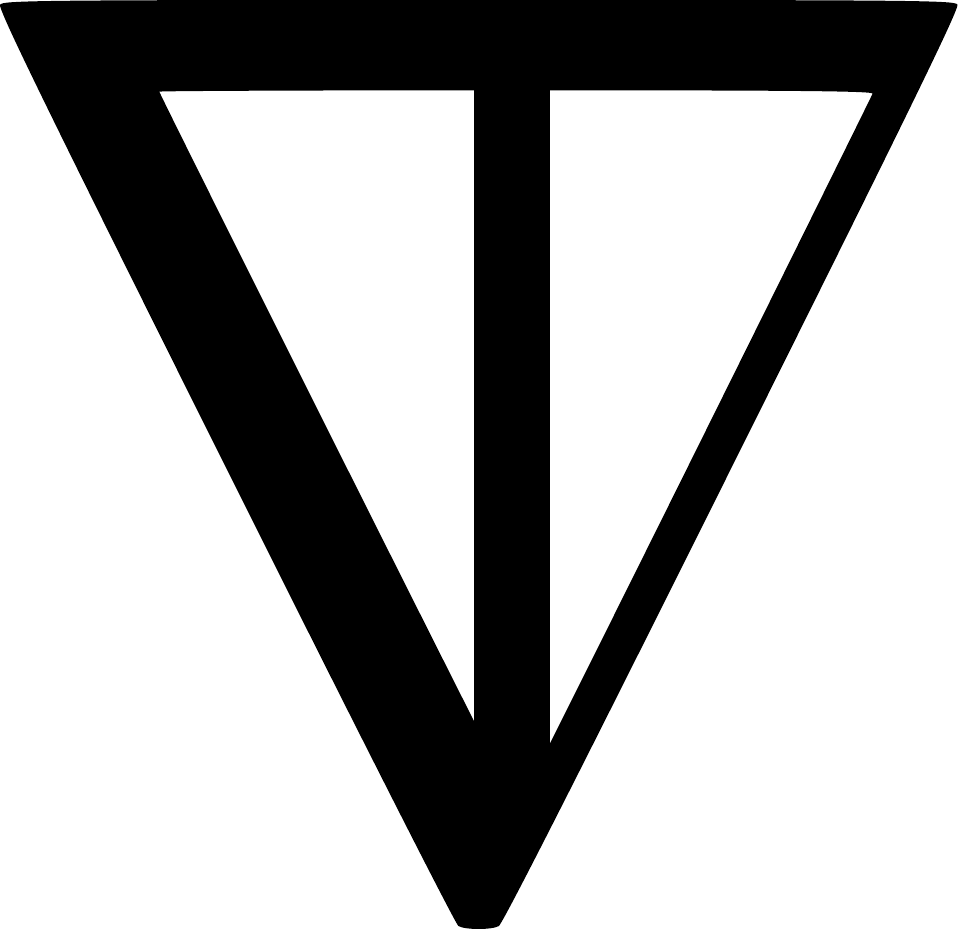}}

\begin{document}

\title[Quantum Gravity via Manifold Positivity]{Quantum Gravity via Manifold Positivity}

\author[Freedman]{Michael Freedman}

\address{}

\email{}

\date{\today, Version 17}

\begin{abstract}
The macroscopic dimensions of space-time should not be input but rather output of a general model for physics. Here, dimensionality arises from a recently discovered mathematical bifurcation: ``positive versus indefinite manifold pairings.'' It is used to build actions on a ``formal chain'' of combinatorial space-times of arbitrary dimension. The context for such actions is 2-field theory where Feynman integrals are not over classical, but previously quantized configurations. A topologically enforced singularity of the action can terminate the dimension at four and, in fact, the final fourth dimension is Lorentzian due to light-like vectors in the four dimensional manifold pairing. Our starting point is the action of ``causal dynamical triangulations'' but in a dimension-agnostic setting. Curiously, some hint of extra compact dimensions emerges from our action.
\end{abstract}

\maketitle

\section{Introduction}\label{sec:intro}

Can one hope to reconstruct the universe from mathematics? What about its most prominent feature, its (at least coarse) 3+1-dimensionality? It is illuminating that in most formalisms, stable bound states do not easily arise in other dimensions \cite{german}\cite{schwarzchild}, so even a very weak ``anthropic principle'' would force 3+1 dimensionality. But to avoid completely the taint of circular reasoning, it would be desirable to construct a dimension-agnostic Lagrangian which can then be calculated to concentrate on ``realistic'' 3+1-dimensional spaces. This paper is an initial step in this direction. In this spirit let us think about building up manifolds (space) of increasing dimension starting with the empty set by using the simplest possible operations: ``cobounding'' and ``doubling along the boundary'' (mirror double). The former is adjoint to integration and the latter generalizes $z\rightarrow z\bar{z}$ on complex numbers.

Thinking of the empty set as having dimension = -1, 
\footnote{In topology it is natural to associate a negative integer as the dimension of the empty set and setting this to be -1 avoids delaying the nontrivial steps of the construction.}
locate a compact 0D manifold $X^0$, i.e. a finite set of points, and write $\partial^{-1}(\varnothing)=X^0$. (Yes, the boundary of a finite point set is the empty set.) Next let $Y^0$ be the union of $X^0$ together with a mirror image copy. Now find an $X^1$ with $\partial(X^1) = Y^0\amalg {Y^0}^{\prime}$; $X^1$ is a cobordism from $Y^0$ to some arbitrary compact 0-manifold ${Y^0}^{\prime}$. Double $X^1$ along its boundary to make $Y^1$ (a collection of circles) and find a surface $X^2$ satisfying $\partial(X^2) = Y^1\amalg {Y^1}^{\prime}$, where ${Y^1}^{\prime}$ is an arbitrary compact 1-manifold. Alternately doubling and cobounding produces manifolds $X^d$ and $Y^d$ of increasingly higher dimension $d$, which we picture as links in a chain $X$ of manifolds $X^0,X^1,\ldots$. The idea is that for an appropriate action, explained below, this process will almost surely get stuck at $X^4$ or more precisely on some measure on the set of possible $X^4$'s which constitutes a nonperturbative quantum gravity. At each step, choice of coboundary $X^d$ is random but NOT uniform over cobounding manifolds, and is modeled after the procedure of ``causal dynamical triangulations'' (CDT) \cite{nonpert} \cite{recon} \cite{quantum} which has been successful in producing phases in which most metrics fluctuate around those with flat space-like leaves and globally are somewhat deSitter-like.

We have been ambiguous about the signature in which these CDT-like constructions will be made. Actually, one beauty of CDT is that there is a well defined Wick rotation so one may pass back and forth between statistical and quantum mechanical interpretations at will. We also will use a topologically flexible version of CDT \cite{nonpert} which grow not just product collars but general manifolds of zero relative Euler characteristic. In fact, while we will arrange to concentrate the measure on manifolds, $X$ and $Y$ are a priori permitted to be singular.

It is hoped that the process sketched above can produce a superposition concentrated near solutions of Einstein's equations on smooth space-times (or in the Euclidean case a probability measure concentrated near manifolds whose metric is proportional to the Ricci tensor --- i.e. ``Einstein.'') This hope is borrowed from the CDT community; our contribution is to treat the process CDT as recursive in dimension and describe a natural action for which the process almost surely terminates with $X^4$. 

In geometry, it is natural to enhance manifolds to local products with small additional dimensions which can collapse without curvature blow-up \cite{polynomial}. Examples of this include Seifert fibered spaces as enhanced surfaces, Nil-bundles, and more generally manifolds with F-structures. Enhancement with Calabi-Yau directions appears similar, since the basic example of C-Y's are resolution of toroidal orbifolds. As explained below, there does appear to be some scope for our construction producing small toroidal directions; but unfortunately these are not adequate for standard model physics. It would be interesting to propose a variant which would generate additional compactified dimensions which concentrate in a useful locality of the infamous ``landscape.''

Using ideas of Connes, it should be possible to give a supersymmetric version of our action, but we will not treat that here.

Let us now discuss the main ingredients for the action $S$; see equation \eqref{eq:saction} for a fuller formulation. We will work with combinatorial $d$-manifolds $X^d$ built up as in \cite{nonpert}\cite{recon} from layers of Lorentzian simplices with space-like edges having length$^2=a$ and time-like edges having length$^2=-\alpha a$. In the CDT literature, $\alpha$ is a constant, but one can be more flexible and regard it as a random variable drawn from some distribution. In the simplest model, $X^d$ is built with a fixed space-like foliation but this should be relaxed \cite{nonpert} to allow certain topology changing singularities at constant time levels. Using Regge calculus, scalar curvature $R$ can be defined and integrated on each $X^d$. Also, the boundary $\partial X^d$ has a distribution valued second fundamental form whose norm squared should be included in $S$. We also permit $X^d$ itself to be singular, i.e. not a manifold with boundary. This requires extending the definition of $R$ to singular contexts. We do not have a specific proposal here, but note that it may be desirable to supress singular spaces within the path integral by choosing the extension so that they are assigned a large action. However, singularities --- at least of the Lorentzian structure of $X$ --- should not be completely supressed. They are required to make contact with the smooth (actually P.L.) theory of manifold pairings. Processes that proceed through such singularities are useful as they ``forget'' details of the causal structure.

Letting $G$ govern the strength of gravity and $\Lambda$ be the bare cosmological constant, we write (schematically):
\begin{equation}\label{eq:S_d}
  S_d^{\textrm{Reg}}(X^d) = -\frac{1}{G}\int_{X^d}R+\delta\int_{\partial X^d}\|2^{\textrm{nd}}\|^2+2\Lambda\textrm{vol}(X^d)
\end{equation}
(When we get to details we will actually double $X$ to $Y$ and use $S_d^{\textrm{Euc}}$(Y) and no boundary term.) The overall action $S$ will include terms $S_d^{\textrm{Reg}}(X^d), d=0,1,2,\ldots,$ a fugacity for metric fluctuations, a volume, and a kinetic term. 

Each $X^d$ may not be a single piecewise Lorentzian ``combinatorial'' manifold, but a superposition. This means that the ``histories'' $X$ over which we integrate to produce a partition function: $Z=\int_{\{X\}}\mathcal{D}X~e^{-iS(X)}$
\footnote{Actually we will work with a Euclidean, Wick rotated version of $S$.}
are not classical but already quantum mechanical objects. (This situation has previously been considered in cosmology \cite{infquant}\cite{babyuniverse} under the name ``third quantization.'') Given a fixed combinatorial $d-1$ manifold $Y^{d-1}$, $X^d$ may be a single manifold with $\partial X^d = Y^d\amalg {Y^d}^{\prime}$, or in the case ${Y^d}^{\prime} = \varnothing,$ $X^d$ is permitted to be a linear combination of combinatorial $d$-manifolds $X^d_i$ with boundaries equal to $Y^{d-1}$ and normalized coefficients $c_1,\ldots,c_n\in\C$. Then $X^d$ means 
\begin{displaymath}
  X^d = \sum_{i=1}^nc_iX^d_i,\hspace{1cm}\sum_{i=1}^n|c_i|^2=1.
\end{displaymath}
We actually permit the case where the sum is infinite and the
coefficients $L^2$-convergent, but less is known mathematically about
pairing $L^2$-combinations.

Finally we come to the pairing $\innerprod{X^d}{X^d}$. In \cite{manifold_pairing},\cite{positivity},\cite{fungroups} the universal manifold pairings were defined and analyzed. Fixing a single closed $d-1$ manifold $Y^{d-1}$, define $\M_{Y^{d-1}}$ to be the $\C$ vector space of finite \footnote{See Appendix A for both finite and $L^2$ sums and pairings.} linear combinations of the cobounding manifolds $\{X^d\}$ with $\partial X^d = Y^{d-1}$. $\M_{Y^{d-1}}$ becomes a Banach space by declaring $\{X^d,\partial X^d = Y^{d-1}\}$ orthonormal.

\begin{figure}[htpb]
  \labellist \small\hair 2pt
    \pinlabel $X^d$ at 12 60
    \pinlabel ${X^d}^{\prime}$ at 12 8
    \pinlabel $Y^{d-1}$ at 55 34
    \pinlabel $f$ at 8 34
  \endlabellist
  \centering
  \includegraphics[width=1in,height=1in]{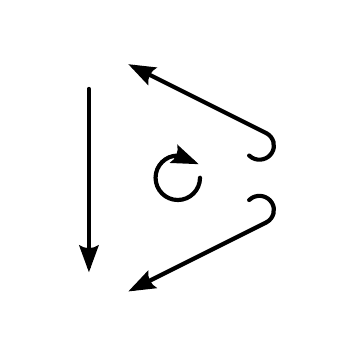}
  \caption{Commutative diagram for equivalent kets.}\label{fig:ket_comm_diag}
\end{figure}
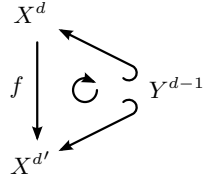

The manifolds $X$ have usually been considered up to diffeomorphism or P.L. equivalence (rel boundary), meaning bounding $X^d$ and ${X^{d}}^{\prime}$ are the same ket if there is a diffeomorphism $f:X^d\rightarrow {X^d}^{\prime}$ extending the identity $\partial X^d = \partial {X^d}^{\prime} = Y^{d-1}$, i.e. if there exists an $f$ making figure \ref{fig:ket_comm_diag} commute.
We will also consider a finer equivalence where $Y, X,$ and $X^{\prime}$ are metric and $f$ required to be an isometry. $\M_{Y}^{\text{iso}}$ will denote finite combinations of isometry classes.

Gluing along the common boundary $Y_{d-1}$ yields (\cite{manifold_pairing},\cite{positivity}) sesquilinear pairings: 
\begin{equation}\label{eq:innerproddef}
  \M_{Y^{d-1}}\times\M_{Y^{d-1}}\xrightarrow{\innerprod{~}{~}}\M_{\varnothing}.
\end{equation}

The main result is that for $d>3$ there are, for certain closed manifolds $Y^{d-1}$, light-like vectors $v\neq 0$ such that $\innerprod{v}{v} = 0$, whereas for $d\leq 3$, $\innerprod{v}{v} \neq 0$ for all $Y^{d-1}$ and all $v\neq 0$. We say the low dimensional pairings are \underline{positive}. (Later, we add a $\hhat{~}$ to the notation for $L^2$-completions and $L^2$-pairings.)

To be more precise about the (2-)action $S$, we need to introduce a little more notation and come to terms with the fact that $X^d\in \M_{Y^{d-1}}$ may be a ``superposition'' of bounding manifolds --- not a classical cobounding manifold. This superposition \underline{inside} the ``path integral'' means that we must work in the context of higher order field theories \cite{infquant}\cite{babyuniverse}. This concept is explained in more detail in Section \ref{sec:chains_action_hamiltonian} and Appendix B. But now let us interrupt our exposition of the technical set up to give in Section \ref{sec:4_D_Uniqueness} some historical perspective on how 4-dimensional spaces have been, up until now, regarded as special. Finally, Section \ref{sec:conclusions} discusses implications and short-comings of our approach. Appendix A is on pairing Hilbert spaces of manifolds and, and Appendix B is on a formalism for higher quantum field theories. I would like to thank I. Klitch, J. Milnor, C. Nayak, and X. Qi for stimulating discussions on the topic of this paper.

\section{4-D Manifolds are Different}\label{sec:4_D_Uniqueness}

$\R^n$ admits a unique smooth (also P.L.) structure for $n\neq 4$ and
by \cite{uncountable} continuum many smooth (P.L.) structures when $n=4$. What is
going on? The revolution in understanding 4-dimensional manifolds circa 1980
lead to three quite distinct perspectives on the question, ``what is
special about $D=4$?'' 
\footnote{Today a similar situation exists in dimension
  three. Three-manifolds admit rather disjoint understandings:
  hyperbolic geometry and Chern-Simons theory linked only weakly by the
  ``volume conjecture.''} 
The three answers may be summarized as:
\begin{itemize}
\item[1.] Topological: 4-2-2=0,
\item[2.] Geometric: $so(4) \simeq so(3)\oplus so(3)$ is reducible,
\item[3.] Analytic: $L^{2,2}$ + C.B. Morrey condition $\Rightarrow$ H\"older continuity.
\end{itemize}
All three answers are essential to the theory of exotic $\R^4$'s.

In smooth and piecewise linear topology, general position is a
powerful tool. It states that after perturbation two submanifolds of
dimension $p$ and $q$ will meet in a submanifold of dimension $d-p-q$,
where $d$ is the ambient dimension. The reader may easily check this
fact for affine subspaces of $\R^d$ and this is essentially the whole
proof since ``submanifold'' is a local notion. Algebraic topology is
dominated by chain complexes: sequences of of modules and boundary
maps --- the latter encoding intersection points. It turns out that the
key player \cite{selfintersect} in cancelling oppositely signed intersection points is
the Whitney disk, a 2-dimensional disk. How a Whitney disk will cross
itself or another Whitney disk is governed by the general position
formula:
\begin{equation}\label{eq:posform}
  \textrm{dim(double pts(Whitney disk))} = d-2-2.
\end{equation}
For $d\geq 5$, Whitney disks are imbedded, allowing cancellation; in
these dimensions, the algebra of chain complexes fully describes
topology \cite{poincare}: ``algebra = topology.'' Dimension 4 is a borderline
case: Whitney disks have isolated point intersections. In this case,
there is a useful topological \cite{4dmanifolds} --- but not
smooth --- technique for achieving cancellation and linking topology to
algebra.

\begin{figure}[htbp]
  \labellist \small\hair 2pt
    \pinlabel $d$ at 0 28
    \pinlabel $p$ at 190 60
    \pinlabel $q$ at 190 -5
    \pinlabel $d$ at 310 33
    \pinlabel $p$ at 495 65
    \pinlabel $q$ at 495 0
    \pinlabel $\textrm{Whitney}$ at 250 50
    \pinlabel $\textrm{trick}$ at 250 38
    \pinlabel $\textrm{pushes across}$ at 250 15
    \pinlabel $\textrm{Whitney disk}$ at 250 0
  \endlabellist
  \centering
  \includegraphics[width=0.8\textwidth]{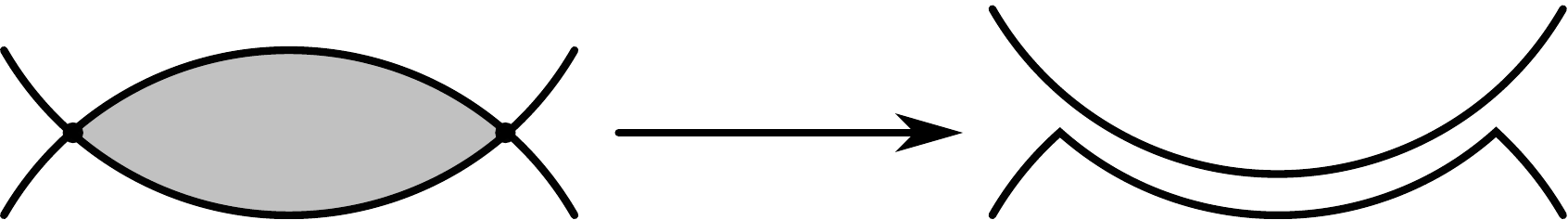}
  \caption{The Whitney trick.}
  \label{fig:whitneytrick}
\end{figure}
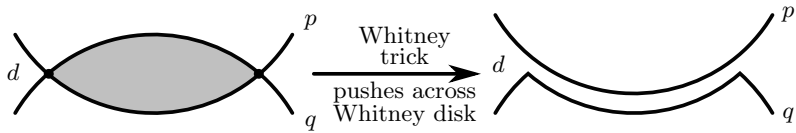
This topological (but not smooth) Whitney trick allows homological
algebra to successfully describe much of 4D topology --- but not
\underline{smooth} topology --- permitting the proliferation of smooth
structures. 

The Lie algebra of the orthogonal group is simple except for $so(4)
\simeq so(3)\oplus so(3)$. Since curvature is a (Lie algebra valued)
2-form within $\Lambda^2(T^{\ast};\textrm{ ad}\mathcal{G})$, the local
identification of 2-forms with skew-symmetric matrices ($so(n)$) allow
curvature --- only in dimension 4 --- to be decomposed according to the
eigenvalues of the Hodge $\ast$ operator into positive and negative
parts, $\Lambda^2 = \Lambda^+\oplus\Lambda^-$. The result is that the
famous anti-self dual Yang-Mills equations and Serberg-Witten
equations can only be formulated in dimension four. One may say that
these equations lead to rather unique theories of \underline{smooth}
four dimensional spaces (including the exotic structures on $\R^4$,) but perhaps cannot explain how this space
emerges in the first place. 

Finally, and most technical, is the analytic answer to the question
``what is special about 4-space?''
This answer dates back to K. Uhlenbeck's and C. Taubes' work \cite{curvature}\cite{gauge} on the
``bubbling'' phenomenon and the existence of solutions to the self
dual equation. In elliptic PDE, the equation itself is made to speak
about the regularity of a weak solution $f$ --- this is the famous
``boot strap.'' If the equation is second order, the ellipticity
condition says that a weak solution in $L^2$ is also in $L^{2,2}$,
i.e. the function and its first two distributional derivatives are in
$L^2$. The ordinary Sobolev imbedding theorem says that
\begin{equation}\label{eq:sobimbed}
  L^{2,2}\subset L^{0,q}\quad\textrm{for}\quad \frac{1}{q} > \frac{1}{2}-\frac{2-0}{d}.
\end{equation}
For dimension $d\leq 3$, such solutions are uniformly continuous and
the corresponding moduli spaces compact. For $d\geq 5$, $L^{2,2}$
formally is too weak to prove regularity. $d=4$ is borderline. If
one adds the ``Morrey'' condition that the $L^{2,2}$ norm of $f$ decays
at least as fast as $r^{\alpha}$ for some $\alpha>0$ on balls of
radius $r$, then $f$ is not only in $L^{0,\infty}$, but is H\"older
continuous. The Uhlenbeck-Taubes bubbling phenomenon happens at those
isolated points where the Morrey condition is unobtainable. ``Bubbling''
is responsible for the noncompact product ends in the moduli spaces of
anti-self-dual connections. Again, perspective 3 addresses how
analysis works on a smooth 4-space, but does not suggest where or how
such a space emerges.

In the last five years, the dimensional dichotomy, already discussed
above --- positive/indefinite manifold pairings --- has emerged. Our
idea is to build the manifold pairing into an action defined on
candidate spaces (actually chains of spaces) and then use this action to construct a quantum
gravity. There are two wrinkles which need to be appreciated from the
start. First, the manifold pairing approach is dimension-agnostic,
so the object that receives a weighting (Euclidean case) or action
(Lorentzian case) is not a single 4-manifold but a ``chain'' starting
with the empty set and proceeding upwards in dimension. Because of the
nature of the paring and the form of the action, the chain almost
surely terminates in dimension 4; this is derived and not assumed. Note that we use the term ``chain'' rather than ``history,''
because we do not want to confuse the recursive dimension raising
processes with the usual notion of time which is an aspect of the
final 4-manifold, and not the way it ``emerged'' from the empty
set. The chain is partially ordered and this order may be conceived as a
fleeting ``pre-time'' or as a second independent direction of
evolution.

The second wrinkle is that manifold pairings (or more precisely their
associated quadratic forms) are defined not on a classical manifold
$M$, but a superposition $\sum a_iM_i$. Chains are formal objects, and
we will sometimes refer to them as such to emphasize that point. This means that the ``path
integral'' is over superpositions. 
Partition functions in a quantum
field theory (QFT) are calculated by integrating over classical objects,
e.g. Brownian paths or connections on a bundle. However, for us the integral
will already be over linearized \footnote{We use ``linearize'' not to
  mean ``to approximate by a linear system,'' but rather ``to replace
  a set by the complex vector space it spans,'' e.g. as in the passage
from a category to a linear category.} objects analogous  to a vector in a
Hilbert space whose kets are Brownian paths or connections. Such
constructions are not unknown in quantum gravity \cite{babyuniverse},\cite{infquant} and have been referred to as ``third quantization'' and
``$n^{\textrm{th}}$ quantization.'' We will introduce here only the
aspects of this formalism which are presently required.

\section{Chains, Action, Hamiltonian}\label{sec:chains_action_hamiltonian}

We now describe the form of a ``2-action'' for quantum gravity in the context of a ``2-quantum field theory'' (2QFT). The essential feature of a 2QFT is a double layer of quantization. This means studying a wave function of wave functions or, via a Wick rotation to a Euclidean action, constructing a measure whose density is $e^{-S^E(\psi)}$. But instead of being a classical state, $\psi$ is a normalized superposition $\psi = \sum a_i\psi_i$ so that $S^E(\psi)$ may be small or vanish due to interference effects from components of $\psi$. This has the consequence in 2-field theory that superpositions, which cancel rather than being unobserved (low amplitude), are instead likely to be observed because their action is small. Most of the formalism of 2-field theory is relegated to Appendix B; here we proceed in a concrete ground-up fashion.

The Hilbert space $\mathcal{A}$ in which we will work has as its ``kets''
\underline{formal chains} which start at the empty set $\varnothing$,
and grow through a process borrowed from CDT, but now in a
dimension-agnostic form. Prominent in the construction is ``mirror
double'' which is a generalization of the norm$^2$ of a complex number
$|z|^2 = z\bbar{z}$. Here $Z\bbar{Z}$ will have the
meaning of gluing $Z$ along a space-like boundary with its mirror
image: $Z\rightarrow Z\cup\bbar{Z}:=Z\bbar{Z}$. On a geometric level,
leaving aside formal combinations, our CDT-like growth process
starting with $d=0$ consists of two cycling steps:
\begin{itemize}
\item[(1)] {\small (Euclidean, dim $d-1$, manifold $Y^{d-1}$)$\xrightarrow{\text{CDT}}$(Lorentzian,
  dim $d$, manifold $X^d$),}
\item[(2)] {\small (Lorentzian, dim $d$, manifold
  $X^d$)$\xrightarrow{\text{mirror double}}$(Euclidean, dim $d$, manifold $Y^d$).}
\end{itemize}
We have used the term manifold loosely. $X$ is permitted singularities in both its causal (Lorentzian) structure and even its manifold structure. Similarly, $Y$ may also be singular. After step (2), $Y^d$ is now allowed to fluctuate to $\tilde{Y}^d$ breaking exact mirror symmetry, then one cycles back to step 1 (with $\tilde{Y}^d$ replacing $Y^{d-1}$). $\tilde{X}^d$ now doubles to $Y^{d+1}$. A chain contains $\varnothing, X^0, X^1, X^2, \ldots$ either terminating with $X^{d^{\prime}}$ for some $d^{\prime}\geq 1$, or continuing indefinitely.

To define a \underline{formal chain}, we introduce formal combinations to the
process. So 
\begin{displaymath}
\varnothing\rightarrow X^0=\sum_ia_i^0X_i^0,\hspace{1cm}\sum_i|a_i^0|^2 = 1.
\end{displaymath}
Since there is no boundary $\partial X_i^0=\varnothing$ to glue along,
mirror double is simply disjoint union with the orientation reversed point set.
The next arrow goes
\begin{displaymath}
  \rightarrow\sum_{i,j}a_i^0\bbar{a_j^0}X_i^0\bbar{X_j^0} :=
  \sum_{i,j}a_i^0\bbar{a_j^0}Y_{i,j}^0 := Y^0.
\end{displaymath}
We now collect terms according to isometry type (in this case number
of (+,-)-points) and write $Y^0 = \sum_{l_0}b_{l_0}^0Y_{l_0}^0$. Next
we would normally permit a topology (actually P.L. structure)
preserving fluctuation $Y^0 = Y^{0,k=0}\rightarrow
Y^{0,k=1}\rightarrow\cdots\rightarrow Y^{0,k^0}:=\tilde{Y}^0$ to
\begin{displaymath}
  \tilde{Y}^0 = \sum_{l_0}b_{l_0}^0\tilde{Y}_{l_0}^0,
\end{displaymath}
but in dimension zero there are no topology preserving fluctuations,
so $Y^0 = \tilde{Y}^0$. The next arrow is
\begin{displaymath}
  b_{l_0}^0Y_{l_0}^0\rightarrow X_{l_0}^1 = \sum_i
  a_{i,l_0}^1X_{i,l_0}^1,\hspace{1cm}a_{i,l_0}^1 =
  \alpha_{i,l_0}b_{l_0}^0,~\sum_i|\alpha_{i,l_0}|^2 = 1\textrm{ for
    all }l_0.
\end{displaymath}
The terms $X_{i,l_0}^1$
are combinatorial Lorentzian 1-manifolds whose simplices have (length$^2$) =
$-a\alpha$: 
\begin{figure}[htpb]
  \labellist \small\hair 2pt
  \pinlabel $\left.\parbox{1pt}{\vspace{64pt}}\right\}X_{l_0}^1$ at 113 42
  \pinlabel $\underbrace{\hspace{80pt}}_{Y_{l_0}^0}$ at 53 -5
  \endlabellist
  \centering
  \includegraphics[width=1.3in]{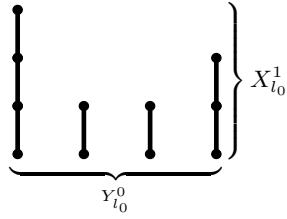}
  \caption{The next arrow.}
  \label{fig:next_arrow}
\end{figure}
Because the $X$ are Lorentzian, in this dimension the $X_{l_0}^1$ must be
P.L. homeomorphic to $Y_{l_0}^0\times I$ union possible additional circles. The only restriction to obtain a non-singular Lorentzian extension is on the Euler characteristic $\mathcal{X}(X_{l_{d-1}}^d) =
\mathcal{X}(Y_{l_{d-1}}^{d-1})$.

The process now continues this cycle:
{\allowdisplaybreaks
  \begin{align*}
    \varnothing \xrightarrow{\textrm{grow}} & X^0=\sum_ia_i^0X_i^0\xrightarrow{\textrm{double}}\sum_{i,j}a_i^0\bbar{a_j^0}X_i^0\bbar{X_j^0}=Y^0=\sum_{l_0}b_{l_0}^0Y_{l_0}^0\\
    &\xrightarrow{\textrm{fluctuate}}\tilde{Y}^0 = \sum_{l_0}b_{l_0}^0Y_{l_0}^0\\
    \xrightarrow{\textrm{grow}} & X^1=\sum_{i,l_0}a_{i,l_0}^1X_{i,l_0}^1\xrightarrow{\textrm{double}}\sum_{i,j,l_0}a_{i,l_0}^1\bbar{a_{j,l_0}^1}X_i^1\bbar{X_j^1}=Y^1=\sum_{l_1}b_{l_1}^1Y_{l_1}^1\\
    &\xrightarrow{\textrm{fluctuate}}\tilde{Y}^1 =
    \sum_{l_1}b_{l_1}^{k_1,1}Y_{l_1}^{k_1,1}\\
    & \vdots\\
    \xrightarrow{\textrm{grow}} & X^4=\sum_{i,l_3}a_{i,l_3}^4X_{i,l_3}^4\xrightarrow{\textrm{double}}\sum_{i,j,l_3}a_{i,l_3}^4\bbar{a_{j,l_3}^4}X_i^4\bbar{X_j^4}=Y^4=\sum_{l_4}b_{l_4}^4Y_{l_4}^4\\
    &\xrightarrow{\textrm{fluctuate}}\tilde{Y}^4 = \sum_{l_4}b_{l_4}^{k_4,4}Y_{l_4,k^4}^{k_4,4}\\
    & \vdots
  \end{align*}
}
where $a_{i,l_3}^4 = \alpha_i^{k_3,3}b_{l_3}^{k_3,3},
\sum_i|\alpha_i^{k_3,3}|^2 = 1$.
Such a process is called a \underline{formal chain}.

There is a general principle: If $\partial X_{l_{d-1}}^d\neq Y_{l_{d-1}}^{d-1}$, i.e. there is a non-empty ``upper boundary,'' then $X_{l_{d-1}}^d$ cannot be a superposition for $d>1$, since there will be no canonical way to identify boundary conditions (except when the boundary has dimension = 0) of the states supposedly in superposition. This severely limits superpositions within formal chains.

Thus a formal chain has sites or ``vertices'' which are formal spaces of increasing dimension and links which can be labeled by: ``grow'', ``double'', or ``fluctuate.'' The word ``formal'' means ``normalized complex-linear combination.'' We argue that the amplitude will concentrate on the special case: ``formal non-singular Lorentzian manifolds'' but a priori one should permit the growth process Euclidean-$d$ $\rightarrow$ Lorentzian-$(d+1)$ to add $d+1$ simplices haphazardly. We permit the $(d+1)$-Lorentzian simplicies to be fitted together without regard to Lorentzian or even manifold structure. The term in our action which we will denote $\int-R$, where $R$ is Regge scalar curvature, should be extended in some (unspecified) fashion to penalize singularities of topology and Lorentzian structure. The role of singular structures will be explained shortly. The action will favor cases in which the $d+1$ simplices are organized into a non-singular Lorentzian manifold with an Einstein metric.

 It is permissible to think of every link in the chain as reversible so that given one chain $c$, it implies many related chains $c^{\prime}$ which simply walk (e.g. randomly) up and down $c$. Such $c^{\prime}$ will have larger action than $c$ and be correspondingly supressed.

\begin{figure}[htbp]
  \centering
  \begin{tabular}[htpb]{ccccccccc}
    & $Y^{-1}$ & $\rightarrow$ & $X^0$ & $\rightarrow$ & $Y^0$ & $\rightarrow$ & $X^1$\\
    \vspace{0.2in}
    & $\varnothing$ & & \includegraphics[width=0.05in]{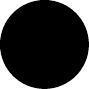} & & \parbox{0.1in}{\includegraphics[width=0.05in]{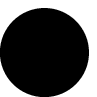}\\\includegraphics[width=0.05in]{figures/dot}} & & \parbox{0.25in}{\rotatebox{90}{\includegraphics[width=0.05in]{figures/stick1}}\\\rotatebox{90}{\includegraphics[width=0.05in]{figures/stick2}}}\\
    $\rightarrow$ & $Y^1$ & $\rightarrow$ & $X^2$ & $\rightarrow$ & $Y^2$ & $\rightarrow$ & $X^3$\\
     & \parbox{0.25in}{\rotatebox{90}{\includegraphics[width=0.15in]{figures/loop2}}\\\includegraphics[width=0.32in]{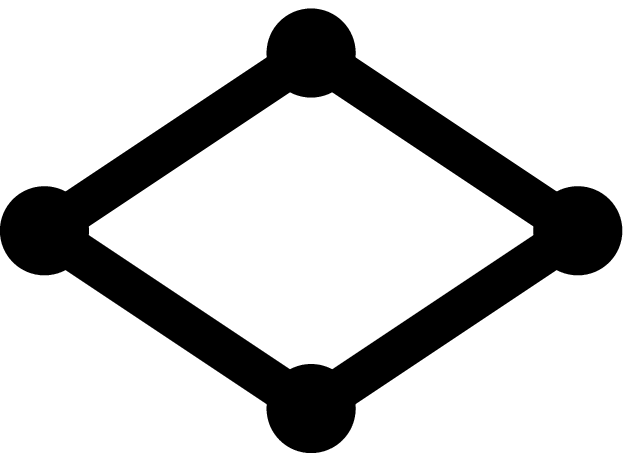}} & & \raisebox{-0.15in}{\includegraphics[width=0.4in]{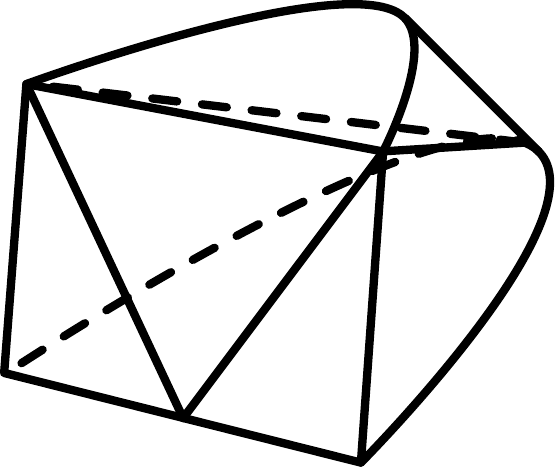}} & & \raisebox{-0.1in}{\includegraphics[width=0.4in,height=0.25in]{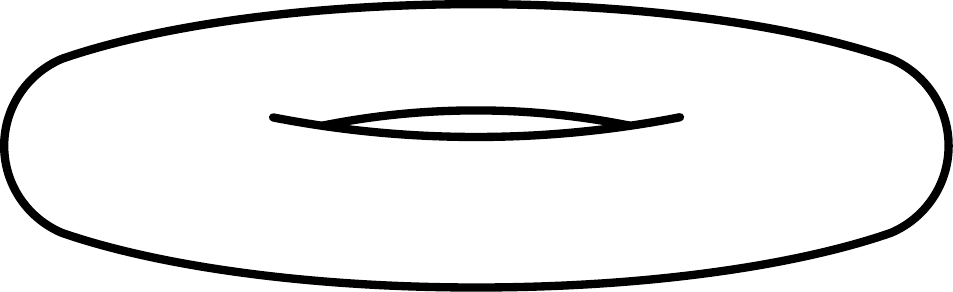}} & & 
  \labellist \small\hair 2pt
  \pinlabel $\alpha$ at -30 190
  \pinlabel $\beta$ at -30 40
  \pinlabel $+$ at 110 115
  \endlabellist
  \raisebox{-0.3in}{\includegraphics[width=0.6in]{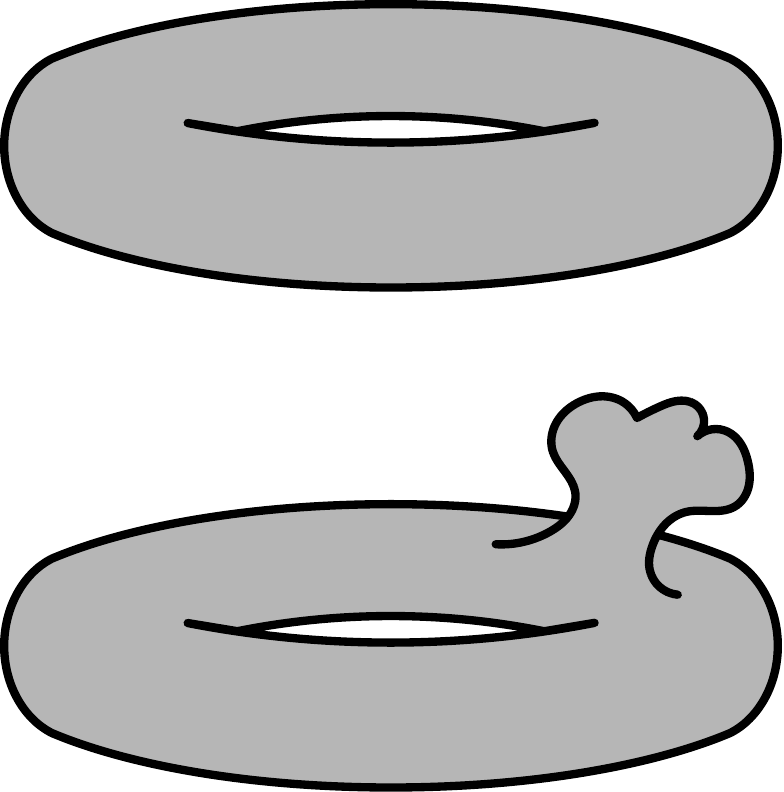}}\vspace{0.2in}\\
  \end{tabular}
  \begin{tabular}[htpb]{cccccc}
  $\rightarrow$ & $Y^3$ & $\rightarrow$ & $X^4$ & $\rightarrow$ & $Y^4$\\
  & \parbox{0.7in}{\hspace{0.15in}$\alpha\bar{\alpha}$ \raisebox{-0.3em}{\includegraphics[width=0.2in]{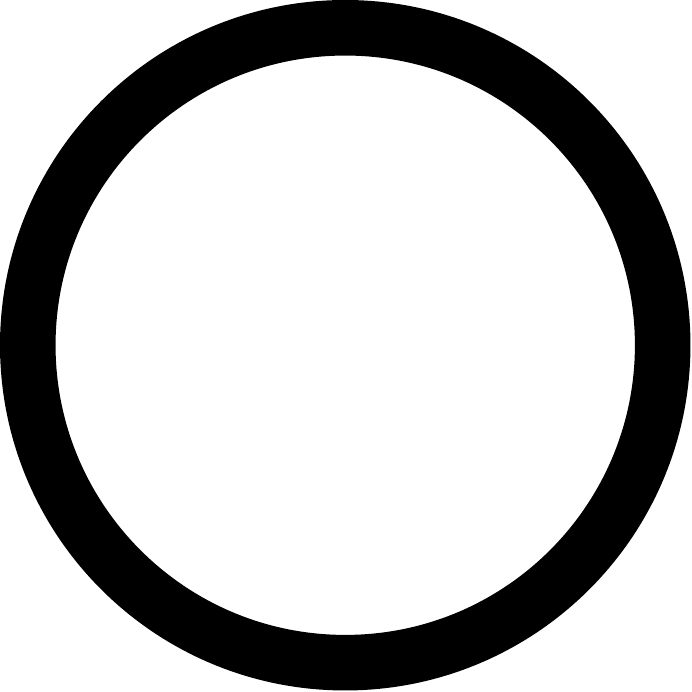}}\\
    + $\alpha\bar{\beta}$ \raisebox{-0.3em}{\includegraphics[width=0.25in]{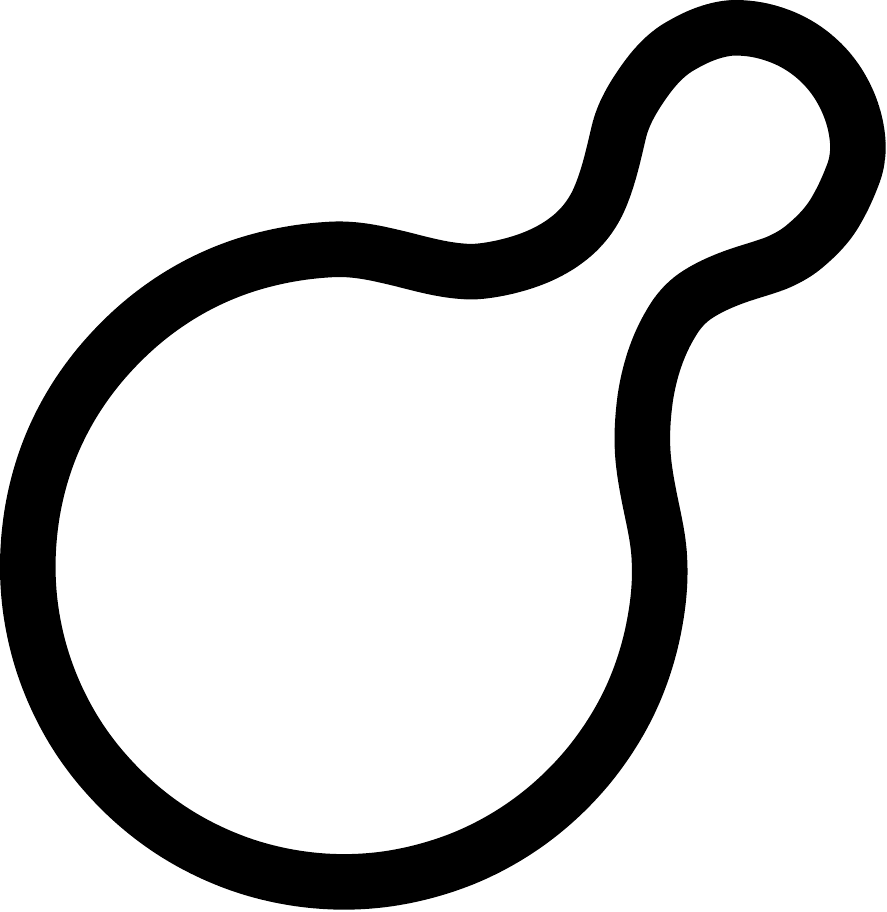}}\\
    + $\beta\bar{\alpha}$ \raisebox{-0.3em}{\rotatebox{90}{\includegraphics[width=0.25in]{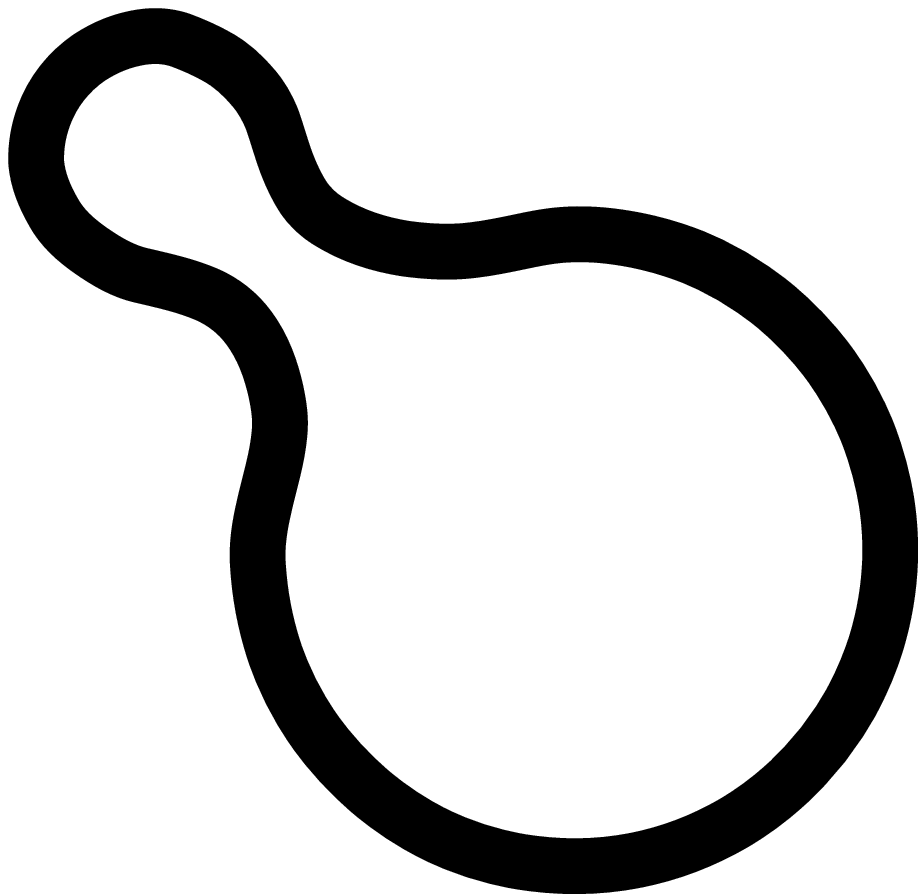}}}\\
    + $\beta\bar{\beta}$ \raisebox{-0.3em}{\includegraphics[width=0.3in]{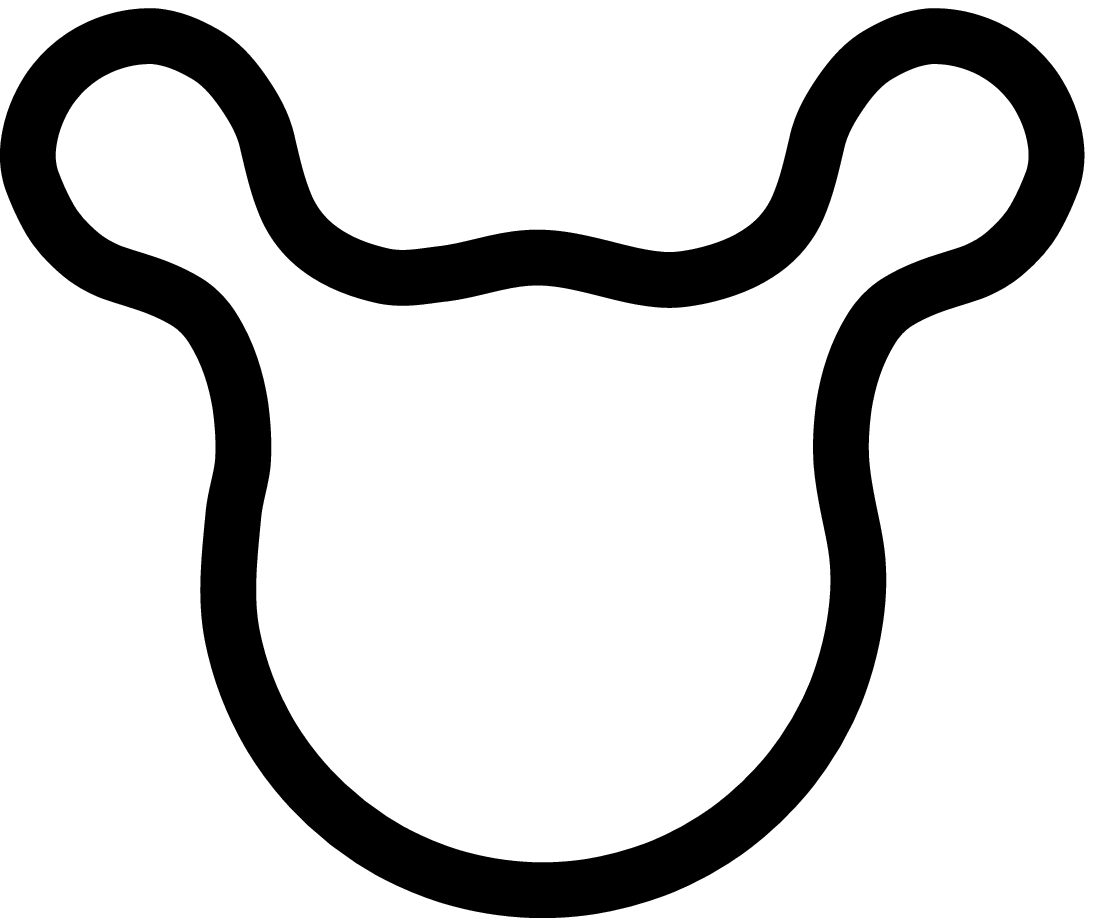}}}
  & 
  & \parbox{0.9in}{
    \hspace{1em}\raisebox{-1.2em}{\includegraphics[width=0.3in]{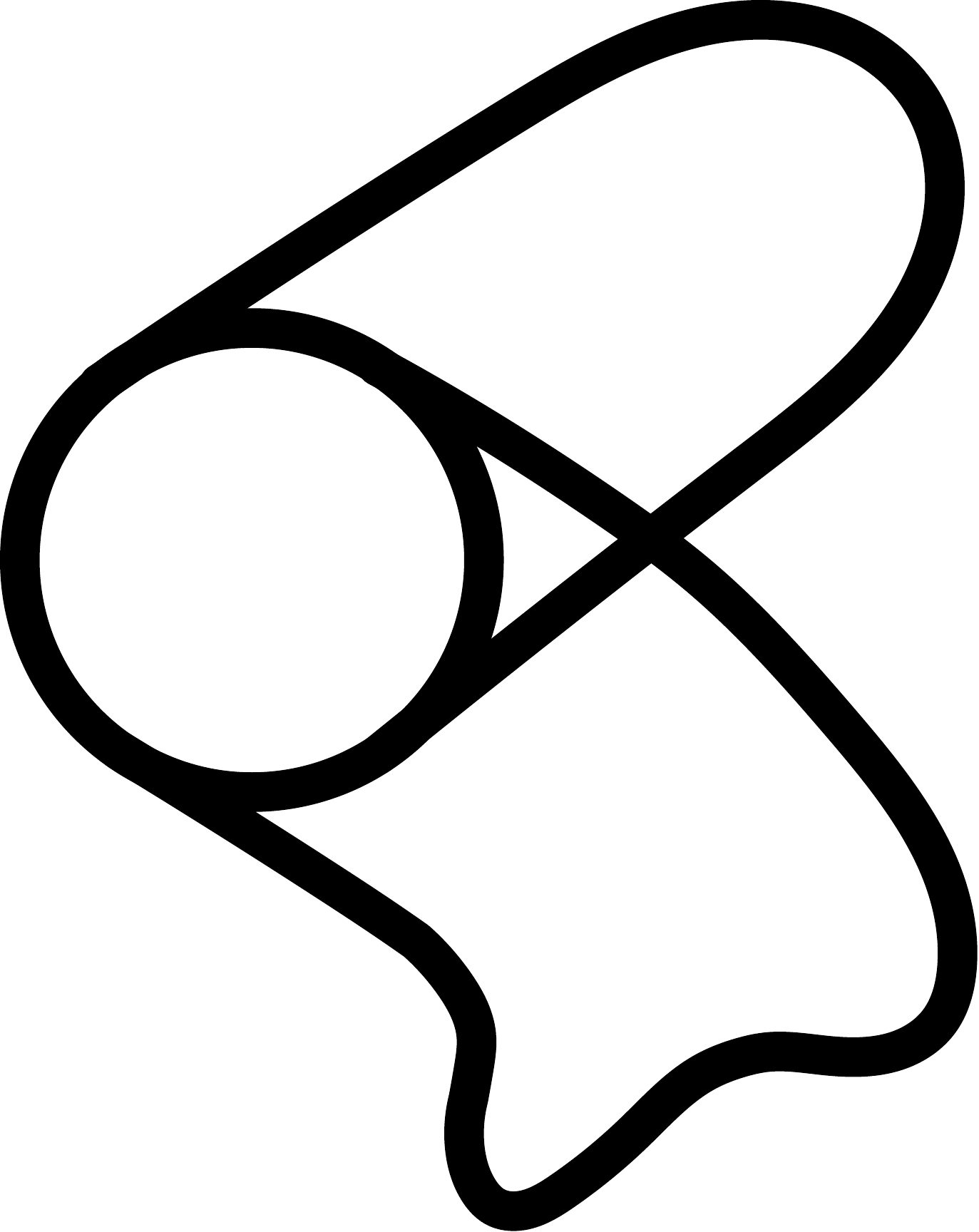}} \parbox{0.1in}{$\hspace{0.7em}\frac{\alpha\bar{\alpha}}{\sqrt{2}}$\vspace{0.3em}\\$-\frac{\alpha\bar{\alpha}}{\sqrt{2}}$}\vspace{0.5em}\\
    + \raisebox{-1.2em}{\includegraphics[width=0.3in]{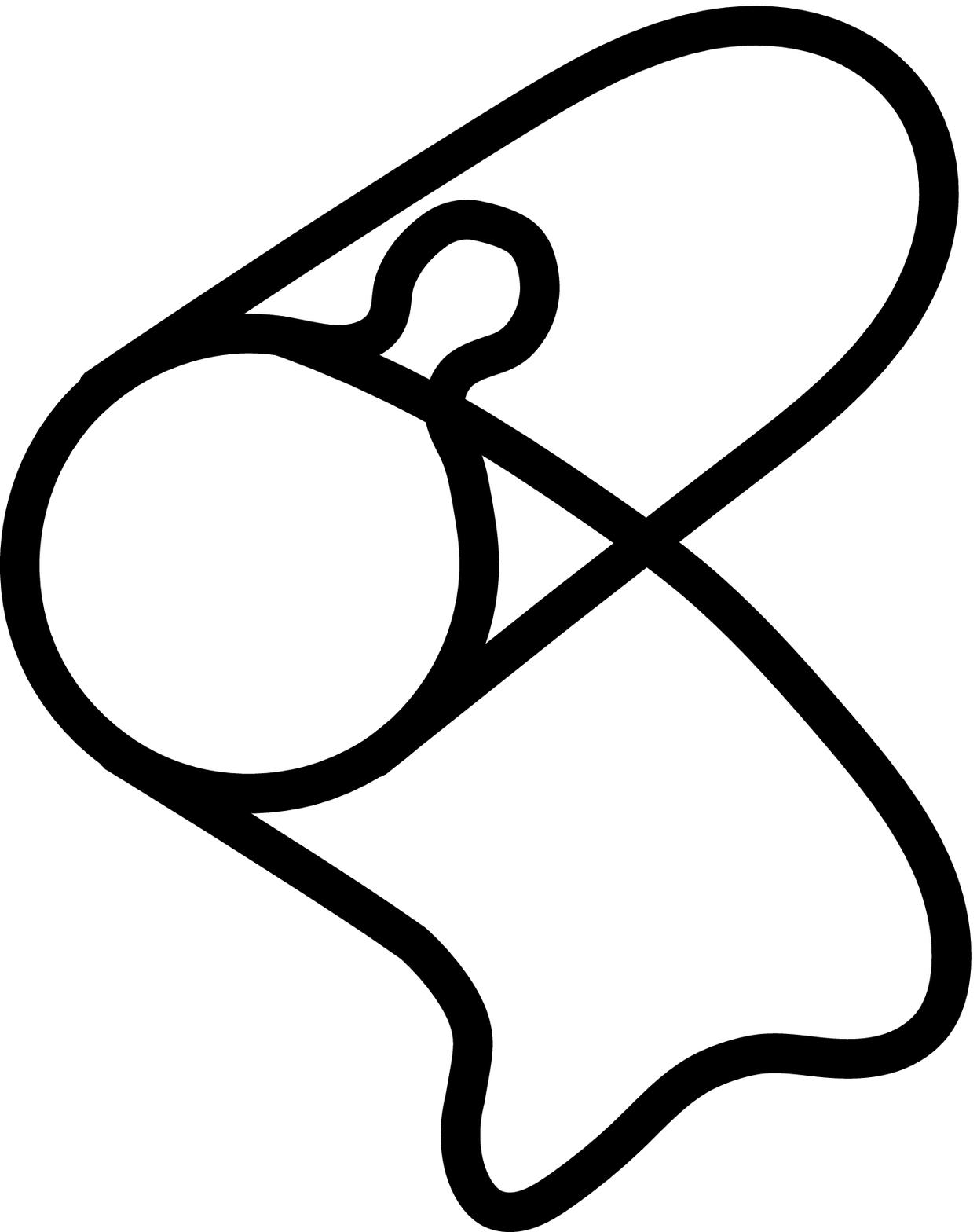}} \parbox{0.1in}{$\hspace{0.7em}\frac{\alpha\bar{\beta}}{\sqrt{2}}$\vspace{0.3em}\\$-\frac{\alpha\bar{\beta}}{\sqrt{2}}$}\vspace{0.5em}\\
    + \raisebox{-1.2em}{\includegraphics[width=0.3in]{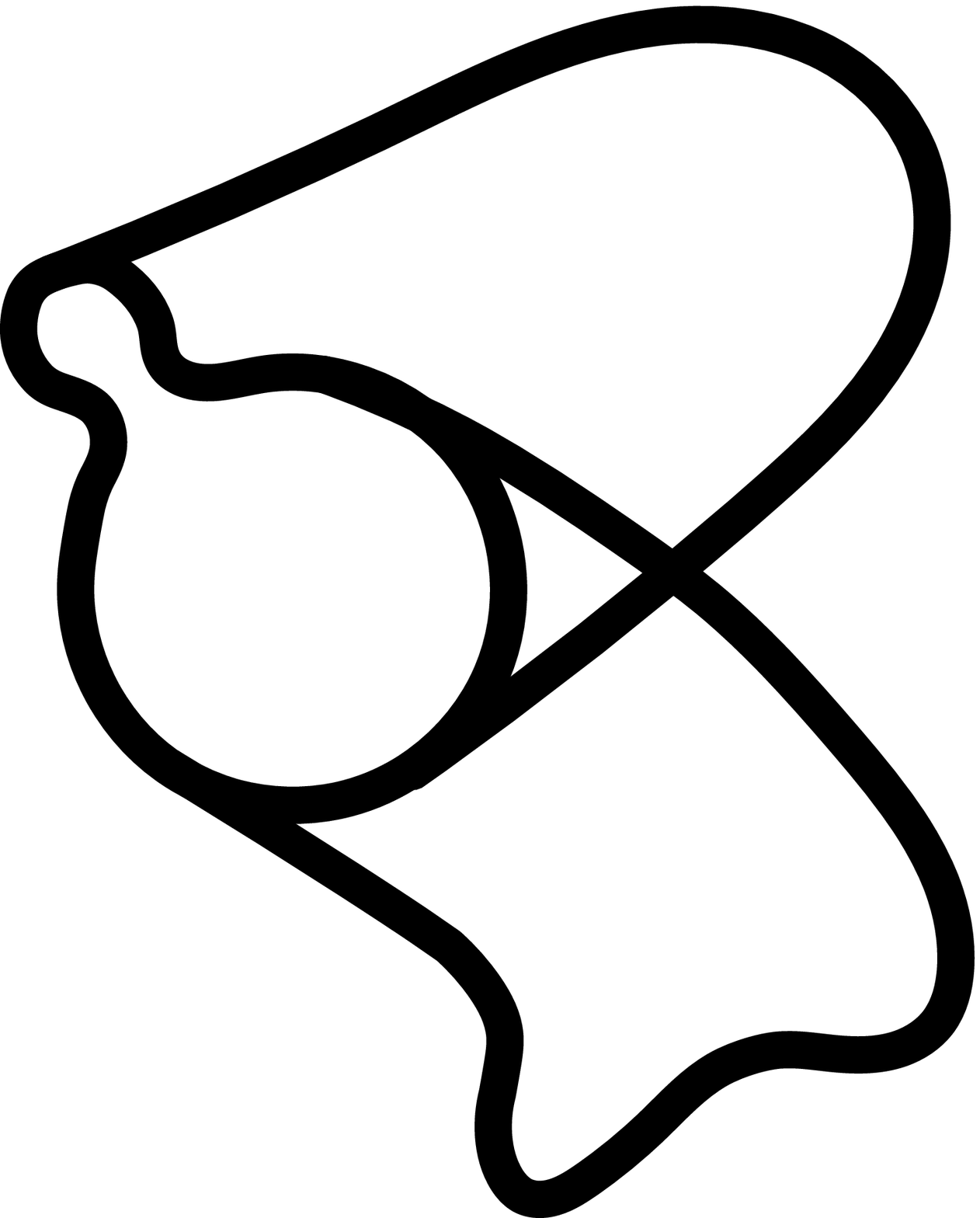}} \parbox{0.1in}{$\hspace{0.7em}\frac{\beta\bar{\alpha}}{\sqrt{2}}$\vspace{0.3em}\\$-\frac{\beta\bar{\alpha}}{\sqrt{2}}$}\vspace{0.5em}\\
    + \raisebox{-1.2em}{\includegraphics[width=0.3in]{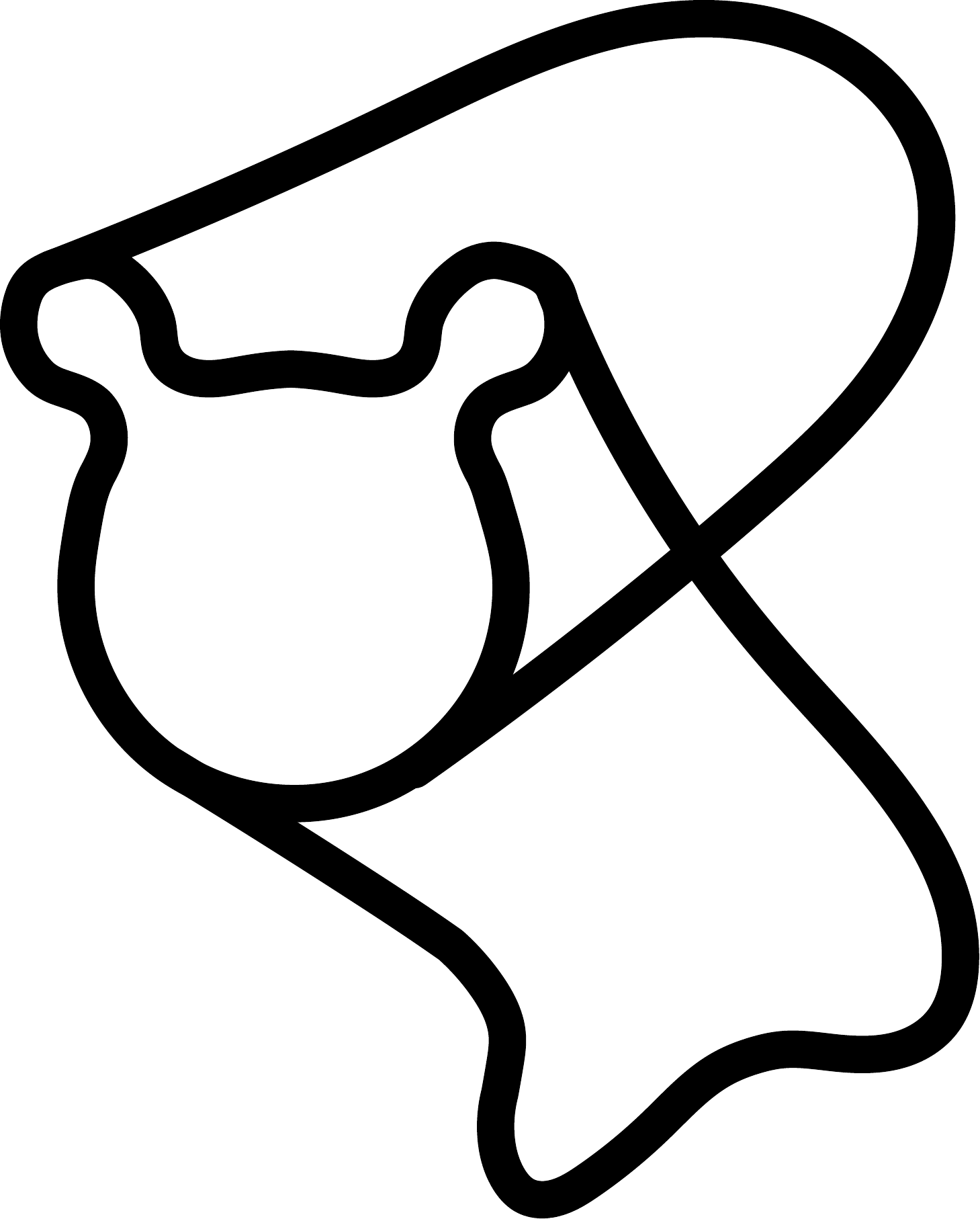}} \parbox{0.1in}{$\hspace{0.7em}\frac{\beta\bar{\beta}}{\sqrt{2}}$\vspace{0.3em}\\$-\frac{\beta\bar{\beta}}{\sqrt{2}}$}}
  &
  & \parbox{0.7in}{
    \raisebox{-0.1em}{\includegraphics[width=0.1in]{figures/dot}} $|\alpha|^4/2$\\
    \raisebox{-0.1em}{\includegraphics[width=0.1in]{figures/dot}} $-|\alpha|^4/2$\\
    \raisebox{-0.1em}{\includegraphics[width=0.1in]{figures/dot}} $-|\alpha|^4/2$\\
    \raisebox{-0.1em}{\includegraphics[width=0.1in]{figures/dot}} $|\alpha|^4/2$\vspace{0.5em}\\
    \raisebox{-0.1em}{\includegraphics[width=0.1in]{figures/dot}} $|\alpha\bar{\beta}|^2/2$\\
    \raisebox{-0.1em}{\includegraphics[width=0.1in]{figures/dot}} $-|\alpha\bar{\beta}|^2/2$\\
    \raisebox{-0.1em}{\includegraphics[width=0.1in]{figures/dot}} $-|\alpha\bar{\beta}|^2/2$\\
    \raisebox{-0.1em}{\includegraphics[width=0.1in]{figures/dot}} $|\alpha\bar{\beta}|^2/2$\vspace{0.5em}\\
    \raisebox{-0.1em}{\includegraphics[width=0.1in]{figures/dot}} $|\beta\bar{\alpha}|^2/2$\\
    \raisebox{-0.1em}{\includegraphics[width=0.1in]{figures/dot}} $-|\beta\bar{\alpha}|^2/2$\\
    \raisebox{-0.1em}{\includegraphics[width=0.1in]{figures/dot}} $-|\beta\bar{\alpha}|^2/2$\\
    \raisebox{-0.1em}{\includegraphics[width=0.1in]{figures/dot}} $|\beta\bar{\alpha}|^2/2$\vspace{0.5em}\\
    \raisebox{-0.1em}{\includegraphics[width=0.1in]{figures/dot}} $|\beta|^4/2$\\
    \raisebox{-0.1em}{\includegraphics[width=0.1in]{figures/dot}} $-|\beta|^4/2$\\
    \raisebox{-0.1em}{\includegraphics[width=0.1in]{figures/dot}} $-|\beta|^4/2$\\
    \raisebox{-0.1em}{\includegraphics[width=0.1in]{figures/dot}} $|\beta|^4/2$\\
  }
  \end{tabular}
  \caption{A formal chain terminating with $Y^4=0$}
  \label{fig:terminating_schematic}
\end{figure}

To summarize: 
\begin{itemize}
\item ``Growth'': Euclidean-$(d-1) \rightarrow$ Lorentzian-$d$ adds Lorentzian $d$ simplices.
\item ``Double'': Lorentzian-$d \rightarrow$ Euclidean-$d$ (Wick rotation).
\item ``Fluctuate'': Euclidean-$d \rightarrow$ Euclidean-$d$ alters the local combinatorial geometry.
\end{itemize}
Fluctuation must be allowed in order to make contact with topological pairing. Once fluctuation is permitted on the Euclidean space, it is perhaps natural to introduce it as well as an additional ``link'' on Lorentzian spaces. For simplicity we have not done this. The action which we describe next is a kind of Einstein-Regge action computed up and down the chain. Figure \ref{fig:terminating_schematic} gives a schematic depiction of a formal chain terminating with $Y^4=0$.

\begin{figure}[htpb]
  \centering
  \begin{tabular}{cm{2.1in}c}
    Component of & & Layer type\\
    \hline
    $X^1$ &
    \begin{center}
      \labellist \small\hair 2pt
      \endlabellist
      \includegraphics[height=0.85in]{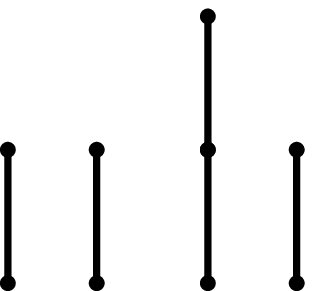}
    \end{center} &
    full layer\\
    $X^2$ & 
    \labellist \small\hair 2pt
    \endlabellist
    \includegraphics[width=2in,height=1in]{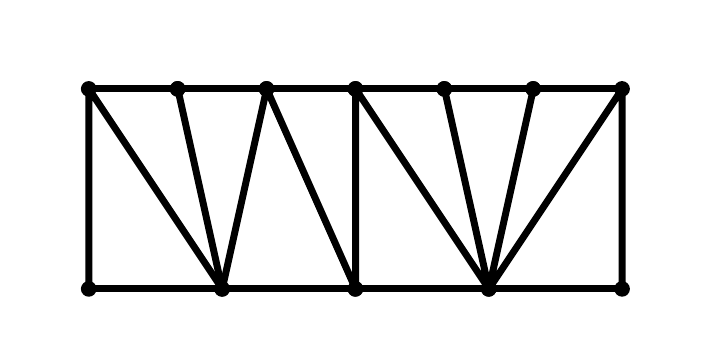} &
    full layer\\
    $X^2$ &
    \labellist \small\hair 2pt
    \endlabellist
    \includegraphics[width=2in,height=1in]{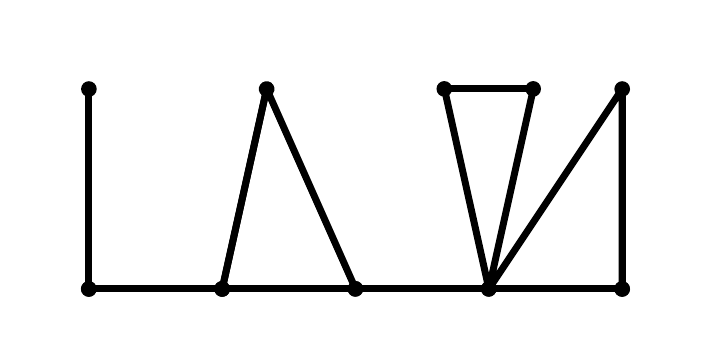} &
    partial layer\\
    $X^3$ &
    \labellist \small\hair 2pt
    \endlabellist
    \includegraphics[width=2in,height=1in]{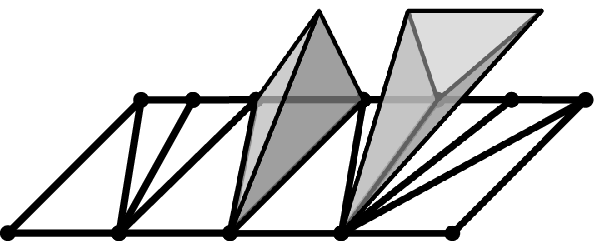} &
    partial layer
  \end{tabular}
  \caption{Aspect ratios for $d=1,2,3$.}
  \label{fig:aspect_ratio}
\end{figure}

The Euclidean sites in a formal chain will concentrate at elements of the $L^2$ Hilbert space $\hhat{\M}_{\varnothing}^{d,\textrm{Euc}}$ spanned by formal $\C$-combinations of simplicial Euclidean metric triangulated $d$-manifolds ($~\hhat{~}$ signifies $L^2$-completion, $d$ = dimension, Euc = Euclidean, Lor = Lorentzian, and $\varnothing$ indicates the empty set), although  formally they are permitted to be more singular (see figure \ref{fig:aspect_ratio}). Similarly, $X^{d+1}$ concentrates in $\hhat{\M}_{Y^d}^{d,\textrm{Lor}}$.

The spaces $X^d$ ``grow'' on $Y^{d-1}$ by adding Lorentzian
simplices. We do not assume $X^d_{i,l_{d-1}}$ are manifolds, but the
action should favor this case. The $Y^d$ are made of ``spatial
doubles'': glued copies of $Y^d$ and $\bbar{Y^d}$ across the
space-like simplices. The fluctuations of Euclidean simplicial
structure ($Y^d\rightarrow \tilde{Y}^d$) are assigned a 
fugacity $f$ which depends on the geometric Pochner move or Euclidean
geometry change occurring at each step $k-1\rightarrow k$, and is to be
extended linearly over superpositions, weighting by $|\text{amplitude}|^2$
\begin{figure}[htpb]
  \centering
  \includegraphics[width=2in,height=1in]{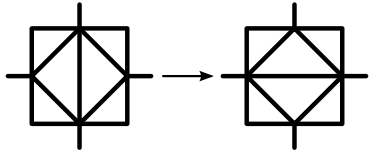}
  \caption{Geometric Pochner move at center in 2D.}
  \label{fig:pocher_move}
\end{figure}

Since reflection inverts the Lorentzian light cones, the components of
the double $Y^d$ only have a canonical Euclidean (simplicial)
structure obtained by Wick rotation of
Lorentzian simplices to Euclidean geometry. The unscaled action term
$S_d$ on $Y^d$ is of the form
\begin{equation}
  \label{eq:unscaled_saction}
  S_d(Y^d) = \int_{Y^d}-\frac{R}{G} + 2\Lambda_dd\text{vol}
\end{equation}
where the integral of scalar curvature $R$ is interpreted
combinatorially \cite{recon} according to Regge calculus. No boundary term
arises since $Y^d$ is doubled. $G$ is Newton's constant
manifesting the strength of gravity, and $\Lambda$ is a bare cosmological
constant. Integrals over superpositions are to be extended linearly
weighting by $|\text{amplitude}|^2$. The total action has the form
\begin{align}
  \label{eq:saction}
  S(Y) &=
  \sum_{d=0}^{\infty}c_d\left[S_d(Y_{l_d,k}^d)
    + \sum_{k=0}^{k_d}f_d(Y^{k-1,d}\rightarrow Y^{k,d})\right]\\\nonumber
  &\qquad\qquad+ \sum_{d=0}^{\infty}\sum_{k=0}^{k_d}g_d|Y^{k,d}|^2 + \textrm{kinetic term}
\end{align}
We define the \underline{volume} term  $|Y^{k,d}|^2:=\sum|b_{l_d}^{k,d}|^2$. Clearly, $S$ depends on constants $G, \Lambda_d, c_d, f_d$, and $g_d$ and the interesting regime appears to be for all $d$, $c_d\cdot f_d\ll g_d$. There are further hidden parameters in each dimension $d$. As in \cite{nonpert}, the time-like edges of all Lorentzian simplices should have (length$^2$) = $-\alpha_da$, where as Euclidean edges have (length$^2) = a$. We wish to take the constants, or random variable, $\alpha_k$ well within the ``C-phase'' of \cite{quantum}, where the CDT growth process produces roughly deSitter-like space-times.  For large values of $c_n$, the least action principle suggests that the measure $e^{-S(Y)}$ will concentrate on formal chains which terminate, i.e. achieve all $b_{l_d}^{k,d}\equiv 0$, in the lowest possible dimension $d$, which is believed to be $d=4$. (See Appendix A for the open mathematical point.) So we expect formal chains to terminate almost surely with a linear combination of $X^4$'s.

If a classical chain $c$ has amplitude $b_i$ in formal chains $\bbar{c}_i$ with normalized amplitudes $a_i$ (computed from the action $S$), then $c$ has probability $\sum_ia_i^2b_i^2$. This and other aspects of the 2-field theory formalism are described in Appendix B.

The CDT growth process adds foliated  layers of $d$ dimensional
Lorentz simplices to an initial $(d-1)$ dimensional Euclidean slice
with aspect ratio length$^2\cdot$(time-like edges)/length$^2\cdot$(space-like
edges) = $-\alpha_d$. In figure \ref{fig:aspect_ratio},
this is illustrated for $d = 1,2,3$ where for $d=2$ both a partial and full
layer is illustrated and for $d=3$ only a partial layer is
shown. After Wick rotation to a Euclidean metric on $Y_{l_d}^{k_d,d}$
the simplices will have a distribution of Euclidean lengths so the
condition for $\alpha_{d+1}$ to be in phase $C$ above will be
different from the numerically determined range in \cite{quantum}.

The definition for $S_d$ (the Einstein Hilbert action with cosmological constant in dimension $d$) should be constructed to give a highly negative result ($R$ near $-\infty$) for $Y^d$, a double of $X^d$, unless $X^d$ is a non-singular Lorentzian $d$-manifold with all spatial boundary. Thus the action blows up unless the constituent components of each $X^d$ are Lorentzian manifolds and so the growth process concentrates on formal chains of manifolds --- not singular spaces --- of increasing dimension.

We have not been explicit on this point until now but as in \cite{nonpert}, our
CDT-like growth process should not be confined to building product
collars, but rather it should allow singularities in the spatial
foliations consistent with the creation of arbitrary $d$ dimensional
cobordisms
$(X^d;X_+^{d-1},X_-^{d-1})$, consistent with the single restriction on
Euler characteristics:
$\mathcal{X}(X^d)=\mathcal{X}(X_{\pm}^{d-1})$. This
condition guarantees a relative reduction of the tangent bundle of
$X^d$ to $SO(d-1,1)$ and thus a $(d-1,1)$ signature pseudo-Riemannian
metric. Fugacities for various foliation singularities must be regulated to obtain the benefits of CDTs, i.e. emergent
geometry on the components of $X^d$ with Hausdorff dimension $\approx
d$.

Topological variability in the CDT growth through the dimensions
allow the second, topological terms $g_d|Y^d|^2$ to vary. Recall
that $X^d$ may be a superposition. A sufficiently large constant $g_d$ in $S$
will, for $d\leq 3$, punish superpositions where ``the collection of
terms into isometry classes'' subsequent to mirror doubling and fluctuation is reinforcing, and encourage
cases where the collection involves cancellation. This can be seen in
Example \ref{ex:1D} below. It is a topological theorem (at least for finite
superpositions) that not until $d=4$
is reached can cancellation be complete.

\begin{example}\label{ex:1D}
  Topological (not Lorentzian) $X^1$ paired with itself.
  \begin{center}
    \begin{displaymath}
      X^1 = \frac{1}{2} \includegraphics[width=0.2in]{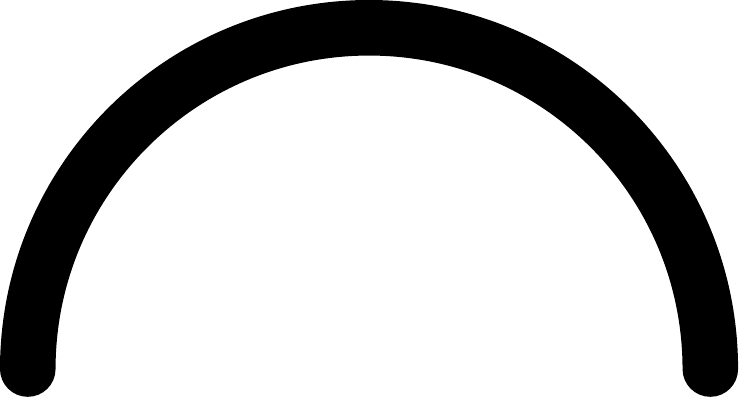} -\frac{1}{2} \includegraphics[width=0.2in]{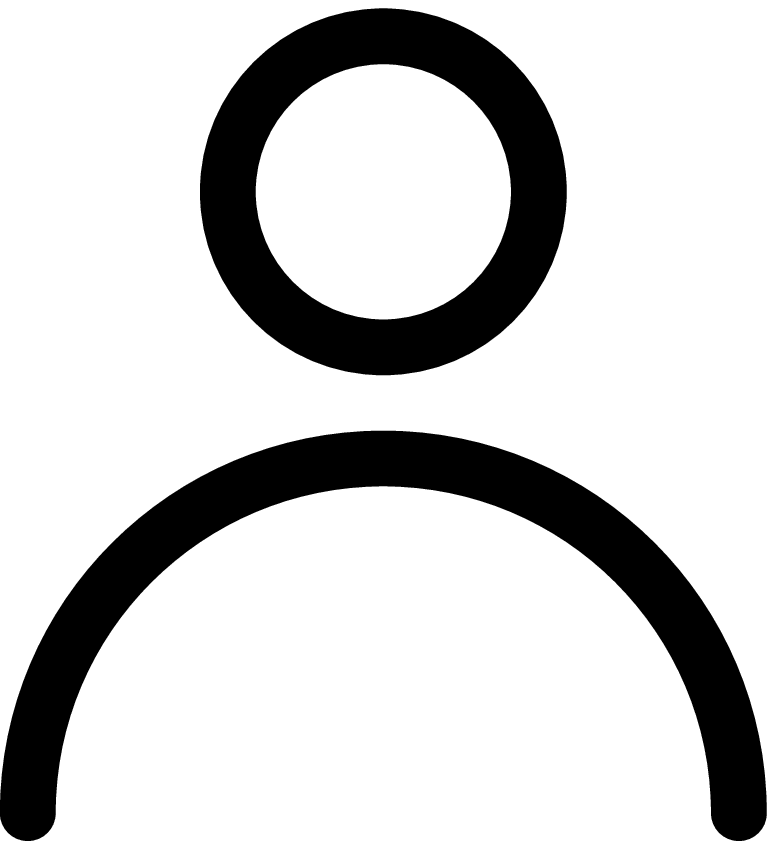} -\frac{1}{2} \includegraphics[width=0.15in]{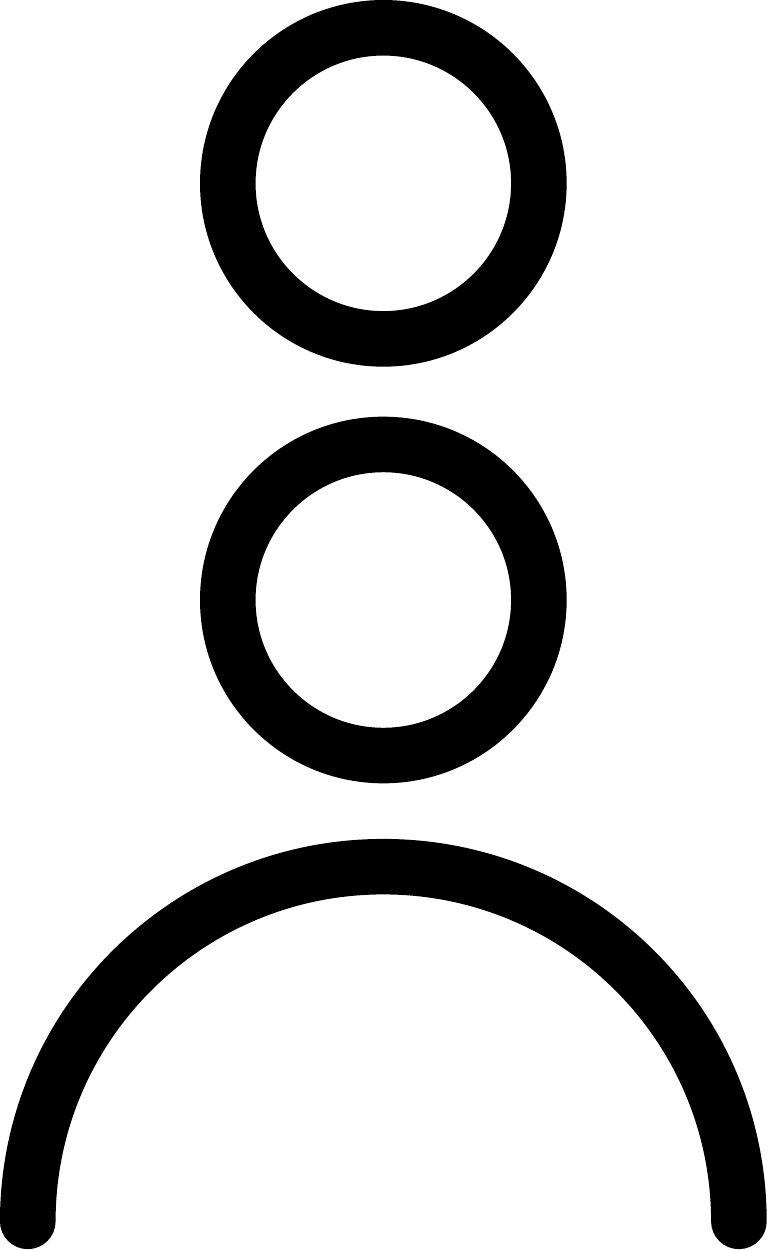} +\frac{1}{2} \includegraphics[width=0.2in]{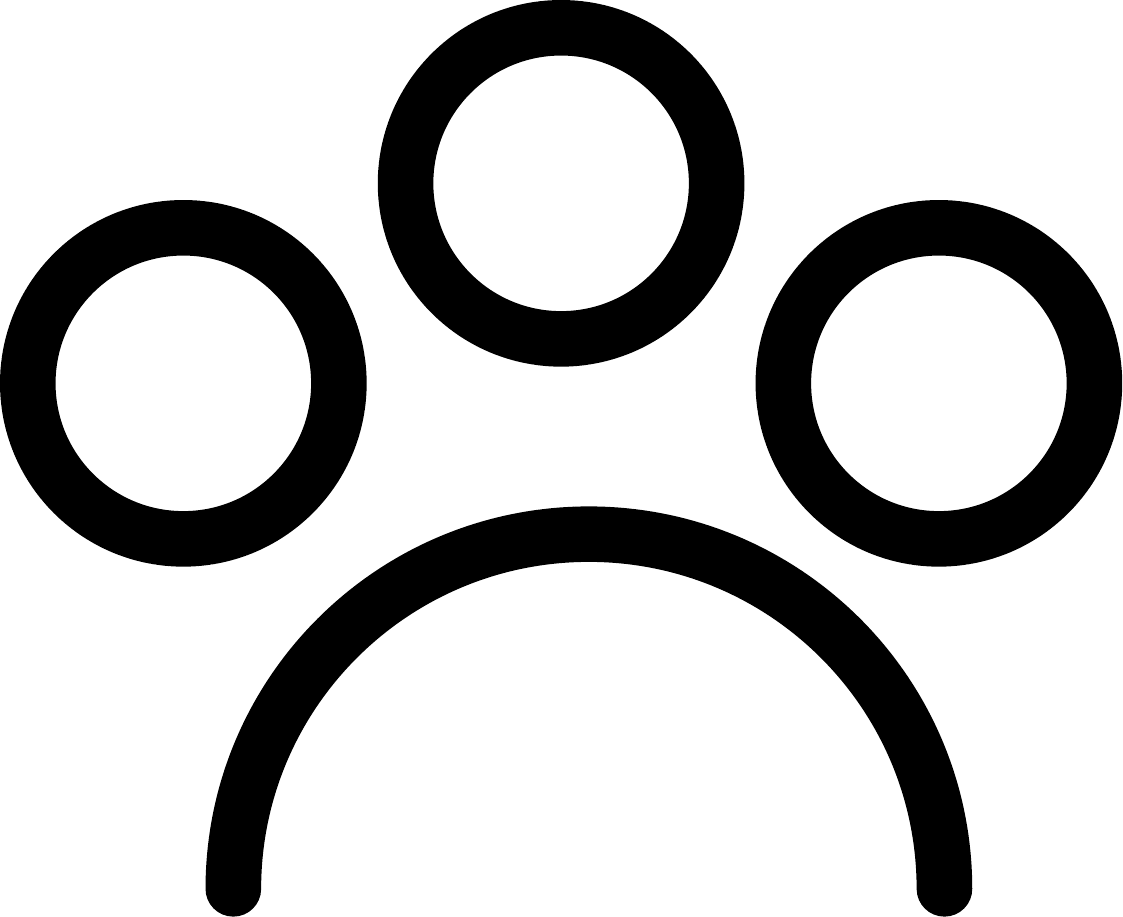}
    \end{displaymath}
    \begin{align*}
      Y^1 =& \frac{1}{4} \raisebox{-0.05in}{\includegraphics[height=0.2in]{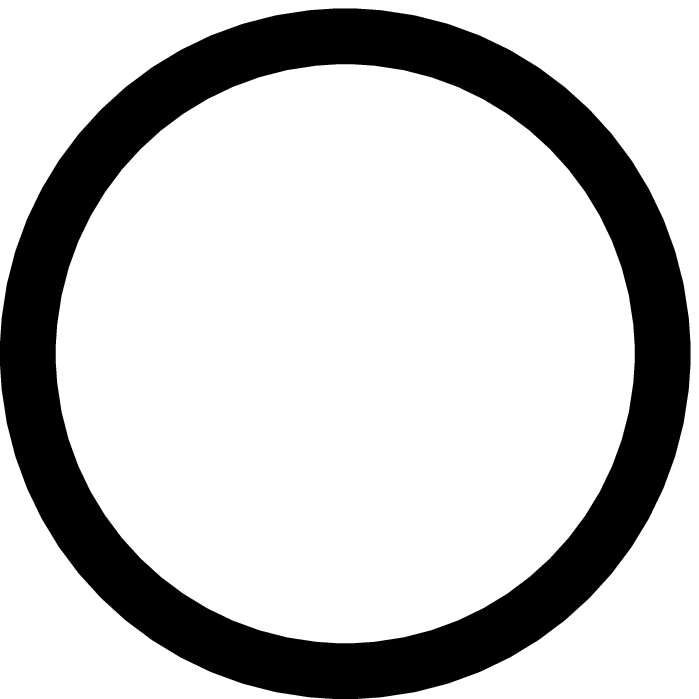}} -\frac{1}{4} \raisebox{-0.05in}{\includegraphics[height=0.2in]{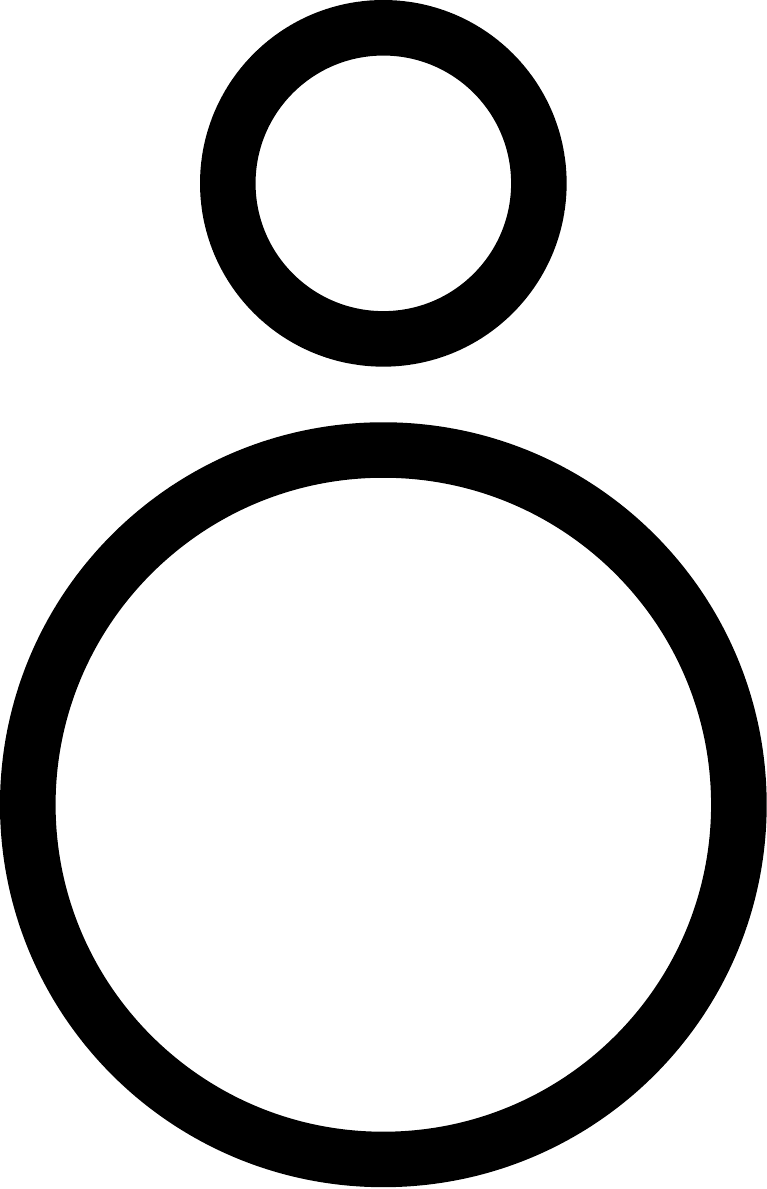}} -\frac{1}{4} \raisebox{-0.05in}{\includegraphics[height=0.25in]{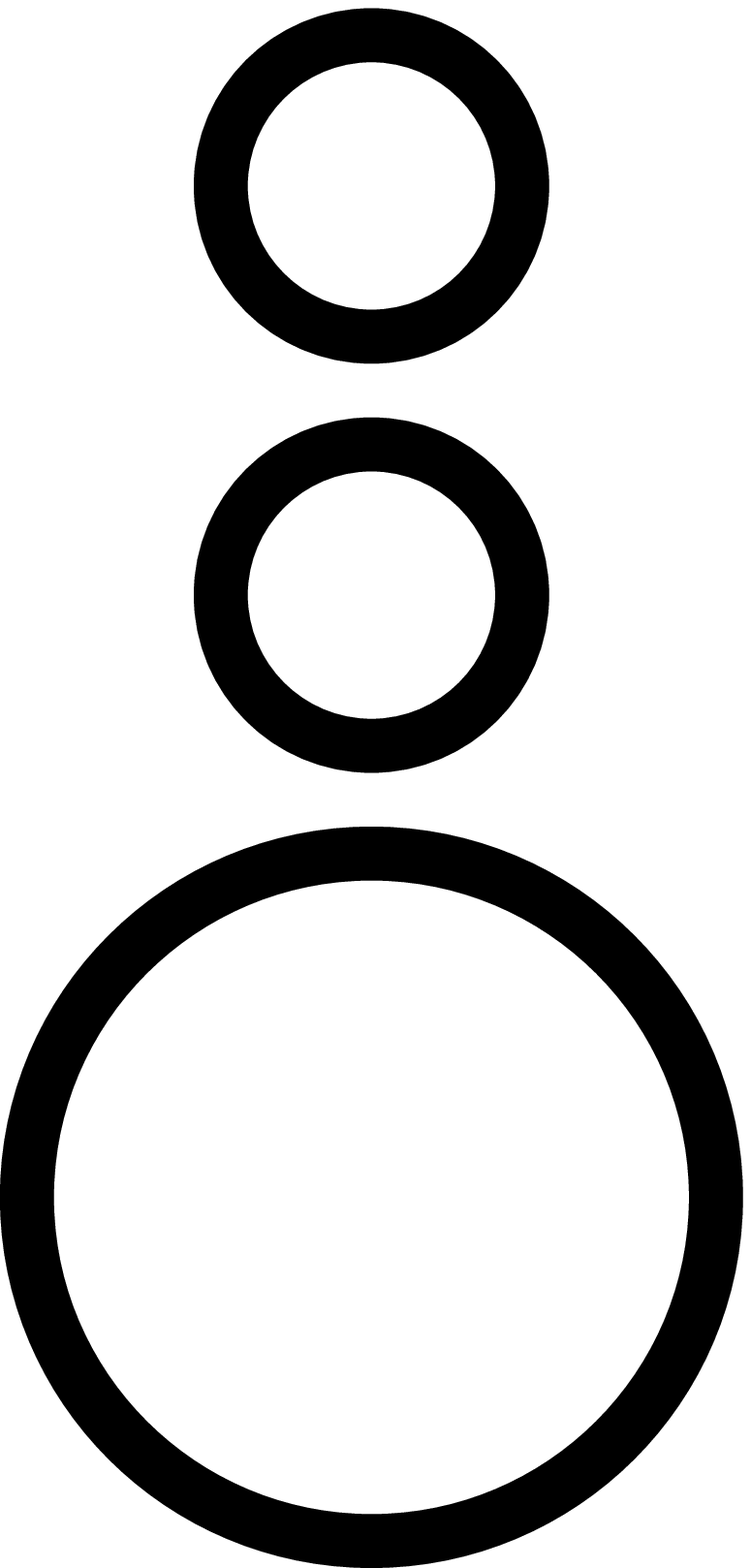}} +\frac{1}{4} \raisebox{-0.05in}{\includegraphics[height=0.2in]{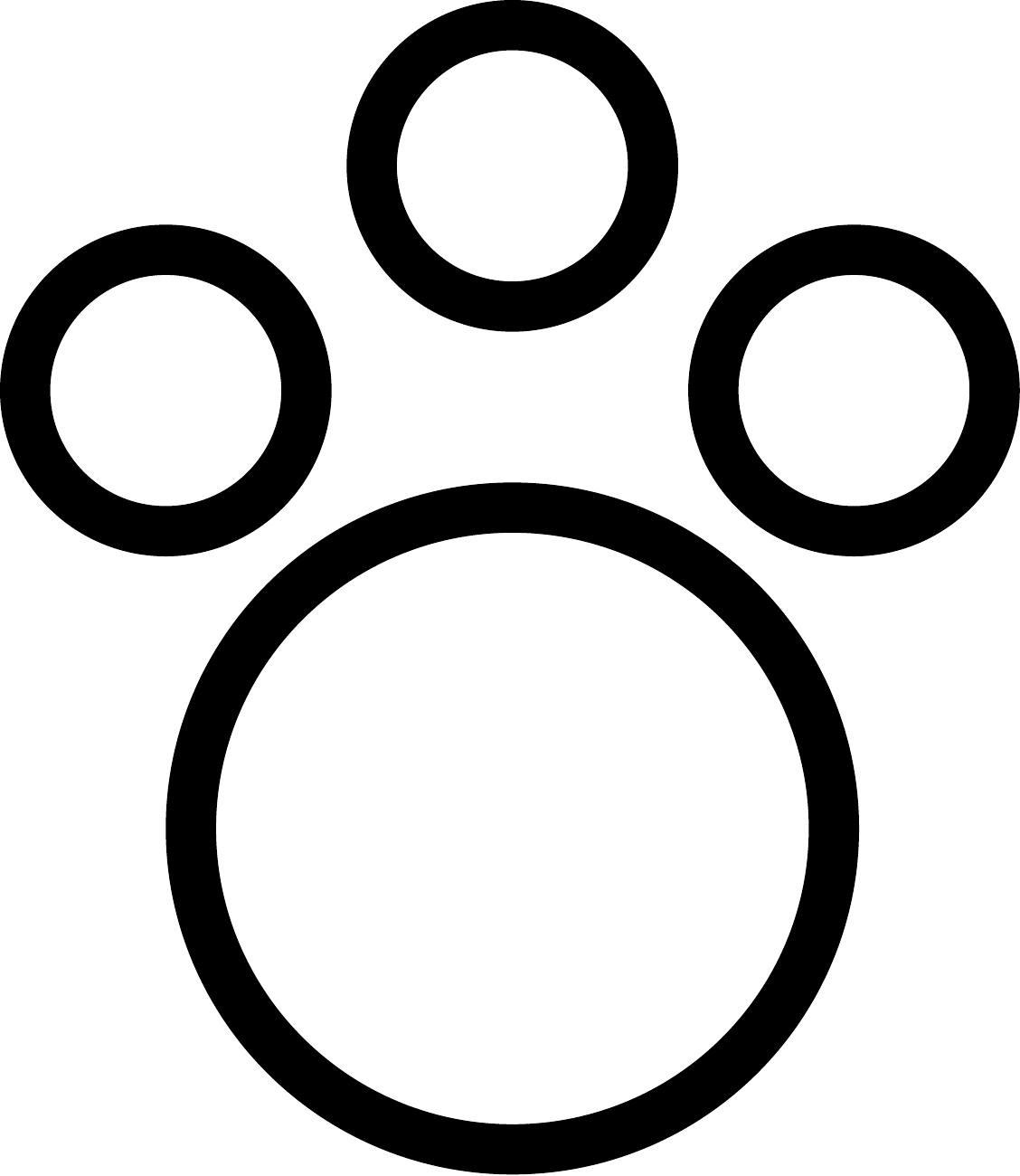}}\\
      & -\frac{1}{4} \raisebox{0.12in}{\rotatebox{180}{\includegraphics[height=0.2in]{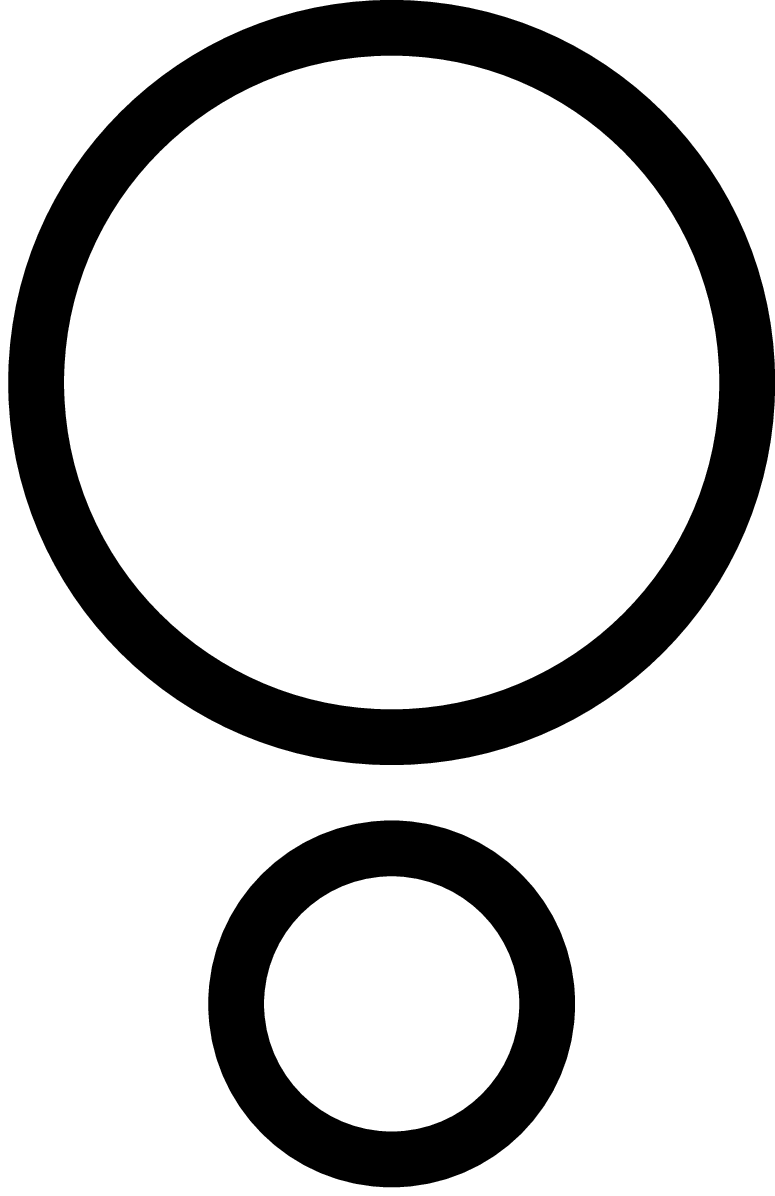}}} +\frac{1}{4} \raisebox{-0.05in}{\includegraphics[height=0.25in]{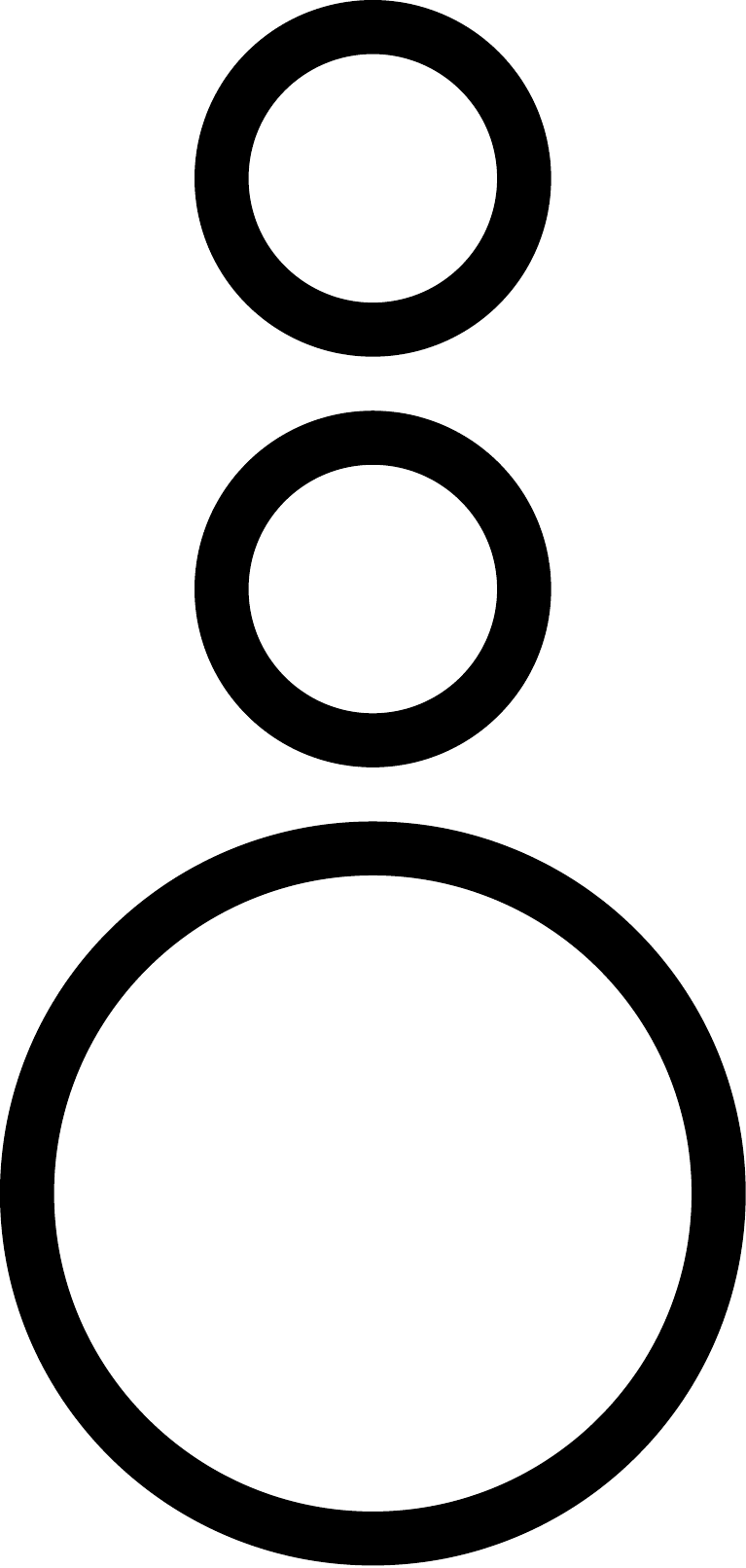}} +\frac{1}{4} \raisebox{-0.05in}{\includegraphics[height=0.2in]{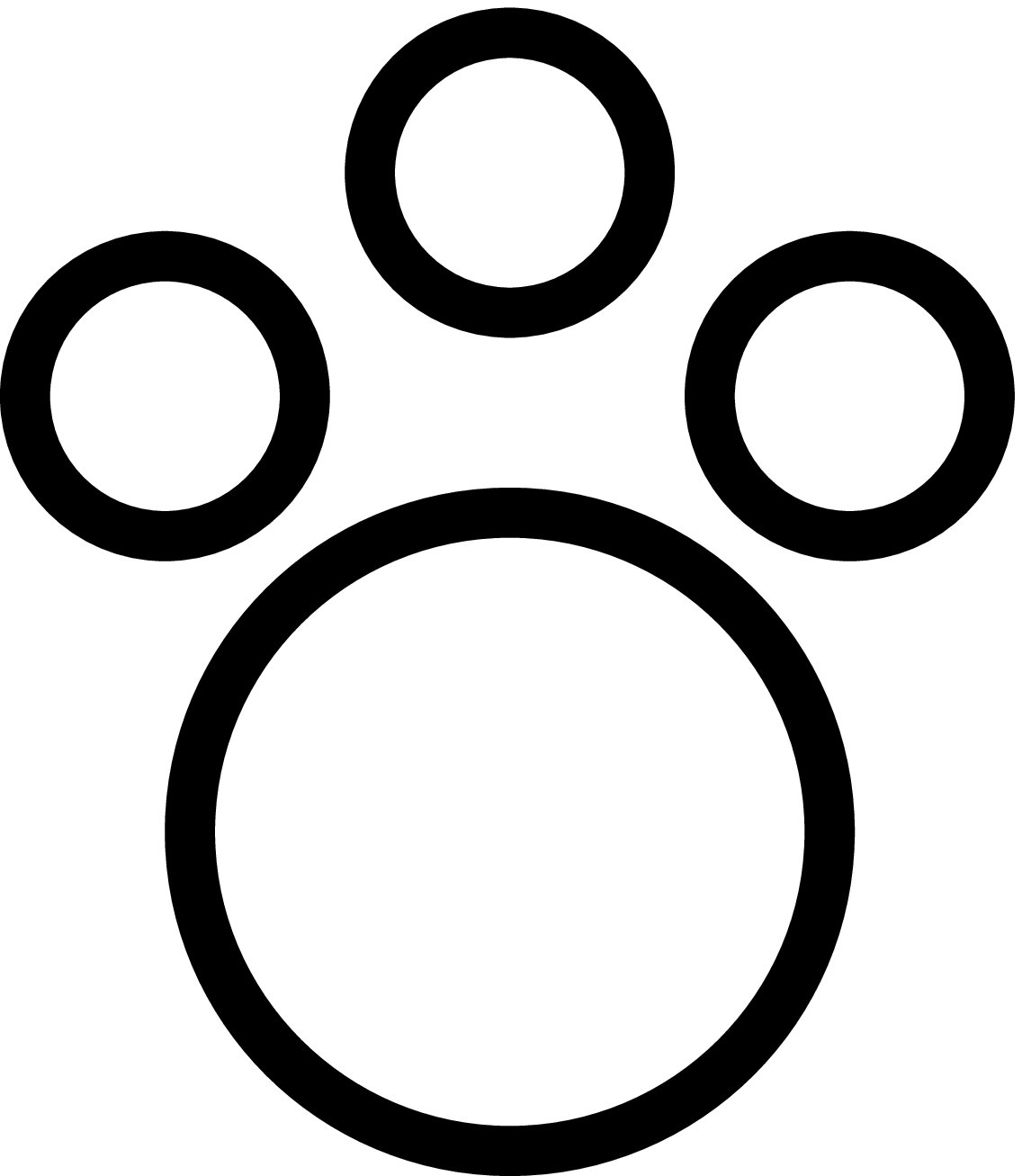}} -\frac{1}{4} \raisebox{-0.07in}{\includegraphics[height=0.25in]{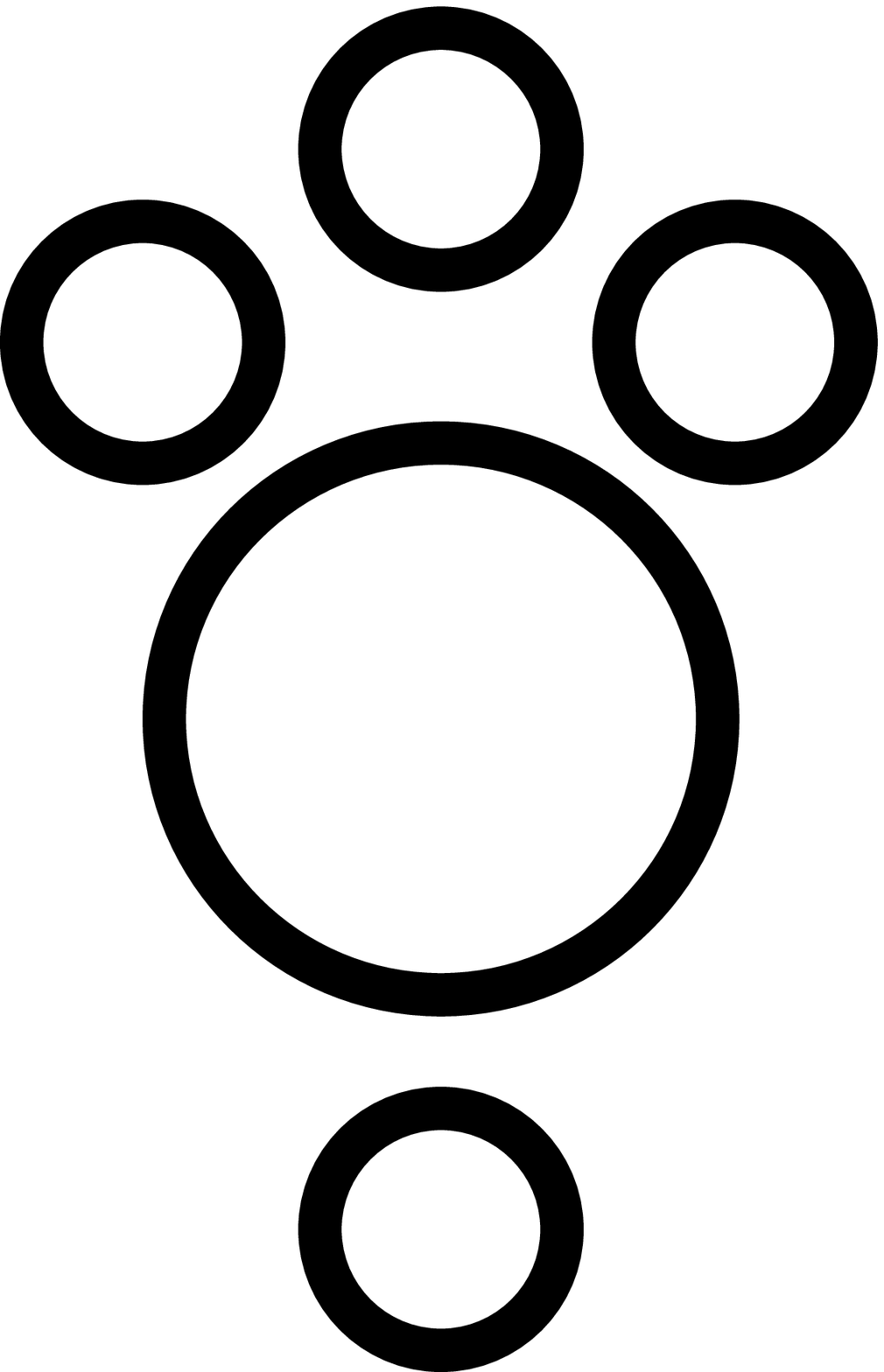}}\\
      & -\frac{1}{4} \raisebox{0.15in}{\rotatebox{180}{\includegraphics[height=0.25in]{figures/circle3}}} +\frac{1}{4} \raisebox{-0.1in}{\includegraphics[height=0.3in]{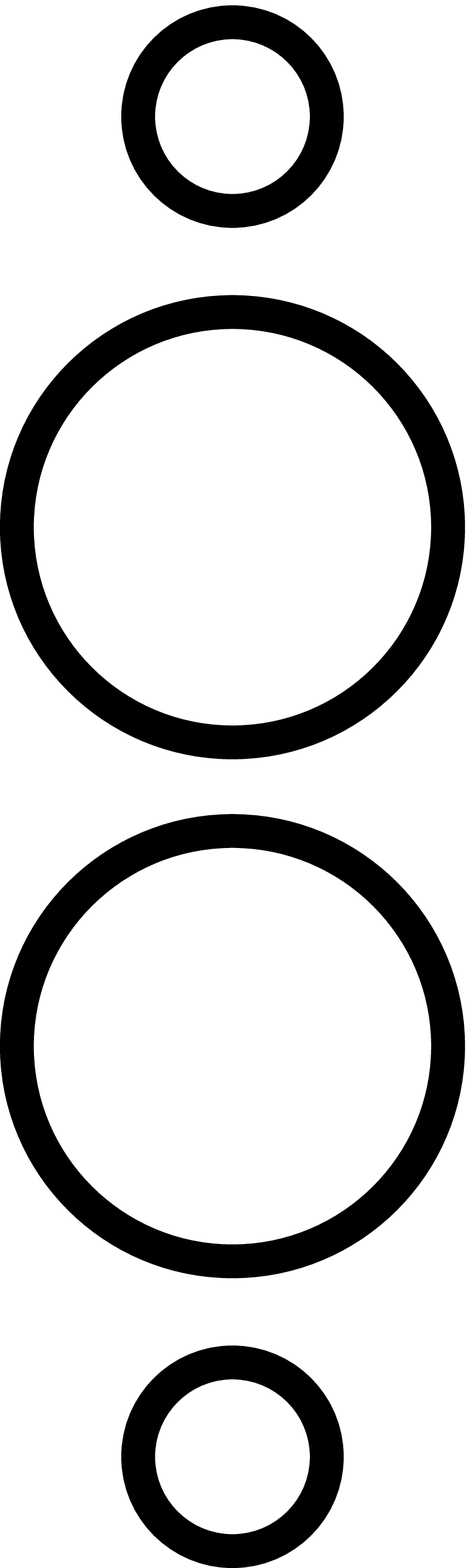}} +\frac{1}{4} \raisebox{-0.13in}{\includegraphics[height=0.35in]{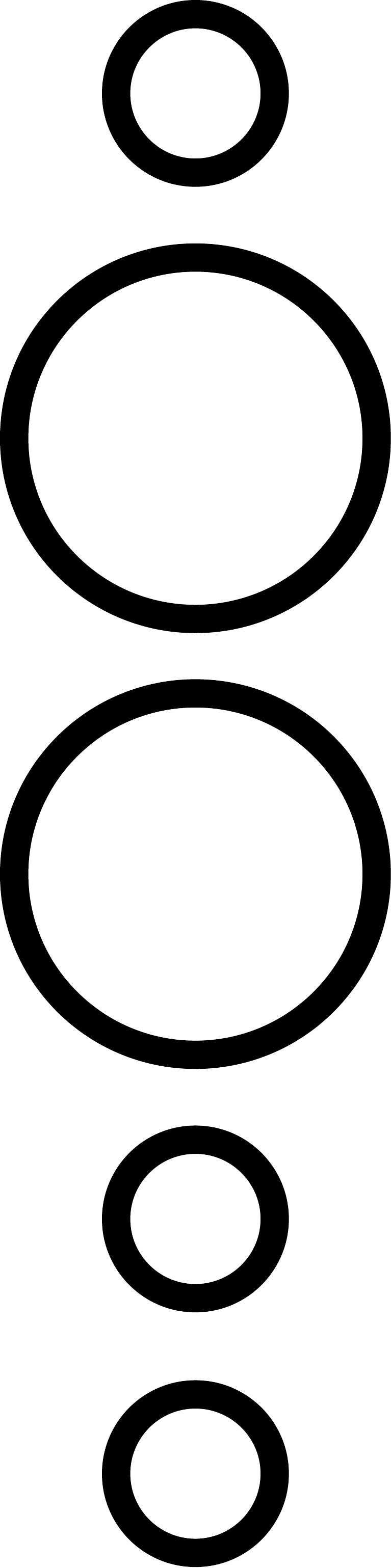}} -\frac{1}{4} \raisebox{-0.1in}{\includegraphics[height=0.3in]{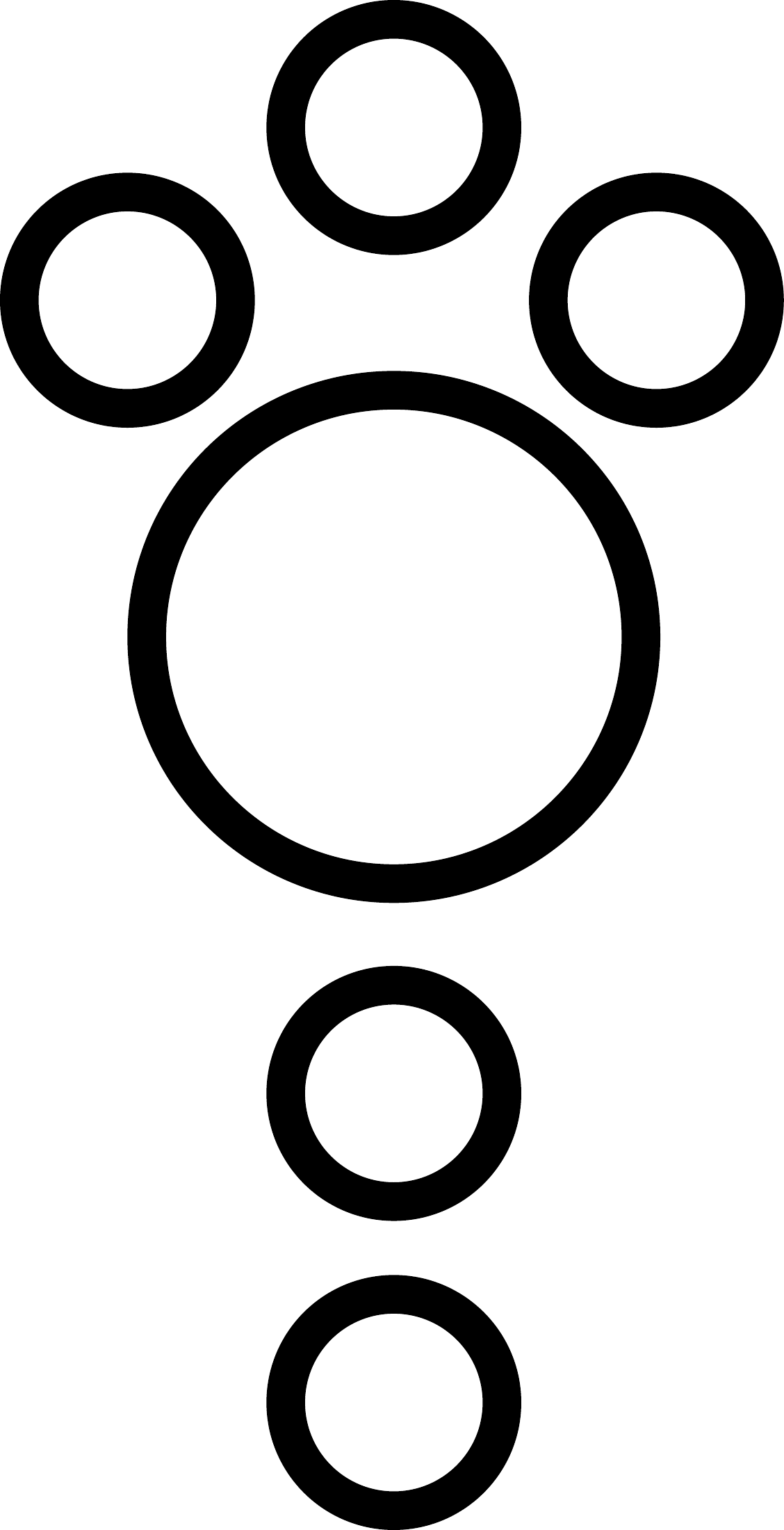}}\\
      & +\frac{1}{4} \raisebox{0.13in}{\rotatebox{180}{\includegraphics[height=0.2in]{figures/circle4}}} -\frac{1}{4} \raisebox{0.18in}{\rotatebox{180}{\includegraphics[height=0.25in]{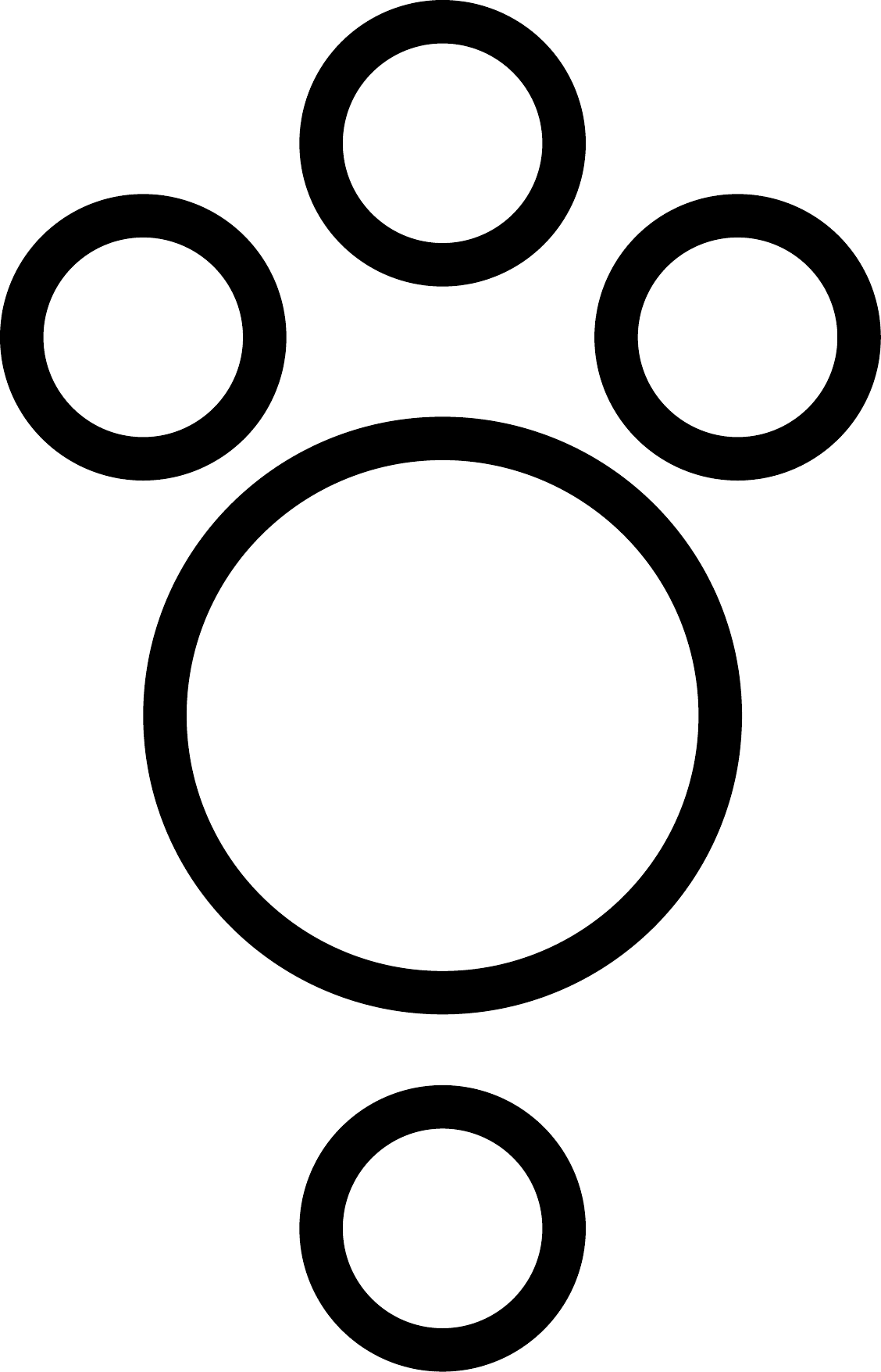}}} -\frac{1}{4} \raisebox{0.2in}{\rotatebox{180}{\includegraphics[height=0.3in]{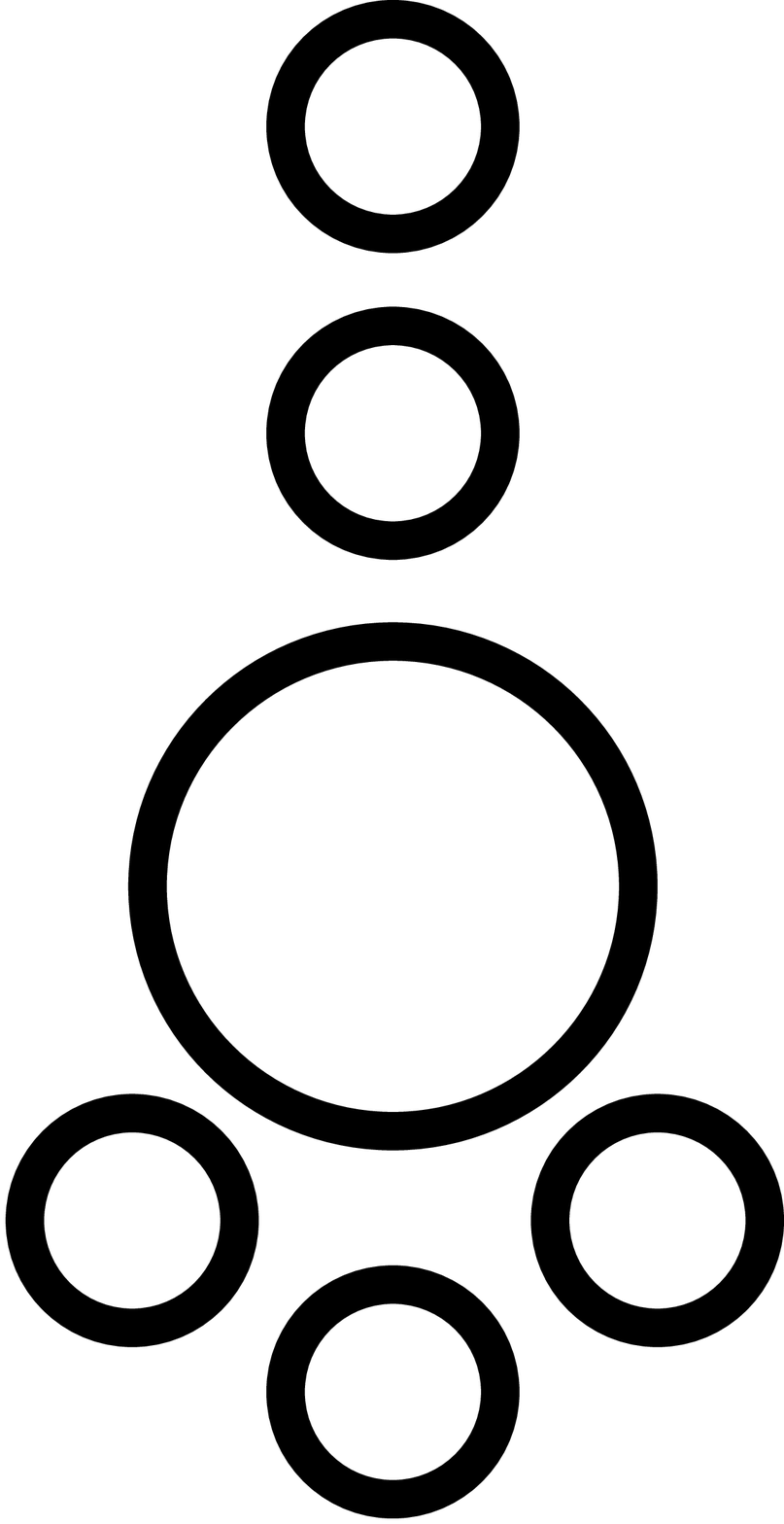}}} +\frac{1}{4} \raisebox{-0.08in}{\includegraphics[height=0.25in]{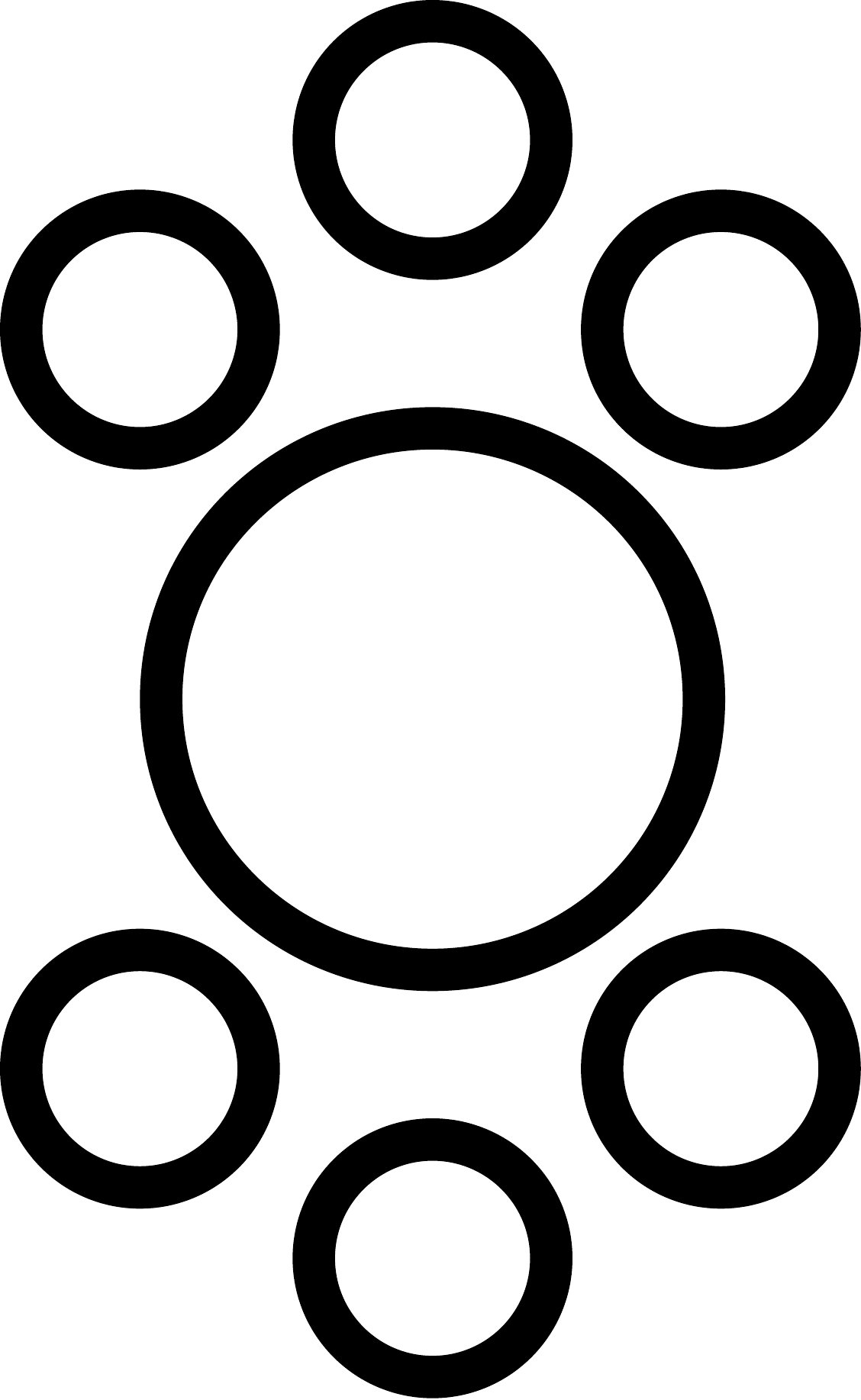}}\\
      =& \frac{1}{4}
      \begin{pmatrix}
        \includegraphics[height=0.2in]{figures/circle}
      \end{pmatrix}
      -\frac{1}{2}
      \begin{pmatrix}
        \includegraphics[height=0.1in]{figures/circle}\\[-0.05in]
        \includegraphics[height=0.1in]{figures/circle}
      \end{pmatrix}
      -\frac{1}{4}
      \begin{pmatrix}
        \includegraphics[height=0.1in]{figures/circle}\\[-0.05in]
        \includegraphics[height=0.1in]{figures/circle}\\[-0.05in]
        \includegraphics[height=0.1in]{figures/circle}
      \end{pmatrix}
      +1
      \begin{pmatrix}
        \includegraphics[height=0.07in]{figures/circle}\\[-0.06in]
        \includegraphics[height=0.07in]{figures/circle}\\[-0.06in]
        \includegraphics[height=0.07in]{figures/circle}\\[-0.06in]
        \includegraphics[height=0.07in]{figures/circle}        
      \end{pmatrix}\\
      & -\frac{1}{4}
      \begin{pmatrix}
        \includegraphics[height=0.07in]{figures/circle} &
        \includegraphics[height=0.07in]{figures/circle}\\[-0.06in]
        \includegraphics[height=0.07in]{figures/circle} &
        \includegraphics[height=0.07in]{figures/circle}\\[-0.06in]
        \includegraphics[height=0.07in]{figures/circle}
      \end{pmatrix}
      -\frac{1}{2}
      \begin{pmatrix}
        \includegraphics[height=0.07in]{figures/circle} &
        \includegraphics[height=0.07in]{figures/circle}\\[-0.06in]
        \includegraphics[height=0.07in]{figures/circle} &        
        \includegraphics[height=0.07in]{figures/circle}\\[-0.06in]
        \includegraphics[height=0.07in]{figures/circle} &
        \includegraphics[height=0.07in]{figures/circle}
      \end{pmatrix}
      +\frac{1}{4}
      \begin{pmatrix}
        \includegraphics[height=0.07in]{figures/circle} &
        \includegraphics[height=0.07in]{figures/circle}\\[-0.06in]
        \includegraphics[height=0.07in]{figures/circle} &        
        \includegraphics[height=0.07in]{figures/circle}\\[-0.06in]
        \includegraphics[height=0.07in]{figures/circle} &
        \includegraphics[height=0.07in]{figures/circle}\\[-0.06in]
        \includegraphics[height=0.07in]{figures/circle}
      \end{pmatrix}
    \end{align*}    
  \end{center}
  Terms have been collected according to the topological (actually smooth) type, in this case increasing $|Y^1|^2$.
  \begin{note}
    In absence of cancellation, $|Y^{1}|^2$ would be equal to $|X^1|^2$. In this example, $|Y^1|^2 = \frac{7}{4}$.
  \end{note}
\end{example}

We must now explain a rather unexpected possibility on which this paper rests. There are closed 3-dimensional
manifolds, of the form $Y^3$, i.e. a double which bound two distinct
4-manifolds $X_1^4$ and $X_2^4$ with $\partial X_1^4 = \partial X_2^4
= Y^3$ so that if we set $X^4$ to be the formal 4-manifold
$X^4=X_1^4-X_2^4$, then 
\begin{displaymath}
  Y^4 = \innerprod{X_1^4-X_2^4}{X_1^4-X_2^4} = X_1^4\bbar{X_1^4} -
  X_1^4\bbar{X_2^4} - X_2^4\bbar{X_1^4} + X_2^4\bbar{X_2^4} = 0\in \M_{\varnothing}^4.
\end{displaymath}
This happens because the four closed 4-manifolds appearing in the final sum are all diffeomorphic (equivalently P.L. homeomorphic.) (Similar examples are constructed in \cite{manifold_pairing}.) For some  $Y^3$ and with little additional work (see Appendix B), we can ensure $Y^3$ is a double and $\mathcal{X}(X_1^4) = \mathcal{X}(X_2^4) = \mathcal{X}(Y^3) = 0$. The final column of figure \ref{fig:terminating_schematic} illustrates such a cancellation ($Y^4 = 0$).

This kind of cancellation occurs in gluing manifolds of dim $d\geq
4$ \cite{fungroups},\cite{manifold_pairing} but does not occur in gluing manifolds of dimension
$d\leq 3$ \cite{positivity}. Appendix B discusses what is known beyond the case of finite combinations considered
loc. cit. in the context of completed pairings $\innerprod{~}{~}^{\wedge}$ on
$L^2$ sequences of amplitude labeled manifolds $X^{\wedge} =
\sum_ia_iX_i, \sum_i|a_i|^2 = 1$.

Because collecting terms may show $Y_{k_4}^4=0\in\M_{\varnothing}^4$, the
chain $Y$ may terminate in dimension 4 (or possibly higher) with
$X^4$. Given that the constants $g_d$ are large, terminating chains are
energetically favorable. If $g_d$ is sufficiently large, we expect
energy to dominate entropy and effectively \underline{all} formal chains
terminate at dimension 4. Thus, even at finite temperature $\beta$, a
basic topological dichotomy may control the macroscopic dimension of
space within this class of models.

We now describe the ``kinetic'' interaction included in \eqref{eq:saction}. While formal smooth 4D spaces may cancel to zero when paired, cancellation never can occur in any dimension if the spaces come equipped with fixed triangulations and if the notion of isomorphism is restricted to a simplicial piecewise linear bijective map (or isometry). (To prove this, apply the discussion of ``graph-pairings'' within \cite{positivity} to the dual graph to the codimension 1 simplicies of the fixed triangulations.) However, the fugacity $f$ for geometric fluctuation $Y_{l_d}^{k-1,d}\rightarrow Y_{l_d}^{k,d}$ softens the pairing and permits cancellation and termination in dimension 4. But, these fluctuations are too much of a good thing as they allow cancellation in lower dimensions as well between terms that, while differing combinatorially, have identical topology and coefficients of opposite sign (see example \ref{ex:cancellation}.) The job of the kinetic term is to prevent, cancellation ($|\tilde{Y}^d| = 0$) for $d<4$. When $d=4$, a new topological phenomenon arises and enforces cancellation for a new reason.

Here is an example of a potential --- fluctuation induced ---
cancellation in dimension $d=1$ in the combinatorial category.

\begin{example}\label{ex:cancellation}
  \begin{eqnarray*}
    X &=&\includegraphics[width=2.5pt]{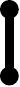}~ - ~\includegraphics[width=2.5pt]{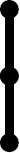}\\
    \innerprod{X}{X} &=& \includegraphics[height=12pt]{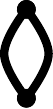}~-~2~\includegraphics[height=12pt]{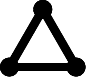}~+~\includegraphics[height=12pt]{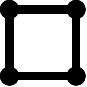}\\
    &
    \begin{matrix}
      \text{{\scriptsize 2 fluctuations}}\vspace{-7pt}\\
      \text{{\Huge $\leadsto$}}
    \end{matrix} & \includegraphics[height=12pt]{figures/loop3.pdf}~-~2~\includegraphics[height=12pt]{figures/loop3.pdf}~+~\includegraphics[height=12pt]{figures/loop3.pdf}~=0\in\M_{\varnothing}^1
  \end{eqnarray*}
\end{example}

This process has the potential to stop the growth of space in
dimension one but may be thwarted by the ``kinetic term'' in $S$. If
component combinatorial spaces $Y_{l_d}^{k,d}$ and
$Y_{l_d}^{k+1,d}$ of a chain $Y$ differ by a geometric Pochner move
(or elementary metric change), we call them ``nearest neighbors.'' The
kinetic term, similar to $-h_d\sigma^x$
in lattice spin models, penalizes disparity in amplitudes $b_{l_d}^{k,d}$
for $Y_{l_d}^{k,d}$ and $b_{l_d}^{k+1,d}$ for $Y_{l_d}^{k+1,d}$
within the formal chain $Y$ by adding 
\begin{equation}
  \label{eq:add_action}
  -2h_d\left|b_{l_d}^{k,d}-b_{l_d}^{k+1,d}\right|^2
\end{equation}
to the action for all nearest neighbor pairs. The strongest kinetic term would be a hard gauge-like constraint requiring terms differing by Pochner moves to have equal phases. The purpose of (earlier) permitting a non-zero amplitude for singular Lorentzian spaces is to allow the kinetic term to act across these and thus stiffen the phase not merely across spaces $X$ with equivalent causal structures, but also between spaces $X_1$ and $X_2$ that are just relatively diffeomorphic, but have unrelated causal structures. Without this, we would encounter even in dimensions 2 and 3 nontrivial light-like vectors like $X_1-X_2$, since the terms $X_i\bbar{X}_j$ are all diffeomorphic for $1\leq i,j\leq 2$.

There is an interesting statistical mechanics problem implicit in the kinetic term. It concerns the stiffness of the space of formal chains under linkages via nearest neighbor components (as above). In a nutshell, if one considers a graph $G$ whose vertices are formal chains and whose edges are induced by (weighted) nearest neighbor occurrences, one asks if the first eigenvalue $\lambda_1$ of the graph Laplacian is positive, i.e. is $G$ gapped or gapless? Is there a region in the large space of model parameters (but constrained by the necessity to lie in ``C-phase'' in all dimensions $d=1,2,3,4$) for which $\lambda_1(G)>0$? In this situation, by setting $h_d$ large enough, the various strands of the chain $Y$ with differing combinatorial geometry but agreeing topology will be so stiffly bound together in phase that any cancellation of the type shown in example \ref{ex:cancellation} would be energetically unfavorable. The formal chain would be forced to develop up to dimension four where $\tilde{Y}^4$ can cancel out for topological reasons.

If it turns out there is no suitable regime in which $\lambda_1(G)>0$,
there  are two possible solutions: (1) to make phase coherence of
nearest neighbors a hard (gauge-like) constraint, or less drastically,
(2) use a non-local kinetic term to stiffen $G$ so that $\lambda_1(G)>0$. 
Non-local means quite
different but P.L. homeomorphic geometries directly interact. In
condensed matter physics, the preference for local interactions is
driven by the ubiquity of charge screening. In constructing a 2-action
for a 2QFT, it was more an aesthetic choice to seek, first, a local
interaction (among formal chains) sufficient to produce a satisfactory
3+1 dimensional phase.

\section{Conclusions}\label{sec:conclusions}

Physics may well be capable of generating in real time any
mathematical tools it requires. This was famously the view of Richard
Feynman. Another view is that new mathematical ideas, in this instance
manifold pairings, may suggest new approaches to physical
problems. Both view may be more or less valid at different times.

This paper begins an exploration of how the positivity of the low
dimensional universal manifold pairing might yield a model for quantum
gravity. The idea is to write a (2-)action $S$ which picks out
something like (3+1)-deSitter space from all possible pseudo metric spaces,
ideally with no assumptions about regularity, dimension, or long scale
structure. We have tried to keep the ingredients abstract
and the action $S$ simple. The results are (only) mildly
encouraging. Enough has been seen to believe that manifold pairings can play a role in quantizing gravity, but it is quite open how best to formulate that role. This paper is a first attempt.

We began with the CDT approach of building P.L. Lorentzian cobordisms
in Euclidean layers but started back at the empty set $\varnothing$,
rather than an initial 3-sphere. To this we add the idea of
superpositions of cobordism, metrical fluctuations, and a doubling
operation $z\rightarrow z\bbar{z}$ modeled on norm square of a complex
number.

Superposition of cobordisms (thought of as paths) is essential to
connect with the idea of manifold pairings. This means that the usual
formalism for integrating over paths is not the correct analog, but
rather one should integrate over superpositions of paths, i.e.
\underline{linearized paths}. This puts us in the realm of 2-field
theory (see Appendix B), which we regard as a bonus. It seems natural
that multiple layers of quantization would be encountered in the trip
back to highest energy.

But overall, we are not completely happy with the notion of a formal chain and the action schema $S$ we have written. Both the formal chain and the action $S$ should be simpler --- more fundamental. Perhaps general partial orders can stand in for Euclidean and Lorentzian (locally) flux simplicial structures (which appear already to assume too much.) Perhaps the fugacity for Euclidean fluctuation can be expressed as a perturbative consequence of \underline{growth}. The goal would be to begin with an elementary combinatorial structure, the complex numbers, and a rather succinct 2-action $S$ and extract space time. Preferably, the dimension $3+1$ would be singled out from all possible $p+q$, whereas we \underline{assume} the form $p+1$. Also, it would be nice to see compactified dimensions emerge.

\begin{figure}[htpb]
  \labellist \small\hair 2pt
  \pinlabel $X^d\left\{\parbox{0.1in}{\vspace{0.18in}}\right.$ at -20 17
  \pinlabel $Y^{d-1}$ at -10 -10
  \pinlabel $Y^{d-1}$ at 570 -10
  \pinlabel $\text{time}$ at 280 15
  \pinlabel $\uparrow$ at 280 30
  \pinlabel $\text{(a)}$ at 117 -35
  \pinlabel $\text{(b)}$ at 443 -35
  \endlabellist
  \centering
  \includegraphics[width=4in]{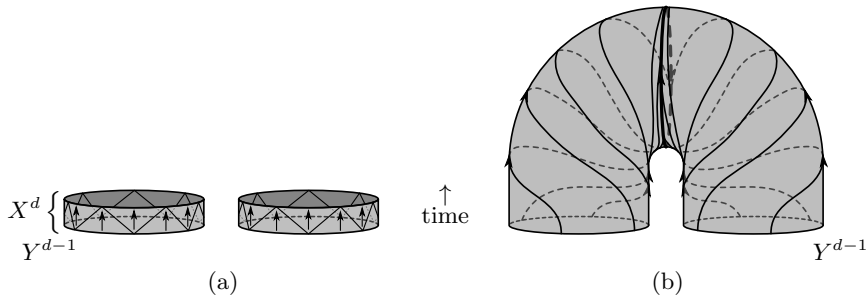}
  \vspace{0.25in}
  \caption{A product (a) and a non-product (b) Lorentzian cobordism.}
  \label{fig:collar}
\end{figure}

Actually, we do see a hint within $S$ (example \ref{ex:cancellation})
that space-time might not be simple deSitter-like but could in some
parameter regimes contain small ``compact'' dimensions. Extra circle
factors are the easiest to understand. There is less action involved
growing a small collar $X^d$ than a cobounding manifold
${X^d}^{\prime}$ with empty ``upper'' boundary (see figure
\ref{fig:collar})

\begin{figure}[htpb]
  \labellist \small\hair 2pt
  \pinlabel $X^d\bar{X}^d\cong Y^{d-1}\times S^1_{\text{small}}$ at 150 -20
  \endlabellist
  \centering
  \includegraphics[height=0.5in]{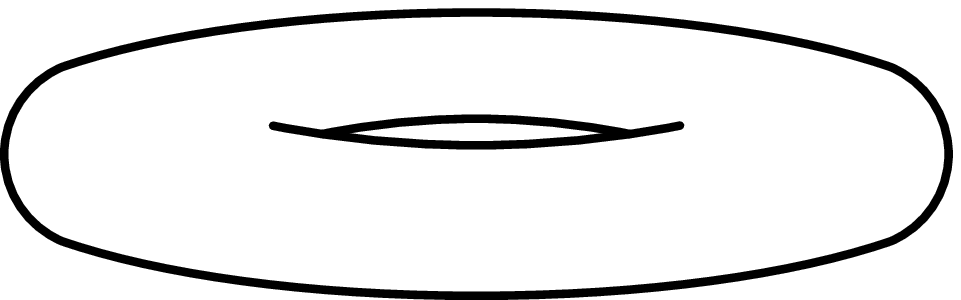}
  \vspace{0.1in}
  \caption{Doubling \ref{fig:collar}(a) above yields a Euclidean manifold with a small circle factor}.
  \label{fig:donuts}
\end{figure}

The cost in action in building compact directions on the way to building a light-like vector $v$ may be small enough that it wins for entropic reasons for some parameter settings of $S$. Although compact tori do not, apparently, lead to the field content required by standard model physics, this way of generating compact tori may be a useful start.

The proposal in this paper may have a falsifiable prediction. In earlier drafts, we hoped to show that $S^3$ has no light-like vector $v$ in its pairing $\innerprod{~}{~}^{\hat{}}_{S^3}$. If this mathematical fact were established, it would seem to exclude $S^3$ as the spatial topology near the ``big bang''. We recently discovered light-like vectors in $\innerprod{~}{~}_{S^3}$ (see Appendix A), but the manifold constituents of $v$ have much more homology than Mazur-like examples (again, see Appendix A) such as $M\# S^1\times S^3$, based on a nontrivial homology sphere $\Sigma:=\partial M$. Thus it is still possible that a non-trivial homology spheres $\Sigma$ may be favored by the action over $S^3$.


\appendix
\renewcommand{\thetheorem}{\Alph{section}.\arabic{theorem}}

\section{Manifold Pairings}

Consider closed oriented $d$-manifolds $\M^d$ of class P.L. (or Diff.) Define $\M_{\varnothing}^{\text{P.L. }d}$ to be the $\C$-vector space consisting of finite linear combinations of P.L. homeomorphisms (alternatively homeomorphism or diffeomorphism) classes of $\M^d$. (Henceforth we treat only the P.L. case.) If $S^{d-1}$ is closed of dimension $d-1$, define $\M_S^d$ to be the $\C$-vector space of finite linear combinations of cobounding $M$, $\partial M = S$, taken up to the equivalence relation of P.L. homeomorphism rel identity$_S$.
\begin{center}
  \labellist \small\hair 2pt
  \pinlabel $M_1$ at 12 60
  \pinlabel $M_2$ at 12 8
  \pinlabel $S$ at 48 33
  \pinlabel $f$ at 8 34
  \pinlabel $\text{inc.}$ at 35 54
  \pinlabel $\text{inc.}$ at 38 14
  \pinlabel $\parbox{1.5in}{$\ket{M_1}=\ket{M_2}$ if and only if there is a P.L. homeomorphism $f$ making this diagram commute}$ at 120 35
  \endlabellist
  \includegraphics[width=1in,height=1in]{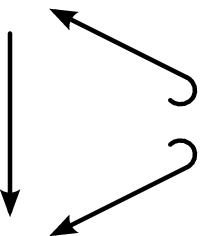}
\end{center}
For each $S$ there is a sesquilinear pairing:
\begin{align*}
  \M_S^d\times \M_S^d & \xrightarrow{\innerprod{~}{~}_S}\M_{\varnothing}^d,\\
  \left(\sum_ia_iM_i,\sum_jb_jN_j\right) & \mapsto \sum_{i,j}a_i\bbar{b_j}M_i\bbar{N_j}
\end{align*}
where $\bbar{~}$ means complex conjugation or orientation reversal according to context.

The literature \cite{manifold_pairing}\cite{positivity}\cite{fungroups} on $\innerprod{~}{~}_S$ may be summarized by:
\begin{theorem} \label{thm:positivity}
  If $d\leq 3$, then for all $S$, $\innerprod{~}{~}_S$ is \underline{positive}, meaning $\innerprod{v}{v}_S=0$ implies $v = 0$. For every $d\geq 4$, there is some $S$ such that for some $v\neq 0\in \M_S^d$, $\innerprod{v}{v}_S = 0$. Such $v$ are called \underline{light-like} and such pairings indefinite. For $d=4$, there are homology spheres $\Sigma^3$ for which $\innerprod{~}{~}_{\Sigma^3}$ is indefinite.
\end{theorem}

In forming superpositions, $L^2$ rather than finite combinations of manifolds would be the more natural setting, so let us see which facts extend formally and which require work. Let a $\hat{~}$ denote $L^2$-completion and let us add hats to the pairing and extend its natural $|~|^2$ evaluation to $\R$.
\begin{displaymath}
  \begin{matrix}
    \M_S^d\times\M_S^d & \xrightarrow{\innerprod{~}{~}_S} & \M_{\varnothing}^d & \xrightarrow{|~|^2} & \R\vspace{10pt}\\ 
    & \raisebox{7pt}{\rotatebox{-90}{$\leadsto$}} & \sum_ic_iY_i & \mapsto & \sum_ic_i\bbar{c_i}w(Y_i)\vspace{10pt}\\
    {\M^{\wedge}}_S^d\times {\M^{\wedge}}_S^d & \xrightarrow{\innerprod{~}{~}_S^{\wedge}} & {\M^{\wedge}}_{\varnothing}^d\cup\infty & \xrightarrow{|~|^2} & \R\cup\infty
  \end{matrix}
\end{displaymath}
where $w$ is a weight function on $\{Y_i\}$, which for convenience we take to be $w(Y_i) = 1$.

Notice that $\innerprod{~}{~}_S^{\wedge}$ may not land in square summable sequences --- hence the symbol $\infty$. For example, let $S = S^1$, the circle, and let
\begin{eqnarray*}
  v &=& \includegraphics[height=30pt]{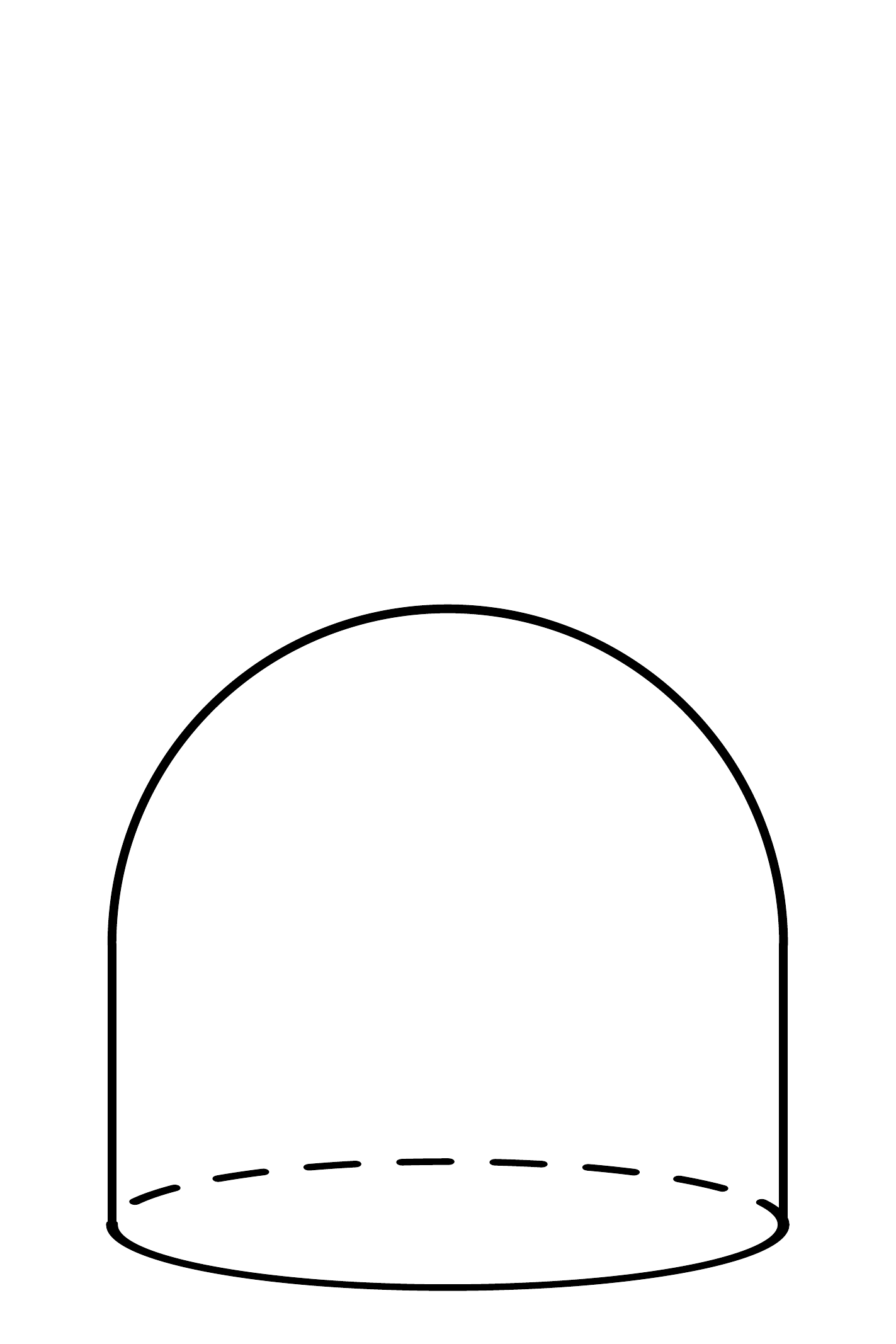} + \frac{1}{2}\includegraphics[height=30pt]{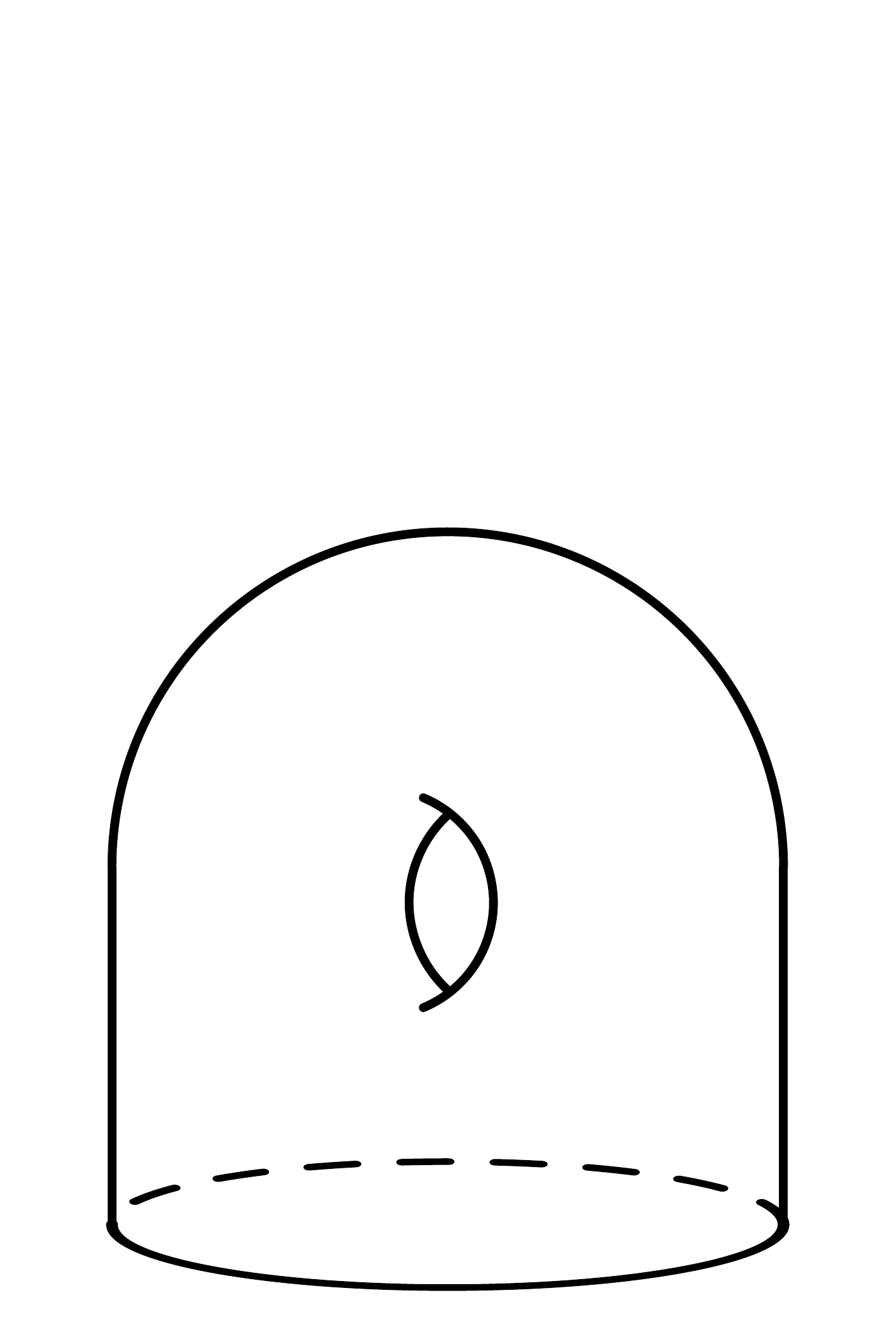} + \frac{1}{3}\includegraphics[height=30pt]{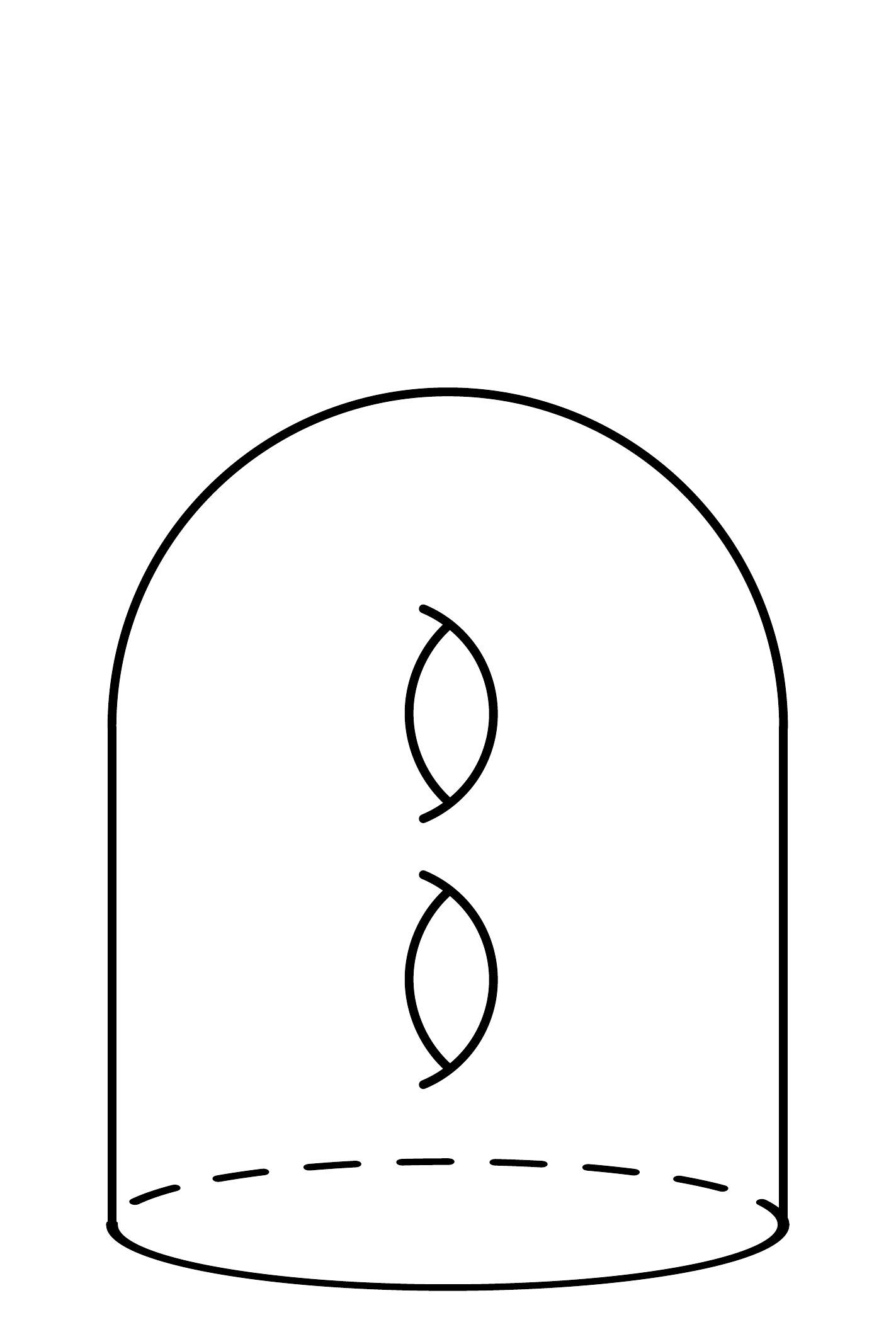}+\frac{1}{4}\includegraphics[height=30pt]{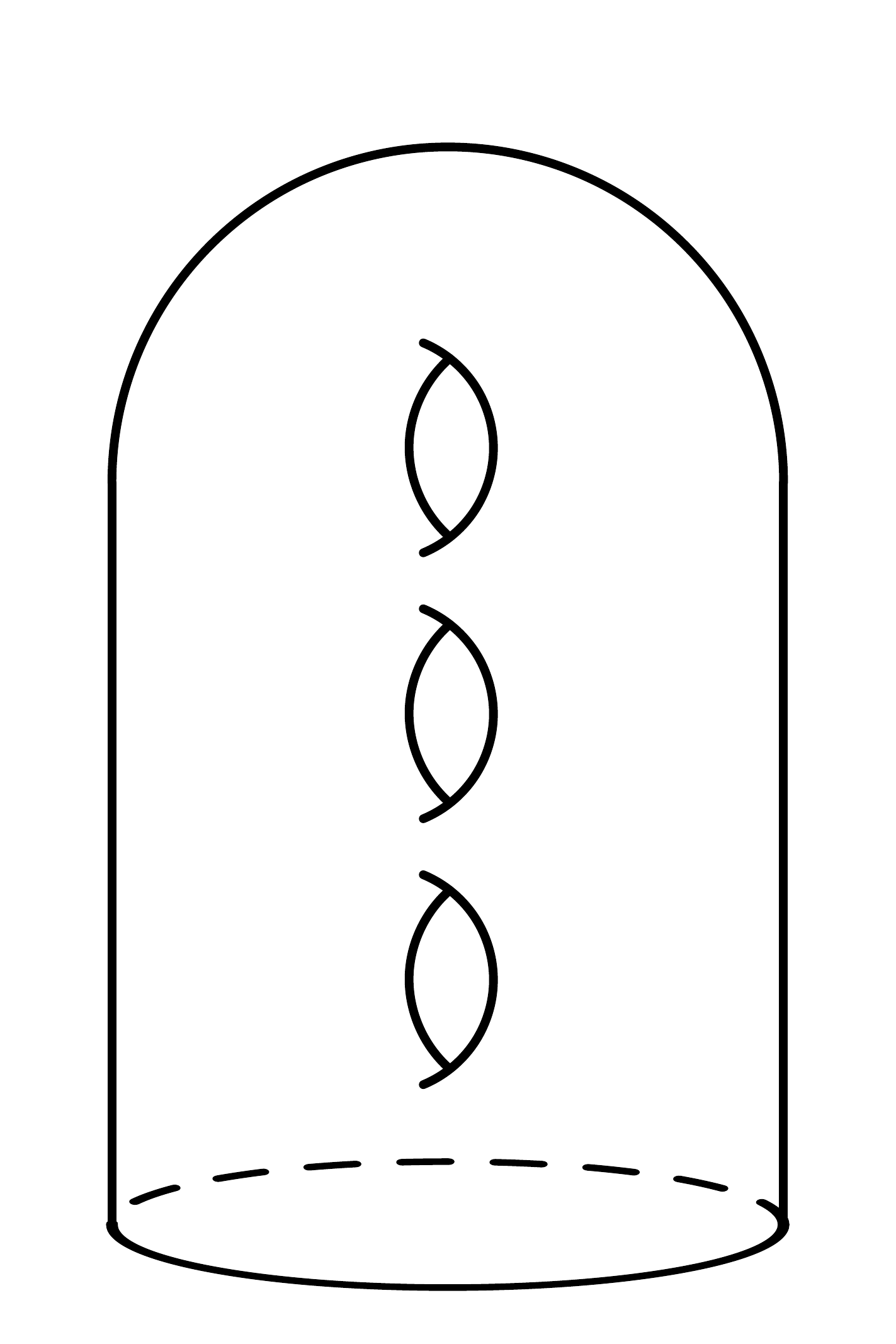}+\ldots\\
  \innerprod{v}{v}^{\hat{}}_S &=& 1\includegraphics[height=30pt]{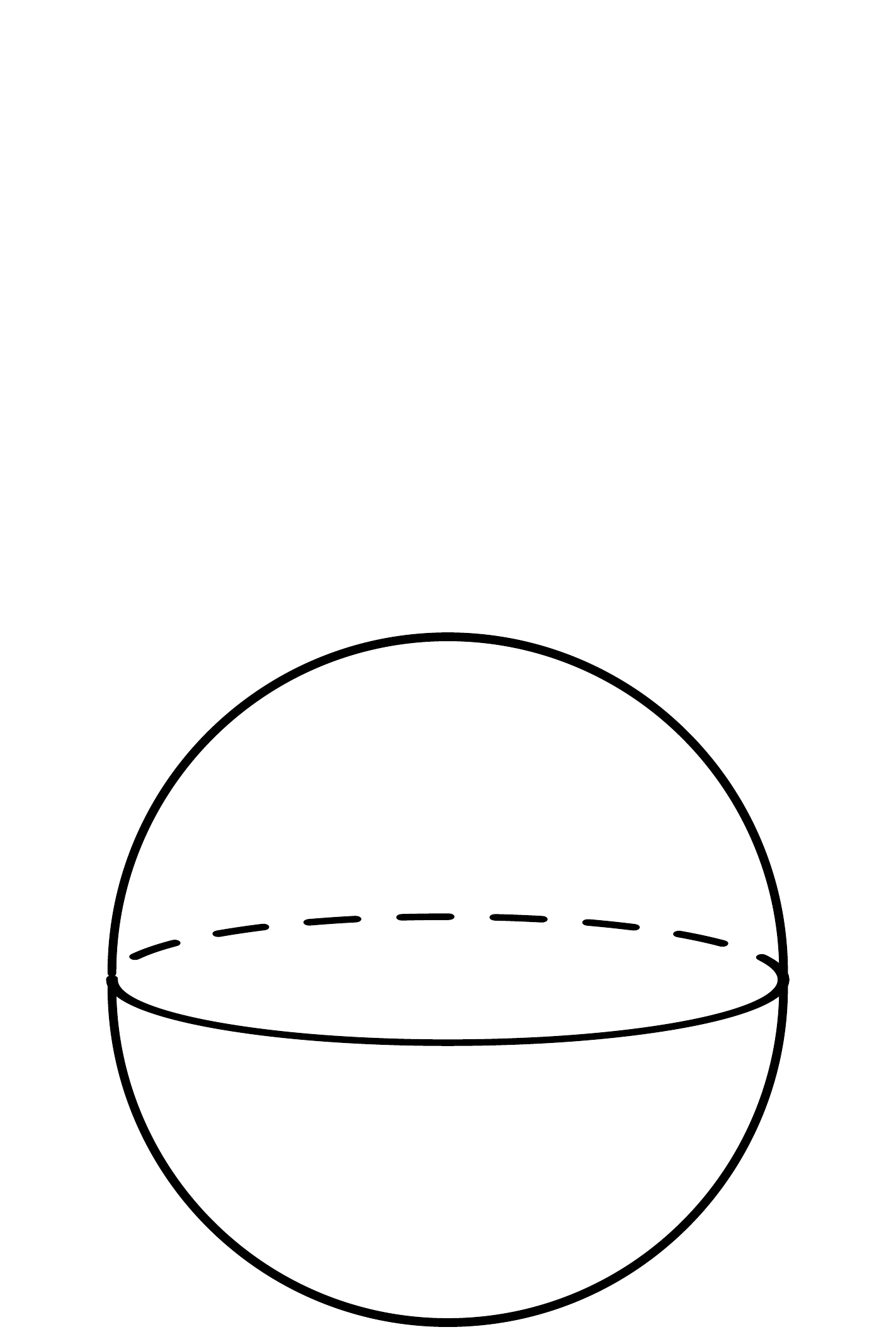}+1\includegraphics[height=30pt]{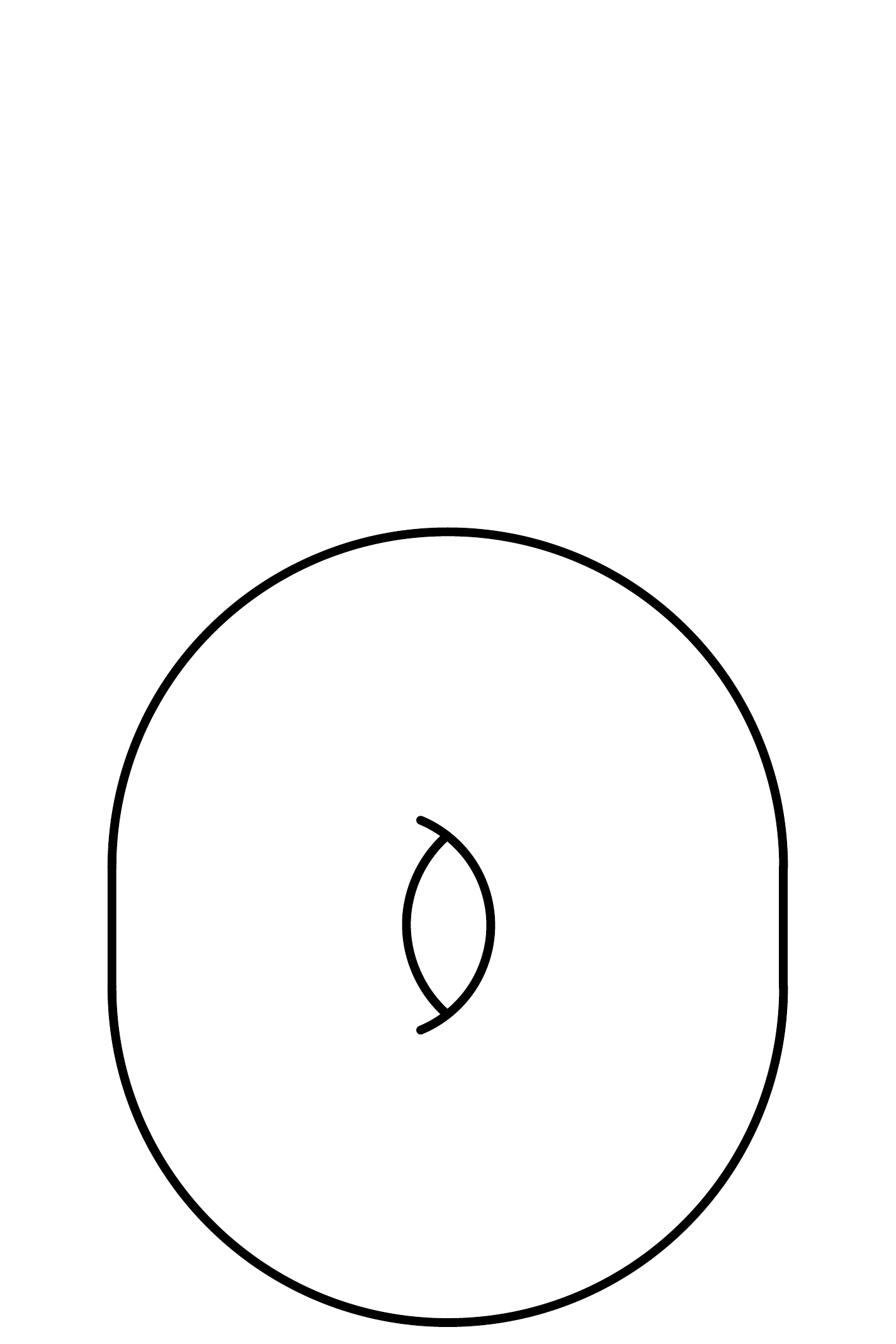}+\frac{11}{12}\includegraphics[height=30pt]{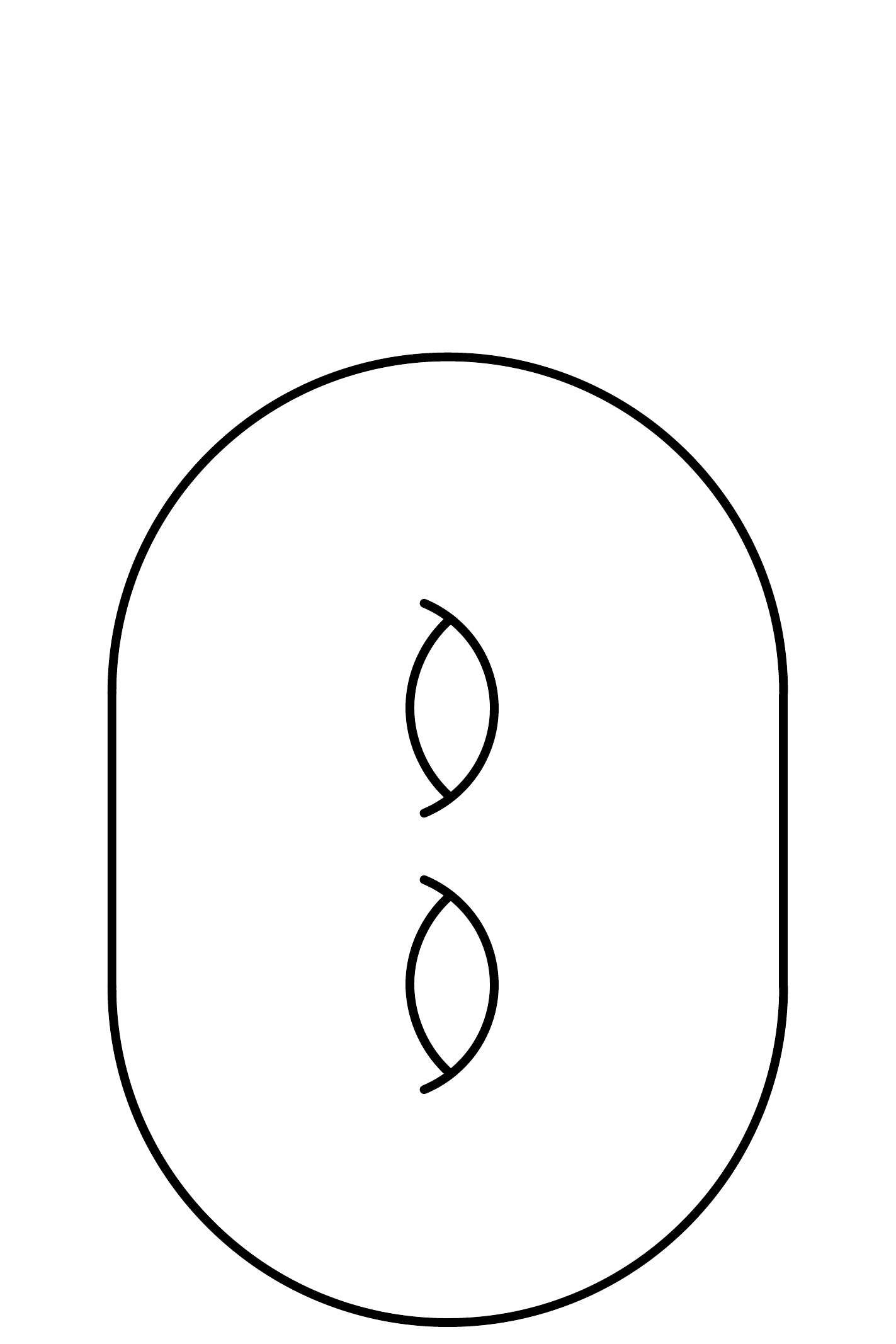}+\frac{17}{18}\includegraphics[height=30pt]{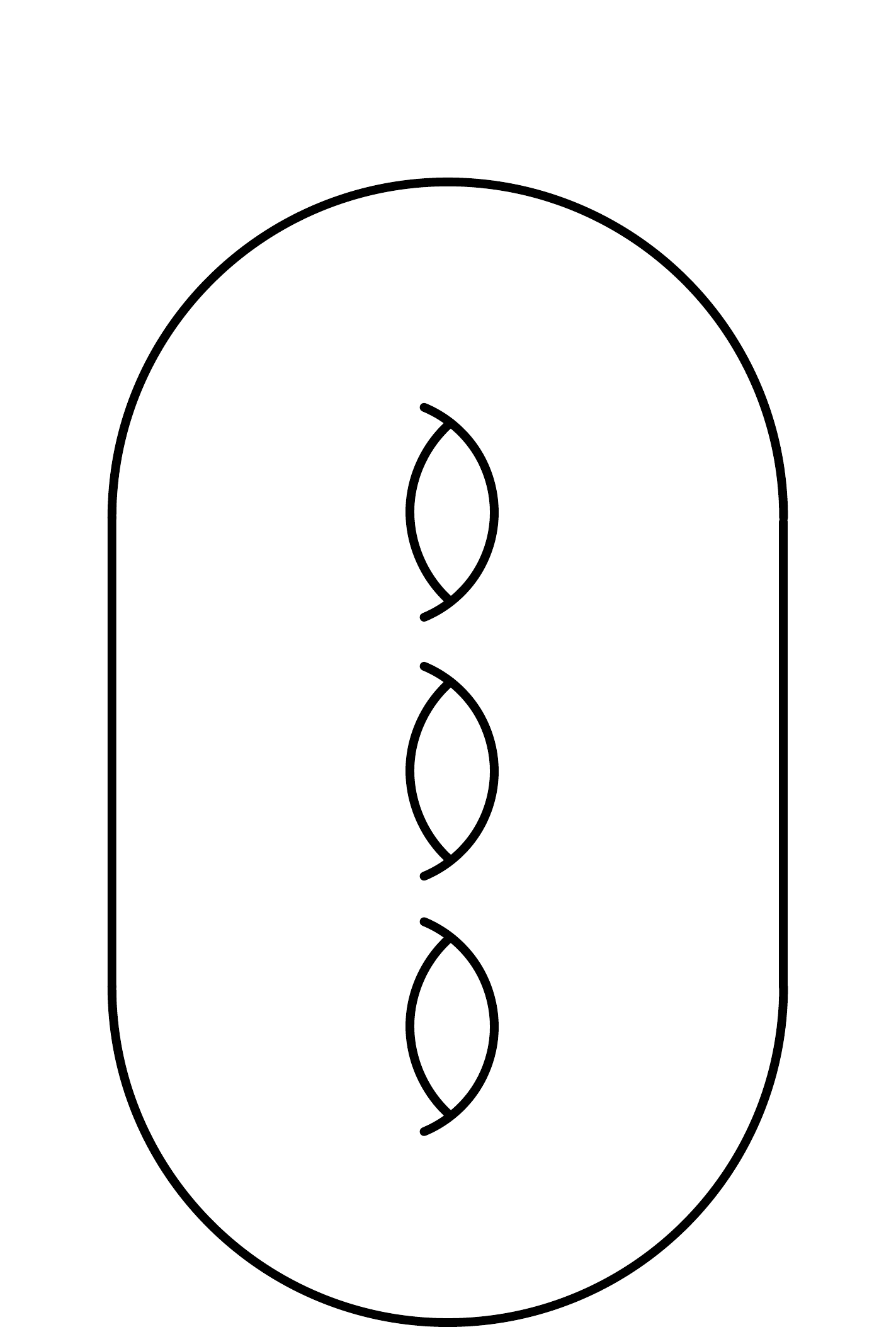}+\ldots
\end{eqnarray*}
a series easily estimated \underline{not} to be square summable.

Further notice that failure to converge is due to sufficient constructive interference. In the above example, $\innerprod{v}{v}$ contains two terms topologically a torus, $\innerprod{\includegraphics[height=30pt]{figures/handle_0.pdf}}{\frac{1}{2}\includegraphics[height=30pt]{figures/handle_1.pdf}}$ and $\innerprod{\frac{1}{2}\includegraphics[height=30pt]{figures/handle_1.pdf}}{\includegraphics[height=30pt]{figures/handle_0.pdf}}$; the coefficients collect to $\frac{1}{2}+\frac{1}{2} = 1$, whose norm squared is 1. Without collection, the contribution to norm squared would be $\frac{1}{2}^2+\frac{1}{2}^2 = \frac{1}{2}$. In fact, it is immediate that if \underline{no} terms in the pairing can be collected (i.e. none are P.L. homeomorphic), then: 
\begin{align*}
  |\innerprod{v}{w}|^2 &= (|\bra{v}|^2)(|\ket{w}|^2)\\
  &:= \left(\sum_ia_i\bbar{a_i}\right)\left(\sum_jb_j\bbar{b_j}\right),
\end{align*}
where $\bra{v} = \sum a_iM_i$ and $\ket{w} = \sum b_jM_j$.

Oppositely, destructive interference reduces $|\innerprod{v}{w}|^2$. In \cite{manifold_pairing}, we found for certain integral homology 3-spheres $\Sigma$ that there were cobounding pairs of homotopy 4-balls $A$ and $B$, $\partial A = \Sigma = \partial B$, so that the following 4-closed manifolds were all (oriented) P.L. homeomorphic to the 4-sphere $S^4$:
\begin{displaymath}
  A\bar{A} \cong A\bar{B} \cong B\bar{A} \cong B\bar{B} \cong S^4
\end{displaymath}
Certainly this means $v = \ket{A}-\ket{B}$ is a light-like vector for $\innerprod{~}{~}_{\Sigma}$.
\begin{displaymath}
  \innerprod{A-B}{A-B} = A\bar{A} - A\bar{B}-B\bar{A}+B\bar{B} = 0\in\M_{\varnothing}^4.
\end{displaymath}

We are not troubled by infinite values for $|\innerprod{~}{~}|^2$ since these will be accorded infinite energy by the action and in our formalism will never be observed. What we would like to know is that Theorem \ref{thm:positivity} for $d\leq 3$ remains valid after completion. Presently, we know this only for $d\leq 2$. For $d=3$, we
\begin{conjecture}
  For all compact 2-dimensional surfaces $S$, the quadratic function on $L^2$ completions, $|\innerprod{~}{~}_{S}^{\wedge}|^2:\M_{S^2}^{\wedge}\rightarrow \R\cup\infty$ has no kernel (i.e. $|\innerprod{v}{v}_S^{\wedge}|^2=0$ implies $v=0$).
\end{conjecture}
\begin{discussion}
  In the original (uncompleted) setting, positivity was proved by producing a (remarkably intricate) ordering of $d$-manifolds ($d\leq 3$) \{P.L. homeo. types of closed $d$-manifold\} $:= \{d\}$ to an ordered set $\Owe_d:\{d\}\xrightarrow{o}\Owe_d$ obeying what is called the \underline{topological Cauchy-Schwartz inequality}: for all $A,B$ with $A\neq B$ and $\partial A = \partial B = S$,
  \begin{displaymath}
    o(A\bar{B}) < \max\{o(A\bar{A}),o(B\bar{B})\}.
  \end{displaymath}
  It is an immediate consequence that, for finite vectors $v_f = \sum_{i=1}^na_iM_i$, $o(M_i\bar{M_j})$ is maximized \underline{only} on the diagonal by terms of the form $a_k\bar{a}_kM_k\bar{M}_k$. Since $a_k\bar{a}_k>0$, these terms cannot cancel when the terms are collected (by P.L. homeomorphism type), thus $\innerprod{v_f}{v_f}\neq 0$, and thus $|\innerprod{v_f}{v_f}|^2\neq 0$.
  
  The argument breaks down for more general $L^2$-convergent sums $v$ because $o(M_i\bar{M}_j)$ may not achieve a maximum at all. If the complexity function \cite{positivity} can be altered to have an ascending chain condition (a.c.c.), 
  all ascending chains have finite length, then the positivity theorem above would automatically extend to the $L^2$-completed pairing:
  \begin{displaymath}
    \hat{\M}_S^d\times\hat{\M}_S^d\xrightarrow{\innerprod{~}{~}_S^{\widehat{~}}}\hat{\M}_{\varnothing}^d\rightarrow \R^+\cup\infty,\hspace{20pt}d=0,1,2,\ldots
  \end{displaymath}
  This is easily done for $d\neq 3$. For example, when $d=0, 1$, and 2, replace the complexity ``number of connected components'' by ``-number of connected components'' and when $d=0$, $|\text{Euler characteristic}|$ by Euler characteristic. So we have:
  \begin{theorem}
    For $d=0,1,$ and 2, $\innerprod{~}{~}^{\wedge}$ is positive, i.e. $\innerprod{v}{v}^{\wedge} = 0$ implies $v=0$.
  \end{theorem}
  
  When $d=3$ our order contains real quantities such as partition functions of graph TQFT \cite{tennets} and finite group TQFT \cite{finitegroup}, which do not lend themselves to an a.c.c. However, the single most important term in the $d=3$ complexity function is -hyperbolic volume. Since the volumes of compact (or even finite volume) hyperbolic manifolds for a well ordered subset of $\R$, the a.c.c. holds and we have:
  \begin{theorem}
    The $L^2$-completed hyperbolic manifold pairing
    \begin{displaymath}
      \hat{\M}_{\text{hyp},S}^3\times\hat{\M}_{\text{hyp},S}^3\rightarrow\hat{\M}_{\text{hyp},\varnothing}^3\rightarrow\R\cup\infty
    \end{displaymath}
    is positive. The subscript ``hyp'' means the ket 3-manifolds $M$ are compact hyperbolic and with totally geodesic boundary = $S$ if $\partial M\neq 0$. The gluings defining the pairing are only homeomorphisms, not necessarily isometries.
  \end{theorem}
\end{discussion}
\begin{remark}
  It is a consequence of Thurston \cite{acylmanifolds} that the geometric hypothesis $M$ is \underline{actually} a topological one: $M$ should be irreducible, boundary irreducible, atoroidal, acylindrical, and with incompressible boundary. Furthermore, if $M$ and $M^{\prime}$ obey these hypotheses with $\partial M = S = \partial M^{\prime}$, then $M\cup_SM^{\prime}$ admits a (unique) hyperbolic metric.
\end{remark}
\begin{remark}
  When the finite group TQFT term plays no role, the conjecture can be proved. This happens when the surface $S=S^2$ or a disjoint union of 2-spheres, or more generally when the kernel $(\pi_1(S)\rightarrow\pi_1(M_i))$ is fixed over all $M_i$ with nonzero coefficients.
\end{remark}

Finally, we prove two theorems about 3-manifolds $S$ for which the $d=4$ pairing is know to contain light-like vectors.
\begin{itemize}
\item[1.] The 3-sphere $S^3$ has a light-like vector in its pairing $\innerprod{~}{~}_{S^3}$.
  \begin{proof}
    According to \cite{fourmanifoldgeometry}, the anti-self-dual Donaldson invariants (ASDD) of closed 4-manifolds $\mathcal{M}$ with $b_2^{\dagger}\geq 3$ are stable with respect to complex blow up, i.e. connected sum with orientation reversed complex projective spaces: $\mathcal{M}\rightarrow\mathcal{M}\sharp\overline{CP}^2_S$.. Since all orientation preserving automorphisms of $S^3$ are isotopic to $id_{S^3}$, $\mathcal{M}\substack{diff\\\cong}\mathcal{M}^{\prime}$ if and only if $(\mathcal{M}_{-},S^3)\substack{diff\\\cong}(\mathcal{M}^{\prime}_{-},S^3)$, the punctured manifolds with boundary are diffeomorphic. Let $\mathcal{M}$ be as above and $\mathcal{M}^{\prime}$ be a smooth closed manifold $s$-cobordant to $\mathcal{M}$, but distinguished from $\mathcal{M}$ by an ASDD. $\mathcal{M}$ may be taken to be a $\mathbb{K}3$ surface and $\mathcal{M}^{\prime}$ its logarithmic transform. Let $\sharp_n (\sharp_{-n})$ denote connected sum with $n$ copies of $\overline{CP}^2$ ($n$ copies of $CP^2$). Define $v_n\in\mcH_{S^3}$ as:
    \begin{displaymath}
      v_n = \mathcal{M}_{-}\sharp_n - \mathcal{M}^{\prime}_{-}\sharp_n.
    \end{displaymath}
    Then $\overline{v}_n = \overline{\mathcal{M}}_{-}\sharp_{-n} - \overline{\mathcal{M}}^{\prime}_{-}\sharp_{-n}$, and so
    \begin{equation}\label{eq:vinnerprod}
      \innerprod{v_n}{v_n} = \mathcal{M}\sharp\mathcal{M}\sharp_n\sharp_{-n} - \mathcal{M}\sharp\mathcal{M}^{\prime}\sharp_n\sharp_{-n} - \mathcal{M}^{\prime}\sharp\mathcal{M}\sharp_n\sharp_{-n} + \mathcal{M}^{\prime}\sharp\mathcal{M}^{\prime}\sharp_n\sharp_{-n}.
    \end{equation}
    
    The four manifolds $\mathcal{M}\sharp\mathcal{M}, \mathcal{M}\sharp\mathcal{M}^{\prime}, \mathcal{M}^{\prime}\sharp\mathcal{M}$, and $\mathcal{M}^{\prime}\sharp\mathcal{M}^{\prime}$ are all $s$-cobordant. Note that $\sharp_n\sharp_{-n}$ is equivalent to connected sum of $\overline{CP}^2, CP^2$ and $(n-1)$ copies of $S^2\times S^2$, and that \cite{topologyfourmanifolds} $s$-cobordism becomes products after a finite stabilization by $S^2\times S^2\times I$. The result is that for $n$ large, the four manifolds in equation \eqref{eq:vinnerprod} are all diffeomorphic. Since $v_n\neq 0$ for all $n$ (by stability of the Donaldson invariants), for $n$ large, $v_n$ is a light-like vector.
  \end{proof}
\item[2.] If some 3-manifold $M$ contains light-like vectors in its pairing, then any 3-manifold of the form $M\# N$ will as well, provided $N$ admits P.L. (or smooth) imbedding $N\subset S^4$ into the 4-sphere.
  \begin{proof}
    Stabilize the terms of a null vector for $M$ as follows: $v=\sum a_iW_i$ to $v^{\prime} = \sum a_i(W_i\natural P)$ where we have taken the boundary connected sum with one of the closed complementary components of $N\subset S^4$: $S^4=P\cup_N Q$. We observe that the composition: 
    \begin{displaymath}
      W_i\hookrightarrow W_i\natural P\hookrightarrow W_i\natural P\cup_{N\backslash B^3}Q \cong W_i,
    \end{displaymath}
    where $\natural$ denotes boundary connected sum, is simply addition of a product collar (an equivalence so
    \begin{align*}
      W_i\natural P\equiv W_j\natural P & ~\Rightarrow~ W_i\natural P\cup_{N\backslash B^3}Q \equiv W_j\natural P\cup_{N\backslash B^3}Q\\
      & ~\Rightarrow~ W_i\equiv W_j.
    \end{align*}
    Thus $v\neq 0$ implies $v^{\prime}\neq 0$. But all terms in $\innerprod{v}{v}$ are each modified by connected sum with $P\bar{P}$, the double of $P$, to yield the corresponding term in $\innerprod{v^{\prime}}{v^{\prime}}$. Thus, term by term, we see $\innerprod{v}{v}=0$ implies $\innerprod{v^{\prime}}{v^{\prime}}=0$.
  \end{proof}
  
  In the construction of formal chains, we encounter 3-manifolds of the form $Y=X\bar{X}$ where $\partial X = T^2$, the 2-torus, with $\innerprod{~}{~}_{X\bar{X}}$ having light-like vectors. Let us understand why this is so. Using the above remark (twice) we can build such $Y$ starting with the Mazur homology 3-sphere $M$ for which light-like vectors were previously found \cite{manifold_pairing}. An example of a $Y = X\bar{X}$, $\partial X = T^2$, can be manufactured as $X\bar{X} = M\#\bar{M}\#(S^1\times S^2)$, where $X = (M\setminus B^3)\cup\text{1-handle}$ and $M$ is the Mazur homology 3-sphere. Because of the connected sum decomposition and the well known facts that $\bar{M}$ and $S^1\times S^2$ is imbedded in $S^4$, $\innerprod{~}{~}_Y$ will have light-like vectors.
  
\item[3.] It is not yet proved that $\innerprod{~}{~}_S$ is positive for any smooth (P.L.) 3-manifold $S$.
\end{itemize}

\section{2-Field Theory (and Higher)}

We briefly explore a formalism for concatenated quantization.
\begin{warning}
  This appendix is schematic, please read skeptically. We intentionally suppress analytic detail to sketch a broad picture. For example, any two linear spaces dense within a third function space are treated interchangeably.
\end{warning}
In the main text, we worked in a Hilbert space $\mcH$ whose kets were formal chains --- object which are built from linear combinations of piecewise linear spaces. Formal chains are themselves closed under $\C$-linear combinations (up to normalization) so $\mcH$ is a ``relinearization'' of an already linear space. It is not a foreign concept. Consider a typical single particle Hilbert space $\mcH = L^2(\R^3)$ that is promoted to (bosonic) Fock space $\mathcal{F}$ via a formal exponentiation, $\mathcal{F} = e^{\mcH}$:
\begin{equation}
  \label{eq:Fock_promo}
  \mathcal{F} = \C\oplus\mcH\oplus\frac{\mcH\otimes\mcH}{2!}\oplus\frac{\mcH\otimes\mcH\otimes\mcH}{3!}\oplus\ldots
\end{equation}
(the denominators are to remind us that symmetrization scales the inner products).

Since \eqref{eq:Fock_promo} describes polynomials in $\mcH$, $\mathcal{F}$ is dense in the linear space of continuous functions (weak topology) on $\mcH$, func$(\mcH,C)$. Furthermore, for wave functions (not basis kets) $\psi_i\in \mcH$, if we relinearize --- notationally place kets around $\psi_i$ ($\ket{\psi_i}$) --- then the expression $\sum_{i=1}^N a_i\ket{\psi_i}\in \C[\mcH]$, the linear space of complex combinations of elements of $\mcH$. Dually, $\sum_{i=1}^N a_i\ket{\psi_i}$ determines a distribution (generalized function) $\sum a_i\delta_{\psi_i}$ on $\mcH$, which again form a dense set in generalized func$(\mcH,C)$. Thus, we regard, for example, $\C[\mcH]\sim\mathcal{F}$ as essentially equivalent since both are dense in func$(\mcH,C)$, and in this sense, view $\mathcal{F}$ as a relinearization of $\mcH$.

A chain, without linear combinations, is analogous to a Feynman diagram, which propogates dynamics in Fock space. A formal chain is a propogation at the next level represented by
\begin{displaymath}
  \mathcal{A}\sim e^{\mathcal{F}}\subset \C^{\mathcal{F}} = \textrm{func}(\mathcal{F},\C).
\end{displaymath}
This is the signature of 2-field theory. In general, $n$-field theory has a Hilbert space at the $n-1$ level above Fock space
\begin{displaymath}
  \mcH = \underbrace{e^{e^{\iddots^{e^{\mathcal{F}}}}}}_{n-1 e\text{'s}},
\end{displaymath}
where unless otherwise noted, parentheses are inserted from top to bottom (e.g. $3^{3^3} = 3^{27}$). Using only dense linear subspaces within functions one may avoid the apparent explosion of cardinality. By passing to appropriate dense subspaces, we can keep all Hilbert spaces separable.

To lay the hierarchical structure bare, we work here with a model case, somewhat simpler than formal chains, in which higher Hilbert spaces are promoted from scalar fields $\phi\in L^2(\R^3,\R)$ which, extending our policy of ignoring \underline{all} analytical distinctions, we may simply write as functions:
\begin{displaymath}
  L^2(\R^3,\C)\sim\C^{\R^3}\hspace{20pt}\text{and}\hspace{20pt}\text{Fock}(L^2(\R^3,\C))\sim e^{\C^{\R^3}}\sim \C^{\C^{\R^3}}
\end{displaymath}
For example, in 2-QFT, operators will act on 2-Fock:
\begin{displaymath}
  \C^{\C^{\C^{\R^3}}}\hspace{20pt}\left(\sim e^{e^{\R^3}}\right)
\end{displaymath}
the linear space spanned by wave functionals of multiparticle wave functions $\psi$, $\psitwo = \sum b_i\ket{\psi_i}$, i.e. \underline{non-linear} functionals of multiparticle wave functionals.

If quantum field theory (QFT) computes some unitary fuzziness around classical trajectories, then it is the purpose of 2-QFT to compute some fuzziness around the unitary evolution of a QFT (which is itself unitary but only at a higher level.) To illustrate the scope of the idea, we will briefly touch on the ``easier'' and ``harder'' case of $n$ quantum mechanics and $n$-string field theory. Regarding the terminology, $n$-QFT with its stratified structure is reminiscent of $n$-categories; we have kept the notation parallel. Finally, note the index $n$ could also run over the ordinals but we have no use for that here.

Possible applications (besides to the body of this paper) include: 1) investigate models at high energy in which unitarity is only emergent, and 2) construct evective hierarchical description of strongly interacting low energy physics.

The constituents of an $n$-QFT are named in table \ref{tab:QFT-constituents}.
\begin{table}[htbp]
  \centering
  \begin{tabular}{|l|l|}
    \hline
    $n$-Hilbert space &\\ \hline
    $n$-Fock space & \\ \hline
    $n$-$c_{\bbk}^+, n$-$c_{\bbk}$ & \parbox{200pt}{$n+1^{\text{st}}$ quantized operators (to be consistent with the terminology of second quantization)}\\ \hline
    $n$-$H$ & $n$-Hamiltonian\\ \hline
    $n$-$U$ & unitary evolution at level $n$\\ \hline
    $n$-$\mathcal{L}$ & $n$-Lagrangian\\ \hline
    $n$-$S$ & $n$-action\\ \hline
    \multicolumn{2}{|c|}{But there are \underline{only} ordinary 1-observables}\\ \hline
  \end{tabular}\vspace{0.5cm}
  \caption{The constituents of an $n$-QFT.}
  \label{tab:QFT-constituents}
\end{table}

Observables are not really constituents wholly within quantum theory, but a bridge to the classical world, and so will be defined on the familiar level. Observables may include field strength (curvature), charge, and momentum.

Using this very crude notation, let's describe the Hilbert space for quantum mechanics, field theory, quantum field theory, string quantum field theory, nonlinear sigma models, and gauge field theory. The Hilbert space for QM is $\C^{\R}$, or more precisely $L^2(\R)$ or $L^2(\R^n) = \otimes_nL^2(\R)$. Now dropping all analytic detail, the space for field theory (FT) is $\R^{\R^3}$ for, say, a real field $\phi\in\R^{\R^3}$. The Hilbert space for QFT is Fock space $\C^{\R^{\R^3}}$, with wave functional $\psi = \sum a_i\ket{\phi_i}, \sum |a_i|^2 = 1$. The Hilbert space for 2-QFT is $\C^{\C^{\R^{\R^3}}}$, with $\psitwo = \sum a_i\ket{\psi_i}, \sum |a_i|^2 = 1$. 

To get string-QFT from QFT, you fiddle around at the ``\underline{top}'' of the tower:
\begin{center}
  \begin{tabular}{c c c}
    $\C^{\R^{\R^3}}$ & $\leadsto$ & $\C^{\R^{M^{S^1}}}$\\
    \parbox{100pt}{Ordinary Fock space of a real scalar field.} & & \parbox{100pt}{Stringy Fock space of a real scalar field.}
  \end{tabular}
\end{center}
$M$ is an 11-manifold, $S^1$ a circle which sweeps out a world sheet $\Sigma$ in time. Both examples can be promoted to the 2-level simply by placing a ``$\C$'' at the lower left of the stack.

Of course, QFT's come in minor variations:
\begin{center}
  \begin{tabular}{c c c c c}
    (a) & $\C^{\R^{\R^3}}$ & $\leadsto$ & $\C^{X^{\R^3}}$ & \parbox{100pt}{$X$, a manifold, is a ``nonlinear sigma model''}\vspace{0.4cm}\\
    (b) & $\C^{\R^{\R^3}}$ & $\leadsto$ & $\C^{\text{sections of a $G$-principle bundle over $\R^3$}}$ & \parbox{100pt}{is a gauge field theory}
  \end{tabular}
\end{center}
Case (a) replacing $\R$ by $X$ promotes a real scalar to a nonlinear sigma model. Case (b) functions are replaced by sections to yield a gauge field theory.

2-field theory adds a $\C$ at the bottom of the tower, so wave functionals are of the form $\psi\in\C^{\R^{\R^3}}$ and 2-wave functionals are of the form $\psitwo\in\C^{\C^{\R^{\R^3}}}$. 3-field theory treats 3-wave functionals $\psithree\in\C^{\C^{\C^{\R^{\R^3}}}}$, and so on.

The usual passage \footnote{Between Hamiltonian and Lagrangian formalisms.} between $H$ and $\mathcal{L}$, the ``path integral formulation of QFT,'' is based on the  ability to \underline{restrict} fields on $\R^4$ to $\R^3\times t$. Let's see how this works set theoretically. On adding a functional level, \underline{inclusion} and \underline{restriction} alternate.
\begin{center}
  \begin{tabular}{rcl l}
    $\R^3\times t$ & $\hookrightarrow$ & $\R^4$ & (inclusion of spaces)\\
    $\R^{\R^3\times t}$ & $\leftarrow$ & $\R^{\R^4}$ & (restriction of fields)\\
    $\C^{\R^{\R^3\times t}}$ & $\hookrightarrow$ & $\C^{\R^{\R^4}}$ & (inclusion of 2-fields)\\
    $\C^{\C^{\R^{\R^3\times t}}}$ & $\leftarrow$ & $\C^{\C^{\R^{\R^4}}}$ & (restriction of 3-fields)
  \end{tabular}
\end{center}

It is important to be able to restrict fields to time slices, but you will notice that the restriction maps exist naturally only for $k$-fields, $k$ odd. However, for $k$ even, it is possible to pass to the linear duals $V\leftrightarrow V^{\ast}$, and ignore the analytic issue of the dual being a much larger space.

All books on QFT derive the evolution $U$ from the Hamiltonian $H$ as a ``path integral'' over fields $\phi$ weighted by $e^{-iS(\phi)}$, $S$ the action of an ordinary Lagrangian, i.e. a 1-Lagrangian. Given, say, a 2-Hamiltonian 2-$H$, there will be a 2-Lagrangian, 2-$\mathcal{L}$, constructed as a ``path integral'' over 2-fields $\phitwo\in\C^{\R^{\R^4}}$ weighted by $e^{-i(2\text{-}S(\phitwo))}$. Formally, this 2-evolution 2-$U$ is \underline{perfectly} unitary. The 2-evolution naturally ``drags along'' an \underline{ordinary} 1-level linear evolution but this is \underline{not} unitary and only becomes unitary in a certain \underline{squeezed} limit (see below). Consider table \ref{tab:higher_field_theory}.
\begin{table}[htbp]
  \centering
  {\footnotesize
  \begin{tabular}{c|c|c}
    \underline{field} & \underline{2-field} & \underline{3-field}\\
    $\phi\in\R^{\R^4}$ & $\phitwo\in\C^{\R^{\R^4}}$ & $\phithree\in\C^{\C^{\R^{\R^4}}}$\\
    &&\\

    \hline

    &&\\
    \underline{wave functional} & \underline{2-wave functional} & \underline{3-wave functional}\\

    \parbox{105pt}{\centering$\psi(\phi)\in\C^{\R^{\R^4}}$\\$\psi = \sum a_i\ket{\phi_i}, \sum|a_i|^2 = 1$} & \parbox{105pt}{\centering$\psitwo(\phitwo)\in\C^{\C^{\R^{\R^4}}}$\\$\psitwo = \sum a_i\ket{\phitwo_i}, \sum|a_i|^2 = 1$} & \parbox{105pt}{\centering$\psithree(\phithree)\in\C^{\R^{\R^4}}$\\$\psithree = \sum a_i\ket{\phithree_i}, \sum|a_i|^2 = 1$}\\
    &&\\

    \hline

    &&\\
    \underline{Fock($H$) = $F$} & \underline{2-Fock($H$) = 2-$F$} & \underline{3-Fock($H$) = 3-$F$}\\

    \parbox{100pt}{$=\C^{\R^{\left.\R^3\right\}H}}$\\$=e^H$\\$=\C\oplus H\oplus (H\oplus_S H)\oplus\ldots$} & \parbox{100pt}{=Fock(Fock($H$))\\$=\C^{\C^{\R^{\R^3}}_{\rotatebox{-20}{$\leftarrow$}}}$\\$=\C\oplus F\oplus (F\oplus_S F)\oplus\ldots$} & \parbox{100pt}{=Fock$^3(H)$\\$=\C^{\C^{\C^{\R^{\R^3}}_{\rotatebox{-20}{$\leftarrow$}}}_{\rotatebox{-20}{$\leftarrow$}}}$\\$=\C\oplus \text{2-}F\oplus (\text{2-}F\oplus_S \text{2-}F)\ldots$}\\
    &&\\

    \hline

    &&\\
    \underline{$\mathcal{L}^{\lambda}(\phi)$ and $S^{\lambda}$} & \underline{2-$\mathcal{L}^{0,\lambda}(\phitwo)$ and 2-$S^{0,\lambda}$} & \underline{3-$\mathcal{L}^{0,\lambda}$}\\
    
    \parbox{120pt}{$\int dx^4\left((\nabla\phi)^2-\frac{m^2}{2}\phi^2-\frac{\lambda}{4!}\phi^4\right)$\\for $\lambda = 0$, solve in $k$-space\\$\ket{0}$ and $c_{k}^{\dagger}$ generate $F$} & \parbox{120pt}{$\int \mathcal{D}\phi\left((\nablabar\phitwo)_{\phi}^2-\frac{m^2}{2}|\phitwo|^2-\frac{\lambda}{4!}|\phitwo|^4\right)$\\{\tiny using translations in $\R^{\R^4}, \bbk\in(\R^{\ast})^{\R^4}$}\\for $\lambda = 0$, solve in $\bbk$-space\\$\ket{\ket{0}}$ and 2-$c_{\bbk}^{\dagger}$ generate 2-$F$,\\$\left[c_{\bbk_i}^{\dagger},c_{\bbk_j}\right]_{\xi} = \delta_{ij}$} & \parbox{120pt}{$\int \mathcal{D}\phitwo\left((\nablabar\phithree)_{\phitwo}^2-\frac{m^2}{2}|\phithree|^2-\frac{\lambda}{4!}|\phithree|^4\right)$\\$\ket{\ket{\ket{0}}}$ and 3-$c_{\bbk}^{\dagger}$ generate 3-$F$\\$\left[c_{\bbk_i}^{\dagger},c_{\bbk_j}\right]_{\xi} = \delta_{ij}$}\\
    &&

  \end{tabular}
}
  \caption{Higher field theory. Arrows indicate levels which may be removed by squeezing higher order wave functionals.}
  \label{tab:higher_field_theory}
\end{table}
Here, $\nablabar$ is the directional derivative at the next level:
\begin{align*}
  \nablabar_{\phi^{\prime}}\phitwo|_{\phi} &= \faktor{(\phitwo(\phi-\Delta\phi^{\prime})-\phitwo(\phi))}{\|\Delta\phi^{\prime}\|_{L^2}}\\
  \left(\nablabar\phitwo|_{\phi}\right)^2 &= \left(\int_{\|\phi^{\prime}\|_{L^2}=1}d\phi^{\prime}~\|\nablabar_{\phi^{\prime}}\phitwo|_{\phi}\|^2\right)^{\frac{1}{2}}.
\end{align*}
Parallel formulae give $\nablabar\phithree|_{\phitwo}$, and so on. We may also introduce a gradient $\nablabar_x$ with fewer parameters (coming from a lower level). In the ``squeezed context'' explained below, $\nablabar_{\phi}$ may be replaced by $\nablabar_x$. Introduce the ``small gradient'' $\nablabar_x$ based on $x\in\R^4$ (not $\R^{\R^4}$) translation. That is, define $\phi_{\Delta x}(x):=\phi(x-\Delta x)$, then define $\nablabar_x\phitwo|_{\phi} = \faktor{(\phitwo(\phi_{\Delta x})-\phitwo(\phi))}{\Delta x}$, where $x\in\R^3$ or $x\in\R^4$, depending on context. Define a family of 2-actions for $c>0$ by
\begin{align*}
  \text{2-}\mathcal{L}^{c,\lambda} = |\nablabar\phitwo|_{\phi}|^2 - \frac{m^2}{2}|\phitwo|^2-\frac{\lambda}{4!}|\phitwo|^4 - c(\langle|\phitwo|^2\rangle-\langle\phitwo\rangle^2),\\
  \text{ where the last term is} ~~c\left(\int\mathcal{D}\phi|\phitwo(\phi)|^2-\left|\int\mathcal{D}\phi\phitwo(\phi)\right|^2\right).
\end{align*}
As $c\rightarrow\infty$, the 2-physics of 2-$\mathcal{L}^{c,\lambda}$ is expected to concentrate on 2-fields, or ``rules,'' $\phitwo$ which are nearly Dirac, i.e. $\phitwo\approx\delta\phi$, for some $\phi$.

In the $c\rightarrow\infty$ limit, only $x\in\R^4$ translations have bounded energy among general variations, so $\nablabar$ is expected to reduce to $\nablabar_x$. This effectively deletes the $\C$ with the arrow next to it in table \ref{tab:higher_field_theory}. Thus, $c\rightarrow\infty$ ``squeezes'' 2-QFT back to ordinary QFT with $\frac{1}{c}$ the small parameter.

Similarly, let us define a 3-action
\begin{displaymath}
  \text{3-}\mathcal{L}^{c,\lambda} = |\nablabar\phithree|_{\phitwo}|^2 - \frac{m^2}{2}|\phithree|^2-\frac{\lambda}{4!}|\phithree|^4-\text{squeezing term},
\end{displaymath}
where the squeezing term --- conceptually --- is given by
\begin{displaymath}
  \text{const}\min_{\phi_0}\underbrace{\int\mathcal{D}\phitwo|\phithree(\phitwo)-\phitwo(\phi_0)|^2}_{f(\phi)}.
\end{displaymath}
Setwise, evaluation includes \{fields\}$\subset\C^{\C^{\{\text{fields}\}}}$ by $\phithree_{\phi}(\phitwo):=\phitwo(\phi)$. An analytically more convenient squeeze term is given by
\begin{displaymath}
  \beta^{\prime}\int\mathcal{D}\phi~e^{-\beta f(\phi)},
\end{displaymath}
where $\beta^{\prime},\beta\gg 0$. As with 2-fields, we now expect that as $\beta\rightarrow\infty$, the ``physics'' of 3-fields will squeeze down to evaluation of 3-fields of the form $\phithree_{\phi}(\phitwo) = \phitwo(\phi)$, i.e. a 1-field $\phi$. It is also expected that $\int\mathcal{D}\phitwo|\nablabar\phithree|_{\phitwo}|^2\leadsto\int dx^4|\nabla\phi|^2$, similarly for the mass and interaction terms.

Since 3 is odd, 3-fields naturally \underline{restrict} to ``time slices'':
\begin{displaymath}
  \C^{\C^{\R^{\R^3\times t}}}\xleftarrow{\text{restriction}}\C^{\C^{\R^{\R^4}}}.
\end{displaymath}
The path integral allows the formal derivation of a unitary evolution 3-$U$ starting from a Hermitian 3-Hamiltonian 3-$H$. This can also be accomplished at the 2-level by passing to linear duals:
\begin{displaymath}
  \left(\C^{\R^{\R^3\times t}}\right)^{\ast}\xleftarrow{\text{restriction}}\left(\C^{\R^{\R^4}}\right)^{\ast}
\end{displaymath}

Two final points should be explained: how the evolution at level $n$ drags along a linear but not-quite-unitary evolution at all levels $m<n$, and what observables in $n$-QFT are. For both of these, we must define the ``ket erasure'' maps $\alpha_n$.

``Erase kets and extend linearly'' defines a linear map:
\begin{align*}
  \raisebox{15pt}{$
    n
    \hspace{-20pt}
    \rotatebox{-45}{$\displaystyle 
      \raisebox{-4pt}{$\left\{\parbox{1pt}{\vspace{50pt}}\right.$}\hspace{-7pt}
      \begin{matrix}
        \rotatebox{45}{$\C^{\R^{\R^3}}$}\\
        \vdots\\
        \rotatebox{45}{$\C$}
      \end{matrix}
      $}
    $} 
  &\xrightarrow{\quad\alpha_n\quad}\hspace{15pt}
  \raisebox{15pt}{$
    n-1
    \hspace{-20pt}
    \rotatebox{-45}{$\displaystyle 
      \raisebox{-4pt}{$\left\{\parbox{1pt}{\vspace{50pt}}\right.$}\hspace{-7pt}
      \begin{matrix}
        \rotatebox{45}{$\C^{\R^{\R^3}}$}\\
        \vdots\\
        \rotatebox{45}{$\C$}
      \end{matrix}
      $}
    $},\\
  \sum a_i\ket{\phi_i} & \xrightarrow{\quad\alpha_n\quad} \sum \tilde{a}_i\phi_i^n,\text{ where }\tilde{a}_i =
  \begin{cases}
    \bar{a}_i & $n$\text{ odd}\\
    a_i & $n$\text{ even}.
  \end{cases}
\end{align*}
There is also the familiar evaluation map $e_{n-2}$, given by
\begin{align*}
  \raisebox{15pt}{$
    n-2
    \hspace{-20pt}
    \rotatebox{-45}{$\displaystyle 
      \raisebox{-4pt}{$\left\{\parbox{1pt}{\vspace{50pt}}\right.$}\hspace{-7pt}
      \begin{matrix}
        \rotatebox{45}{$\C^{\R^{\R^3}}$}\\
        \vdots\\
        \rotatebox{45}{$\C$}
      \end{matrix}
      $}
    $}
  &\xrightarrow{\quad e_{n-2}\quad}\hspace{15pt}
  \raisebox{15pt}{$
    n
    \hspace{-20pt}
    \rotatebox{-45}{$\displaystyle 
      \raisebox{-4pt}{$\left\{\parbox{1pt}{\vspace{50pt}}\right.$}\hspace{-7pt}
      \begin{matrix}
        \rotatebox{45}{$\C^{\R^{\R^3}}$}\\
        \vdots\\
        \rotatebox{45}{$\C$}
      \end{matrix}
      $}
    $},\\
  e_{n-2}\phi_0^{n-2}(\phi^{n-1}) &= \phi^{n-1}(\phi_0^{n-2})
\end{align*} 

Formally, $\alpha_{n-1}\circ\alpha_n\circ e_{n-2} = \text{id}_{n-2}$, up to an infinite constant.
\begin{proof}
  If $\phithree(\phitwo) = \phitwo(\phi_0)$, then $\phithree = \sum_i\phitwo_i(\phi_0)\ket{\phitwo_i}$, and so 
  \begin{displaymath}
    \alpha_2\phithree = \sum_i\bar{\phitwo}_i(\phi_0)\phitwo_i = \sum_ib_{i0}\phitwo_i,
  \end{displaymath}
  where we have written $\phitwo_i = \sum_jb_{ij}\ket{\phi_j}$. Then
  \begin{align*}
    \alpha_1\alpha_2\phitwo &= \sum_{i,j}b_{i0}\bar{b}_{ij}\phi_j\\
    &=\sum_i b_{i0}\bar{b}_{i0}\phi_0 + \sum_{i,j\neq 0}b_{i0}\bar{b}_{ij}\phi_j\\
    &=\infty(\phi_0)+\sum_{j\neq 0}0\phi_j,
  \end{align*}
  where zero on the last line comes from the symmetry of the sum.
\end{proof}

Measurement will merely be by a Hermitian operator $\Owe$ on ordinary Fock space $F = \C^{\R^{\R^3}}$. The protocol is ``reduce, then observe'': $\psi^n\xrightarrow{\alpha_n}\psi^{n-1}\rightarrow\cdots\rightarrow\psi^1$, and observe $\lambda_i$ of $\Owe$ with probability $|a_i|^2$, where $\psi^1 = \sum a_i\psi_i^1$, where $\{\psi_i^1\}$ is an eigen-basis for $\Owe$. Suppose $n$ is odd (and if not, pass to the dual). Then the successive evaluation maps promote $\psi^1$ back to level $n$ where $n$-$U$ evolves the promoted wave function until the next measurement by some $\Owe^{\prime}$ also acting on ordinary Fock space $F=\C^{\R^{\R^3}}$. If the level $n$-evolution is sufficiently squeezed, then $n$-$U$ evolves very nearly within evaluation subspace $F\subset n$-$F$ and exact unitarity on $n$-$F$ implies that a nearly exact unitarity will be observed on $F$.

\underline{Final notes and examples}: The level 2 creation operators, 2-$c^{\dagger}_{\bbk}$, create a set of states of varying particle numbers, e.g. the set may contain a scalar, a singleton of momentum $k$, linear combinations of pairs $(k^{\prime}\otimes_Sk^{\prime\prime})$, and so on. In other words, 2-$c_{\bbk}$ creates an arbitrary element of Fock space. 3-$c_{\bbk}$ creates sets of sets of states, i.e. an arbitrary element in 2-Fock space, and so on.

Unitarity of the $U$ is derived from the Lagrangian $\mathcal{L}$: $S = \int\mathcal{L}~~~$ reverses sign (via complex conjugation) with reversal of orientation of slab $X\times [0,1]$:
\begin{displaymath}
  U_{ij} = \int_0^1e^{iS} = \bbar{\int_1^0e^{iS}} = \bbar{U}_{ji}^{-1}.
\end{displaymath}
This argument is formally identical at level $n$.

Although this appendix has focused on $n$-QFT, one may promote the discussion to 2-string field theory or, in the other direction, cut the discussion down to $n$-quantum mechanics $n$-QM. By linearizing the top of the tower, we can produce 2-string FT:
\begin{eqnarray*}
  \int_{\text{all 2D field theories}}S &\longrightarrow& \int_{\text{all 2D 2-field theories}}\text{2-}S
\end{eqnarray*}
\begin{note}
  Among 2D field theories are nonlinear sigma-models of the form: (a string action, $S$)$\in\R^{M^{\Sigma}}$. Similarly, among 2D 2-field theories are function on nonlinear sigma-modules of the form: (a 2-string action, 2-$S$)$\in\C^{\R^{M^{\Sigma}}}$. Presumably, these may be important in evaluating the integral perturbatively but are not exhaustive.
\end{note}
Now for 2-QM: consider a wave function $\psi\in H = \C^{\R^{\text{pt.}}}$ and a 2-wave function $\psitwo\in$2-$\mathcal{H} = \C^{\C^{\R}}$. To get a picture of how 2-QM can work, consider as a model for part of 2-$\mcH$ consisting of $\psitwo\in$2-$\mcH$, made from just two Dirac functions,
\begin{displaymath}
  \psitwo = \frac{\sqrt{2}}{2}\ket{\psi_1}+\frac{\sqrt{2}}{2}\ket{\psi_2},
\end{displaymath}
where we think of $\psi_i$ as the amplitude for particle $i$ in position $x_i$.

Choose a 2-Hamiltonian analogous to an ordinary Hamiltonian for a ``molecule'' moving in potential:
\begin{displaymath}
  \text{2-}H = \frac{1}{2}p_{\delta x_1}^2 + \frac{1}{2}p_{\delta x_2}^2 + V(x_1-x_2) + \frac{x_1^2}{2} + \frac{x_2^2}{2} + \frac{\lambda}{4!}x_1^4 + \frac{\lambda}{4!}x_2^4,
\end{displaymath}
where $p_{\delta x_i} = i\partial_{x_i}$ acts inside kets, so for example, $p_{\delta x_1+\delta x_2}\psitwo = \frac{\sqrt{2}}{2}\ket{i\partial_{x_1}\psi_1}+\frac{\sqrt{2}}{2}\ket{i\partial_{x_2}\psi_2}$.

Passing to a center of mass coordinate $\frac{x_1+x_2}{2}$, in the case where $\lambda = 0$, we have that
\begin{displaymath}
  \text{2-}H = \frac{1}{2}p_{\delta x_1+\delta x_2}^2 + \left(\frac{x_1+x_2}{2}\right)^2 + \frac{1}{2}p_{\delta x_1-\delta x_2}^2 + \left(\frac{x_1-x_2}{2}\right)^2 + V(x_1-x_2),
\end{displaymath}
so the center of mass is still SHO, and the evolution is actually unitary at the 1-level.

If $\lambda\neq 0$, the center of mass wave function at the 1-level is induced by ket erasure:
\begin{displaymath}
  \phi(c) = \frac{\int dx_1[\phi_1(x_1)+\phi_2(2c-x_1)]}{\text{norm}}
\end{displaymath}
does not evolve unitarily. I would like to thank Israel Klitch for suggesting this example.

Formal manipulations in 2-QFT, e.g. of (perturbed) Gaussian integrals, at higher levels will produce analogs of many familiar calculational features such as 2-ghosts, 2-Hubbard Stratonovich, and 2-perturbative expansions.

\bibliographystyle{plain}
\bibliography{QGviaMP}

\end{document}